\documentclass[11pt]{article}

\usepackage[margin=1in]{geometry}
\usepackage[utf8]{inputenc}
\usepackage[T1]{fontenc}
\usepackage{lmodern}
\usepackage{microtype}
\usepackage{amsmath,amssymb,amsthm,mathtools}
\usepackage[dvipsnames]{xcolor}
\usepackage{todonotes}
\usepackage{enumitem}
\usepackage{needspace}
\usepackage{xspace}
\usepackage{tikz}
\usepackage{algorithm}
\usepackage[noend]{algpseudocode}
\usepackage[colorlinks=true,linkcolor=BrickRed,citecolor=OliveGreen,urlcolor=MidnightBlue]{hyperref}
\usepackage{aliascnt}
\usepackage[nameinlink,noabbrev]{cleveref}
\makeatletter
\providecommand{\theHALG@line}{}
\renewcommand{\theHALG@line}{\thealgorithm.\arabic{ALG@line}}
\makeatother

\theoremstyle{plain}
\newtheorem{theorem}{Theorem}
\newaliascnt{lemma}{theorem}
\newtheorem{lemma}[lemma]{Lemma}
\aliascntresetthe{lemma}
\newaliascnt{proposition}{theorem}
\newtheorem{proposition}[proposition]{Proposition}
\aliascntresetthe{proposition}
\newaliascnt{corollary}{theorem}
\newtheorem{corollary}[corollary]{Corollary}
\aliascntresetthe{corollary}
\theoremstyle{definition}
\newaliascnt{definition}{theorem}
\newtheorem{definition}[definition]{Definition}
\aliascntresetthe{definition}
\newaliascnt{problem}{theorem}
\newtheorem{problem}[problem]{Problem}
\aliascntresetthe{problem}
\theoremstyle{plain}
\newaliascnt{observation}{theorem}
\newtheorem{observation}[observation]{Observation}
\aliascntresetthe{observation}
\theoremstyle{definition}
\newaliascnt{remark}{theorem}

\aliascntresetthe{remark}
\theoremstyle{plain}
\newaliascnt{claim}{theorem}
\newtheorem{claim}[claim]{Claim}
\aliascntresetthe{claim}

\crefname{theorem}{Theorem}{Theorems}
\crefname{lemma}{Lemma}{Lemmas}
\crefname{proposition}{Proposition}{Propositions}
\crefname{corollary}{Corollary}{Corollaries}
\crefname{definition}{Definition}{Definitions}
\crefname{problem}{Problem}{Problems}
\crefname{observation}{Observation}{Observations}
\crefname{remark}{Remark}{Remarks}
\crefname{claim}{Claim}{Claims}
\crefname{section}{Section}{Sections}
\crefname{subsection}{Section}{Sections}
\crefname{subsubsection}{Section}{Sections}
\crefname{appendix}{Appendix}{Appendices}
\crefname{algorithm}{Algorithm}{Algorithms}
\crefname{figure}{Figure}{Figures}
\crefname{line}{Line}{Lines}

\usepackage{thmtools}

\newcommand{\dist}{\operatorname{dist}}

\newcommand{\eps}{\varepsilon}
\newcommand{\polylog}{\operatorname{polylog}}
\newcommand{\TC}{\operatorname{TC}}
\newcommand{\core}{\textnormal{\textsc{BoundedSSSP}}\xspace}
\newcommand{\batchdijkstra}{\textnormal{\textsc{BatchedDijkstra}}\xspace}

\newcommand{\apsp}{\textnormal{\textsc{ApproxAPSP}}\xspace}

\newcommand{\Rin}{R^-}
\newcommand{\Rout}{R^+}
\newcommand{\Rboth}{R^{\pm}}
\newcommand{\E}{\mathbb{E}}

\definecolor{earlyblue}{RGB}{51,102,153}
\definecolor{lategreen}{RGB}{49,130,89}
\definecolor{boxgray}{RGB}{247,248,250}

\usepackage{todonotes}

\definecolor{garytodo}{HTML}{EAF3FF}
\definecolor{sepidehtodo}{HTML}{F2ECFA}
\definecolor{jakubtodo}{HTML}{FFF3E4}

\title{Semi-Streaming Reachability and Shortest Paths \\ in $n^{1/3+o(1)}$ Passes}
\providecommand{\email}[1]{\href{mailto:#1}{\nolinkurl{#1}}}
\author{Gary Hoppenworth\thanks{Toyota Technological Institute at Chicago; \email{garytho@ttic.edu}. This work was done while the author was an intern at Microsoft Research.}
  \and George Z.\ Li\thanks{Carnegie Mellon University; \email{gzli@andrew.cmu.edu}. This work was done while the author was an intern at Microsoft Research.}
  \and Sepideh Mahabadi\thanks{Microsoft Research; \email{smahabadi@microsoft.com}.}
  \and Jakub Tarnawski\thanks{Microsoft Research; \email{jatarnaw@microsoft.com}.}}
\date{}
\begin{document}
\pagenumbering{roman}
\maketitle
\thispagestyle{empty}
\begin{abstract}

 In the semi-streaming model, the edges of an $n$-vertex graph arrive in a stream, and an algorithm makes multiple passes over that stream while using only $\widetilde{O}(n)$ space. The number of passes needed to compute single-source reachability in directed graphs is a long-standing open question in streaming algorithms, going back to Henzinger, Raghavan, and Rajagopalan~\cite{HRR98}. 
The best known semi-streaming algorithm uses  $n^{1/2+o(1)}$ passes due to Jambulapati, Liu, and Sidford~\cite{JLS19, AJJST22}, while the best lower bound is $\Omega(\log n/\log\log n)$ passes due to Guruswami and Onak~\cite{GO16}. 

We give streaming algorithms for computing single-source reachability and shortest
paths on $n$-vertex directed graphs using $\widetilde{O}(n)$ space and $n^{1/3+o(1)}$ 
passes. Our algorithms are based on \emph{implicit hopsets}, which are large hopsets that can be represented implicitly in $\widetilde{O}(n)$ space. We introduce this notion and give new constructions for implicit hopsets with small hopbound.
Along the way, we obtain new work--depth tradeoffs for parallel  shortest paths in directed graphs, improving the recent result of Ashvinkumar, Bernstein, Probst Gutenberg, and Saranurak~\cite{ABGS}.

\end{abstract}
\clearpage
\tableofcontents
\clearpage
\pagenumbering{arabic}
\section{Introduction}
\label{sec:introduction}

In the \emph{graph streaming model}, the edges of an
input graph arrive one at a time in an arbitrary order, and a
streaming algorithm can make multiple passes over this stream~\cite{HRR98,McG14}. A particularly common setting in graph streaming is the semi-streaming model~\cite{feigenbaum2005graph}, where we are allowed $\widetilde O(n)=O(n\polylog n)$ space to process an $n$-vertex graph, and our goal is to use as few passes as possible.

One of the earliest problems studied in  graph streaming is \textbf{single-source
reachability}: given a directed graph $G=(V,E)$ and a
source $s\in V$, find all vertices reachable from $s$ in $G$~\cite{HRR98}. In the streaming model, we can solve this problem using breadth-first search (BFS) in $O(n)$ space and $n-1$ passes. 
To see why: 
\begin{itemize}
    \item Let $S$ be the set of vertices we have reached so far, starting with $S = \{s\}$.
    \item  In each pass over the graph stream,   add to $S$ all vertices $v$ that have an edge $(u,v)$ with $u\in S$.
\end{itemize}
After $i$ passes over the graph stream, $S$ contains all vertices within distance $i$ of $s$. Thus BFS takes at most $D\le n-1$ passes, where $D$ is the \emph{diameter} of $G$.\footnote{We define the diameter of a directed graph as the largest (finite) distance between two vertices in the graph.}

A natural way to improve this algorithm is to reduce the diameter of $G$. We can do so by adding extra edges to $G$, as long as these edges do not change its reachability. A set of such edges is called a \emph{shortcut set}~\cite{UY91,Tho92}.

\begin{definition}[Shortcut set]
\label{def:shortcut-set}
Given a directed graph $G$, a set of edges
$H$ in the transitive closure is a \emph{$D$-diameter shortcut set} of $G$ if the
graph $G\cup H$ has diameter at most $D$.
\end{definition}

Shortcut sets give a simple recipe for faster reachability algorithms: compute a $D$-diameter 
shortcut set $H$, and then run BFS on $G\cup H$. In the semi-streaming model, if $H$ has $\widetilde O(n)$ edges, then we can store
$H$ in memory and run this BFS in $D$ passes. 
This approach has proven useful far beyond streaming reachability. Shortcut sets and their distance version, hopsets~\cite{Cohen00}, have been used extensively across many
models of computation for both directed and undirected graphs,
including parallel~\cite{UY91,KS,Cohen00,MPVX15,EN16,EM21,Fineman18,JLS19,CFR,CF23},
distributed~\cite{Nan14,HKN16,Elk17},
streaming~\cite{HKN16,EN16,Elk17,AJJST22,ET22},
dynamic~\cite{HKN14,BGWN20,LN22}, and others~\cite{CDKL19,Naz19,DN19}.

We can compute a shortcut set $H$ with $\widetilde O(n)$ edges and diameter $n^{1/2+o(1)}$ in the semi-streaming model using $n^{1/2+o(1)}$ passes. As an immediate consequence, we get an algorithm for streaming reachability that takes $n^{1/2+o(1)}$ passes~\cite{AJJST22}. This shortcut set construction was originally developed for parallel reachability by Jambulapati, Liu, and Sidford (JLS)~\cite{JLS19}, building on a breakthrough of Fineman~\cite{Fineman18}. The JLS shortcut set remains the best shortcut set construction we know of that can be computed efficiently.

As a result, semi-streaming algorithms for reachability and shortest paths remain stuck at $n^{1/2+o(1)}$ passes. 
Obtaining new upper bounds for streaming reachability is an important open problem in the area. As the authors of~\cite{ABCGLR26} recently pointed out, improving the $n^{1/2+o(1)}$ pass complexity for single-source reachability would be a breakthrough in streaming algorithms.

\subsection{Our results}
\label{sec:contributions}

Our main result is a new semi-streaming algorithm that computes single-source reachability and
shortest paths using $\widetilde{O}(n)$ space and $n^{1/3+o(1)}$ passes. We state our result in terms of shortest paths, which captures reachability as a special case.

\begin{restatable}[Streaming shortest paths]{theorem}{streamingthm}
\label{cor:streaming-exact-sssp}
There is a randomized streaming algorithm that, with high probability,
computes exact single-source shortest paths using $\widetilde O(n)$ space and $n^{1/3+o(1)}$ passes in $n$-vertex directed graphs with nonnegative, polynomially bounded
integer edge weights.
\end{restatable}

We present the proof of \cref{cor:streaming-exact-sssp} in \Cref{sec:streaming-reductions}. 
As an immediate consequence of  \cref{cor:streaming-exact-sssp} and existing reductions in the streaming algorithm literature, we obtain the following corollary.

\begin{corollary}
\label{cor:applications}
There exist randomized semi-streaming
algorithms that, with high probability, solve
\begin{itemize}[itemsep=1pt]
 \item exact maximum cardinality matching in bipartite graphs in
 $n^{2/3+o(1)}$ passes,
 \item $(1+\eps)$-approximate monotonicity testing in
 $n^{1/3+o(1)}\cdot\operatorname{poly}(1/\eps)$ passes,  and
 \item topological sorting and strongly connected components in
 $n^{1/3+o(1)}$ passes.
\end{itemize}
\end{corollary}

This corollary follows from reductions to single-source reachability due to~\cite{AJJST22, ABCGLR26, Sch08}.

As a byproduct of our techniques, we also obtain improved algorithms for \emph{parallel} single-source shortest paths in directed graphs. Let $m$ be the number of edges and let $\omega$ be the exponent of matrix multiplication~\cite{OpenAI26,Shp26}. We prove the following work-depth tradeoff.

\begin{restatable}[Parallel shortest paths]{theorem}{parallelthm}
\label{thm:exact-tradeoff}
There is a randomized parallel algorithm that, given a directed graph
with polynomially bounded nonnegative integer weights and a source $s$,
computes a shortest-path tree rooted at $s$ with high
probability, using
\[
 \widetilde O(m)\text{ work}
 \qquad\text{and}\qquad
 \left(\frac{n^\omega}{m}\right)^{1/(2\omega-2)}n^{o(1)}
 \text{ depth}.
\]
\end{restatable}
\Cref{thm:exact-tradeoff} improves on the recent work of~\cite{ABGS}, which obtains depth $(n^3/m)^{1/4}n^{o(1)}$ with
$\widetilde O(m)$ work. Their bound is our bound with $\omega$ replaced
by~$3$.
More generally, for every $x\in[n]$, our algorithm uses
$\widetilde O(m+nx^{\omega-1})$ work and $(n/x)^{1/2}\,n^{o(1)}$ depth.
This matches the best bounds for parallel
reachability, also due to~\cite{ABGS}. We complete the proof of \Cref{thm:exact-tradeoff} in \Cref{sec:parallel}.

\subsection{Implicit shortcut sets and hopsets}
\label{sec:implicit-intro}

Our  algorithm for streaming reachability  emerges from the following observation. In the streaming model, an 
algorithm does not need to store all the edges of a shortcut set $H$, as long
as it can recover the edges  it currently needs from a compact
description of $H$ in memory.
We formalize this observation with the new notions of
\emph{implicit shortcut sets} and \emph{implicit hopsets}. These are
ordinary shortcut sets and hopsets that we can store implicitly in
$\widetilde O(n)$ space, even when they have far more edges. We first
explain implicit shortcut sets, which suffice for reachability, and
then implicit hopsets, which we need for shortest paths.

\begin{definition}[Implicit shortcut set]
\label{def:implicit-shortcut-set}
Given an $n$-vertex directed graph $G=(V,E)$ and parameters $D\ge1$ and
$x\in[n]$, a \emph{$(D,x)$-implicit shortcut set} of
$G$ consists of a set of edges $H_0\subseteq\TC(G)$\footnote{We use $\TC(G)$ to denote the transitive closure of $G$.} and a collection
$\mathcal B$ of vertex sets $B\subseteq V$, called \emph{blocks}, such
that:
\begin{enumerate}[label=(\arabic*),itemsep=1pt]
 \item $|H_0|+\sum_{B\in\mathcal B}|B|=\widetilde O(n)$,
 \item $|B|\le x$ for all $B\in\mathcal B$, and
 \item $H:=H_0\cup\bigcup_{B\in\mathcal B}\TC(G[B])$ is a $D$-diameter
 shortcut set of $G$.
\end{enumerate}
\end{definition}

The edge set $H_0$ and the blocks in $\mathcal{B}$ fit in memory, even though the
shortcut set $H$ can contain $|H| \ge nx$ edges, well beyond the
$\widetilde{O}(n)$ memory budget. To use the edges of $H$
inside a block $B$, we can load the subgraph $G[B]$ into memory in one
pass over the stream and compute $\TC(G[B])$ without any additional passes.

The notion of an implicit shortcut set is new, but we can obtain good
implicit shortcut sets from the recent parallel reachability algorithm
of~\cite{ABGS}. Their algorithm follows the JLS algorithm, which
recursively partitions $G$ into smaller subgraphs to construct its
shortcut set. Once the subgraphs of $G$ are sufficiently small,
\cite{ABGS} directly adds their full transitive closures to its
shortcut set. This algorithm naturally yields an implicit shortcut set
with the following bound.

\begin{restatable}[Implied by~\cite{ABGS}]{theorem}{abgsimplicitshortcut}
\label{thm:abgs-implicit-shortcut}
For every $n$-vertex directed graph $G$ and every $x\in[n]$, there
exists a $(D, x)$-implicit shortcut set of $G$ with diameter
$D = (n/x)^{1/2}\,n^{o(1)}$.
\end{restatable}

In \cref{sec:overview-reachability}, we sketch how a semi-streaming
algorithm can use the implicit shortcut set of
\cref{thm:abgs-implicit-shortcut} with $x=n^{1/3}$ to compute
single-source reachability in $n^{1/3+o(1)}$ passes, assuming that this
implicit shortcut set is already stored in memory.

\paragraph{Implicit hopsets.}
For shortest paths, we need hopsets, the distance analogue of shortcut
sets. Informally, a set $H$ of weighted
edges is an \emph{$(h,\eps)$-hopset} of a weighted graph $G$ if adding
$H$ to $G$ does not decrease any distance, and for every pair of
vertices $(u,v) \in \TC(G)$,
the graph $G\cup H$ contains a path from $u$ to $v$
with at most $h$ edges whose length is at most
$(1+\eps)\dist_G(u,v)$.

To extend our algorithm from reachability to shortest paths, we
develop a corresponding notion of implicit hopsets. For a weighted
graph $G'$, the \emph{weighted transitive closure} $\TC_w(G')$ contains
an edge $(u,v)$ of weight $\dist_{G'}(u,v)$ for every pair
$(u,v)\in\TC(G')$.

\begin{definition}[Implicit hopset]
\label{def:implicit-hopset-intro}
Let $G=(V,E,w)$ be an $n$-vertex weighted directed graph, and let
$h\ge1$, $\eps\in(0,1)$, and $x\in[n]$. An \emph{$(h,\eps,x)$-implicit
hopset} of $G$ consists of a set $H_0$ of weighted edges between pairs
in $\TC(G)$ and a collection $\mathcal B$ of vertex sets $B\subseteq V$,
called \emph{blocks}, such that:
\begin{enumerate}[label=(\arabic*),itemsep=1pt]
 \item $|H_0|+\sum_{B\in\mathcal B}|B|=\widetilde O(n)$,
 \item $|B|\le x$ for every $B\in\mathcal B$, and
 \item $H:=H_0\cup\bigcup_{B\in\mathcal B}\TC_w\bigl((G\cup H_0)[B]\bigr)$
 is an $(h,\eps)$-hopset of $G$.\footnote{See
 \cref{def:hopset} in \cref{sec:prelim} for the formal definition of
 an $(h,\eps)$-hopset.}
\end{enumerate}
\end{definition}

As we discuss in \cref{sec:overview-hopset}, the existing hopset constructions
of~\cite{CFR} and~\cite{ABGS} do not directly yield good implicit hopsets. Thus, to obtain our streaming shortest paths upper bound, we prove
the following result.

\begin{restatable}[Implicit hopsets]{theorem}{implicithopsets}
\label{thm:implicit-hopset-intro}
For every $n$-vertex directed graph $G$ with positive, polynomially
bounded integer edge weights, every $x\in[n]$, and every
$\eps\in(0,1)$ with $\eps\ge\log^{-c}n$ for a  constant $c>0$,
there exists an
$(h,\eps,x)$-implicit hopset of $G$ with hopbound
$h=(n/x)^{1/2}\,n^{o(1)}$.
\end{restatable}

Our construction of these implicit hopsets is based on the hopset of~\cite{CFR}, but with several key changes that 
overcome a technical barrier in~\cite{ABGS}.\footnote{See
\cite[Remark~6.8]{ABGS} for a discussion of this barrier.}

\subsection{Related work}
\label{sec:related-work}
Streaming reachability has also been studied extensively from the lower bound side, and we now review existing lower bounds. 
Guruswami and Onak~\cite{GO16} showed that any $p$-pass streaming
algorithm for $s$-$t$ reachability requires
$n^{1+\Omega(1/p)}/p^{O(1)}$ space. Thus semi-streaming algorithms for
reachability need $\Omega(\log n/\log\log n)$ passes. With fewer
passes, the space lower bounds become nearly quadratic. Assadi and
Raz~\cite{AR20} showed that two-pass algorithms for
$s$-$t$ reachability require $n^{2-o(1)}$ space. Subsequently,  Chen, Kol, Paramonov,
Saxena, Song, and Yu~\cite{CKPSSY21} extended this bound to algorithms with $o(\sqrt{\log n})$ passes.

Shortcut sets and hopsets have also been studied as combinatorial objects. Thorup~\cite{Tho92} conjectured that every directed graph
has a shortcut set with at most $m$ edges and polylogarithmic diameter,
but Hesse~\cite{Hesse03} disproved this conjecture. Later work proved
stronger lower bounds~\cite{HP18,LVWX22,HXX25}, and shortcut sets with
$O(n)$ edges are now known to require diameter
$\Omega(n^{1/4})$~\cite{BH23,VXX24}. On the upper bound side, Kogan and
Parter~\cite{KP22} constructed shortcut sets with $\widetilde O(n)$ edges
and diameter $\widetilde O(n^{1/3})$, and Bernstein and Wein~\cite{BW23}
constructed $(1+\eps)$-approximate hopsets with $\widetilde O(n)$ edges and
hopbound $\widetilde O(n^{1/3})$. These constructions are sequential, and the fastest known 
algorithms for them take $\widetilde O(mn^{1/3})$ time for shortcut
sets~\cite{KP22a,KP23a} and $mn^{2/3+o(1)}$ time for
hopsets~\cite{BDHNR26}. Thus they do not give efficient parallel or
streaming algorithms.

\subsection*{Lean formalization}

Our main result, \cref{cor:streaming-exact-sssp}, is formalized in Lean. The formalization is available under
\\
\url{https://github.com/garytho/semi-streaming-directed-shortest-paths}.

\subsection*{Organization}
In \Cref{sec:technical-overview,sec:prelim} we provide our technical overview and preliminaries. In \Cref{sec:streaming-reductions}, we present our streaming algorithm (\cref{cor:streaming-exact-sssp}), though many of the lemmas we use are proved elsewhere in the paper. In \Cref{sec:sequential}, we present our hopset construction in the sequential setting. In \Cref{sec:parallel}, we extend this construction to the parallel setting, proving \cref{thm:exact-tradeoff}. In \Cref{sec:streaming-closures}, we extend this construction to the streaming setting.
\section{Technical Overview}
\label{sec:technical-overview}

In \Cref{sec:overview-reachability,sec:overview-sssp}, we show how to use implicit shortcut sets and implicit hopsets to compute streaming reachability and shortest paths. In \cref{sec:overview-hopset}, we explain how we construct our implicit hopsets and compare our construction with prior work.

\subsection{Streaming reachability}
\label{sec:overview-reachability}

In this subsection, we demonstrate how to use implicit shortcut sets to obtain a streaming algorithm for single-source reachability.\footnote{We do not actually construct and use implicit shortcut sets in our formal results, since we focus on the more general problem of single-source shortest paths, which only requires implicit hopsets. The purpose of this subsection is mainly to build intuition about implicit shortcut sets and hopsets. 
}
Fix $x\in[n]$, and suppose that we have already
computed the  $(D, x)$-implicit shortcut set $(H_0,\mathcal B)$ of $G$ of \cref{thm:abgs-implicit-shortcut} 
with diameter $D=(n/x)^{1/2}n^{o(1)}$
 and stored it in
memory. 
Let $H := H_0 \cup \bigcup_{B \in \mathcal{B}} \TC(G[B])$ be the shortcut set represented by $(H_0,\mathcal B)$.
Note that only $H_0$ and the blocks are stored in memory. The
shortcut set $H$ itself does not fit in memory, since it can contain
$|H|\ge nx$ edges.

\paragraph{Simulating BFS in $G\cup H$.}
Since $G\cup H$ has diameter at most $D$, we can find all reachable
vertices by running BFS in $G\cup H$ for $D$ rounds. The difficulty with this plan is that most edges in $H$ are not stored in memory. 

Let $S$ be the set of vertices found so far, starting with
$S=\{s\}$. In each round, we add to $S$ every vertex that has an edge
of $G\cup H$ coming from $S$. We can easily process  edges in $G \cup H_0$ in one pass over the graph stream, since $H_0$ is stored in memory. However, processing the transitive closure edges $\TC(G[B])$ in blocks $B \in \mathcal{B}$ poses a challenge. For each block $B \in \mathcal{B}$, we must process every edge in $\TC(G[B])$ from $B \cap S$ to $B \setminus S$. Thus for each block $B$, our concrete goal is to compute the set $U$ of
\emph{newly reachable vertices} in $B$, which are the vertices of
$B\setminus S$ that are reachable from $B\cap S$ in $G[B]$.

\paragraph{A first attempt: load each block in full.}
The simplest way to find $U$ is to load $G[B]$ into memory in one
pass, using at most $|E(G[B])| \le x^2$ space, and compute $U$ there.
We only need to do this after a new vertex of $B$ joins $S$, since otherwise $U$ is empty.
This gives the following algorithm. In each of the $D$ rounds of our
BFS, we use one pass to add to $S$ every vertex that has an edge of
$G\cup H_0$ coming from $S$. We also
load the block subgraph $G[B]$ of each block $B$ that gained a new
vertex since we last loaded it, and we add the newly reachable vertices
$U$ of $B$ to $S$. With $O(n)$ space, one pass can load the subgraphs
of $n/x^2$ blocks. Thus a round in which we load $k$ blocks takes
$O(1+kx^2/n)$ passes.

Each vertex joins $S$ once, so we load each
block at most $|B|$ times, and we perform at most
$\sum_{B\in\mathcal B}|B|=\widetilde O(n)$ loads in total. Summing over
the $D$ rounds, the whole algorithm uses
\[
 O(D)+\widetilde O(x^2)=(n/x)^{1/2}n^{o(1)}+\widetilde O(x^2)
\]
passes. This is $n^{2/5+o(1)}$ passes for $x=n^{1/5}$, which already beats
the $n^{1/2+o(1)}$ passes of JLS. 

\paragraph{Loading a smaller subgraph.}
We can improve this scheme by loading only part of $G[B]$. For an
integer $\tau\ge1$, a \emph{$\tau$-skeleton} of a graph is a subgraph that keeps $\tau$ 
arbitrary outgoing edges at each vertex and drops the rest. We can build a $\tau$-skeleton of $G[B]$ in
one pass using $O(\tau|B|)$ space, by keeping the first $\tau$ edges out
of each vertex that we see in the stream.

To find $U$, we first delete the edges of $G[B]$ that go into $S$,
since they only lead to vertices that we have already found. We then
build a $\tau$-skeleton $F \subseteq G[B]$ of the remaining graph, and let $U_F$ be the set
of vertices of $B\setminus S$ that are reachable from $S\cap B$ in $F$.
Since $F$ is a subgraph of $G[B]$, we have $U_F\subseteq U$. The next lemma says that if the skeleton finds only a few new vertices, then it  finds all of them.

\begin{lemma}[Skeletons find all new vertices]
\label{lem:overview-skeleton}
If $|U_F|<\tau$, then $U_F=U$.
\end{lemma}
\begin{proof}
Suppose that some $v\in U$ is not in $U_F$. Then some path in $G[B]$ goes
from $S\cap B$ to $v$, and without loss of generality it does not use any edges going into $S$. Let $(a, b)$ be the first edge on this path such that
$a$ is reachable from $S\cap B$ in $F$ but $b$ is not. The skeleton $F$ omits
the edge $(a, b)$, which means that it keeps $\tau$ other outgoing edges at $a$. These edges 
go to $\tau$ distinct vertices outside $S$, all of which are reachable from $S\cap B$ in $F$. Then $|U_F|\ge\tau$.
\end{proof}

We do not know $|U|$ in advance, so we use the standard doubling trick: we build a $\tau$-skeleton for $\tau = 1, 2, 4, \dots$ until $|U_F| < \tau$, at which point $U_F = U$ by \cref{lem:overview-skeleton}. This procedure builds $O(\log x)$ skeletons of total size $O((|U|+1)\,x)$ over $O(\log x)$ passes. Thus instead of paying $O(x^2)$ space to load $G[B]$, we pay $O(x)$ space for each newly reachable vertex in $U$ that we discover in $G[B]$. 

\paragraph{Putting it together.}
We run the BFS simulation above and use the skeleton subroutine to
discover the newly reachable vertices $U$ in the block subgraphs
$G[B]$. In each round, we process all blocks that gained a new vertex
since we last processed them. We run the doubling tricks of these
blocks side by side, in phases $i=0,1,\dots,O(\log x)$. In phase $i$,
we build a $2^i$-skeleton for each block whose search has not finished
yet. The search of a block finishes in the first phase $i$ in which its
skeleton finds $|U_F|<2^i$ new vertices, since then $U_F=U$. In one
pass, we can build skeletons of total size up to $n$ edges using $O(n)$
space. Thus a phase takes one pass, plus one more pass for every $n$
skeleton edges that it builds.

Each round has $O(\log x)$ phases, which take $O(D\log x)$ passes over
all $D$ rounds. To bound the number of skeleton edges, recall that we
pay $O(x)$ space for each new vertex $u \in U$ that we discover in a
block subgraph $G[B]$. Each vertex in $G$ joins $S$ at most once.
Then the skeletons that we build have total size
$O(x\sum_{B\in\mathcal B}|B|)=\widetilde O(nx)$, which takes
$\widetilde O(x)$ additional passes. Altogether, the algorithm uses
\[
 O(D\log x)+\widetilde O(x)=(n/x)^{1/2}n^{o(1)}+\widetilde O(x)
\]
passes. This gives our desired $n^{1/3+o(1)}$ pass complexity by taking $x=n^{1/3}$.

\subsection{Streaming shortest paths}
\label{sec:overview-sssp}

We now turn to shortest paths. We assume that we have already
computed an implicit hopset and stored it in memory, and our goal is to
show how to use it to compute $(1+\eps)$-approximate distances from
$s$. Fix $x$ and $\eps$, and let $(H_0,\mathcal B)$ be an  
$(h,\eps,x)$-implicit hopset of $G$ with $h=(n/x)^{1/2}n^{o(1)}$, as in \cref{thm:implicit-hopset-intro}. Let $H:=H_0\cup\bigcup_{B\in\mathcal B}\TC_w\bigl((G\cup H_0)[B]\bigr)$ be the hopset 
represented by $(H_0,\mathcal B)$.

A shortcut set $H$ lets us run BFS in $G\cup H$ with few passes. With an 
implicit hopset, however, neither Dijkstra's algorithm nor
Bellman--Ford directly gives few passes. Dijkstra's algorithm handles
one distance value at a time, so it may need $n$ rounds. Bellman--Ford needs only $h$ rounds in $G\cup H$. But the distance
estimate of a vertex can improve in many rounds, and each improvement
would force us to look at its blocks again. Instead we  adapt  a standard technique from  graph algorithms. As before, the main difficulty will be that the edges in hopset $H$ are not stored in memory.

\paragraph{Scaling and rounding.}
We use the standard scaling and rounding reduction of Klein
and Subramanian~\cite{KS}. 
At a high level, this reduction rescales the edge weights in $G$ and $H$ and then rounds them  to integer values. 
Let $\bar G$ and $\bar H$ denote the rounded versions of $G$ and $H$, respectively. 
This operation ensures that distances in $\bar G \cup \bar H$ are a $(1+\varepsilon)$-approximation of distances in $G \cup H$. 

Scaling and rounding reduces our problem to the simpler problem of computing \textit{distance-bounded} single-source shortest paths. Specifically,
we only need to compute shortest paths from $s$ in $\bar G\cup\bar H$
up to distance $R:=O(h/\eps)$. We can do this using a modified version of Dijkstra's algorithm.

\paragraph{Dijkstra's algorithm in few rounds.}
Since the weights of $\bar G\cup\bar H$ are positive integers, we can
run a batched version of  Dijkstra's algorithm. As in the standard Dijkstra's algorithm, every vertex~$v$ maintains a distance estimate $d(v)$. Initially, we let $d(s) = 0$ and $d(v) = \infty$ for $v \ne s$. In round $r = 0, \dots, R$, we \emph{settle} every vertex $u$ with $d(u)=r$ (i.e., we finalize its distance estimate). Then we \emph{relax} the outgoing edges of $u$: for each edge
$(u, v)$, we set $d(v)\gets\min\{d(v),d(u)+w(u,v)\}$. Round $r$ settles exactly the vertices at distance
$r$ from $s$ in $\bar G \cup \bar H$.

Relaxing the edges of $\bar G$ takes one pass per round, or $R+1$
passes in total. The main obstacle remains handling the edges of $\bar H$ that are not stored in memory.

\paragraph{A first attempt: load each block in full.}
As in \cref{sec:overview-reachability}, we could try loading the block subgraph $(G\cup H_0)[B]$ whenever a vertex $u\in B$ settles. This would allow us to compute distances in $(G\cup H_0)[B]$ and relax all edges of $\bar H$ out of $u$. This takes $O(R)+\widetilde O(x^2)$
passes, or $n^{2/5+o(1)}/\eps$ passes for $x=n^{1/5}$.

The first attempt runs Dijkstra's algorithm exactly as described above.
When a vertex $u\in B$ settles, we relax all edges of 
$\bar H$ inside $B$ out of $u$. This requires the distances from $u$ to all vertices of
$B$, and computing these distances requires the whole block subgraph
$(G\cup H_0)[B]$. To improve the first attempt, we change when the
edges of $\bar H$ are used. We still relax the edges of $\bar G$ out
of $u$ when $u$ settles, but we no longer relax the edges of $\bar H$
out of $u$. Instead, in each round $r$, we search each block for the
vertices that settle in round $r$ through an edge of $\bar H$. Such an
edge starts at a vertex that settled in an earlier round, possibly
long before round $r$. Thus in each round $r$, we look for the unsettled
vertices that have distance exactly $r$. This change lets us
load a small subgraph of each block and pay only for the new vertex
distances that we settle.

\paragraph{The search problem inside a block.}
We now formulate the search problem that a block must solve in each
round, analogous to finding the newly reachable vertices $U$ in the
reachability algorithm.

Fix a block $B$. For $u,v\in B$, let $\dist_B(u,v)$ be the distance
from $u$ to $v$ in the block subgraph $(G\cup H_0)[B]$, which is the (unscaled) 
weight of the edge $(u, v)$ in $\bar H$.\footnote{Here, we
ignore some technical details related to edge rounding.
\Cref{sec:core-problem,sec:core-algorithm} explain how the algorithm handles them.} In
each round $r$, our goal is to find the vertices of $B$ that settle in
round $r$ with a distance that comes from an edge of $\bar H$ inside
$B$.

These vertices have a simple description. For each $v \in B$, define the \emph{block distance} of $v$ in $B$ as
\[
 d_B(v)=\min_{\text{settled } u\in B} \bigl(d(u)+\dist_B(u,v)\bigr).
\]
Then an edge of
$\bar H$ inside $B$ gives $v$ distance $r$ in round $r$ exactly when $v$
has not settled yet and $d_B(v)=r$.
So for each block $B \in \mathcal{B}$, our concrete goal is to compute the set $U$ of \emph{newly settled vertices} in $B$, which are the unsettled vertices $v \in B$ with $d_B(v) = r$.

\paragraph{Loading a smaller subgraph.}
We search for $U$ in a small subgraph instead of the whole block
subgraph.
We define a
\emph{$\tau$-skeleton} of a weighted graph as a subgraph that keeps the $\tau$
\emph{lightest} outgoing edges at each vertex and drops the rest.
First, we delete every edge $(a,b)$ of $(G\cup H_0)[B]$ that
goes into a settled vertex $b$.
Then we build a $\tau$-skeleton $F$ of the resulting graph. Finally, we run
Dijkstra's algorithm on $F$ in memory, starting each settled vertex $u$
of $B$ at distance $d(u)$.

Let $U_F$ be the set of unsettled vertices
that get distance at most $r$ in $F$. Since $F$ is a subgraph of $(G\cup H_0)[B]$,
distances in $F$ are never smaller than block distances, so
$U_F\subseteq U$. As in \cref{sec:overview-reachability}, if the
skeleton finds few new vertices, then it finds all of them.

\begin{lemma}[Skeletons find all new vertices, weighted version]
\label{lem:overview-distance-skeleton}
If $|U_F|<\tau$, then $U_F=U$.
\end{lemma}
\begin{proof}[Proof sketch]
The proof is the same as for \cref{lem:overview-skeleton}, except that
we follow a shortest path instead of an arbitrary path. If $F$ misses
some $v\in U$, let $(a,b)$ be the first edge missing from $F$ on a
shortest path that gives $v$ its block distance. Then $F$ keeps $\tau$
edges out of $a$, no heavier than $(a,b)$, to $\tau$ distinct unsettled
vertices, which all get distance at most $d_B(v)=r$ in $F$. We conclude that 
$|U_F|\ge\tau$.
\end{proof}

\paragraph{Putting it together.}
We find $U$ with $\tau$-skeletons, using the same doubling trick as in
\cref{sec:overview-reachability}: we build a $\tau$-skeleton for
$\tau=1,2,4,\dots$ until $|U_F|<\tau$. By
\cref{lem:overview-distance-skeleton}, using $\tau$-skeletons allows us
to find the $|U|$ newly settled vertices of a block using
$O((|U|+1)\,x)$ space.

In each round of our Dijkstra's algorithm, we search every block $B \in \mathcal B$ for newly settled vertices. 
We begin by building a $1$-skeleton for every block $B \in \mathcal{B}$. These $1$-skeletons each take $O(|B|)$ space, so the $1$-skeletons of all blocks take $\sum_{B \in \mathcal{B}} |B| = \widetilde O(n)$ space in total and fit in one pass. We build higher-level $\tau$-skeletons only for blocks where our search finds newly settled vertices. Each vertex is newly settled at most once. Then over all rounds, these skeletons take
$O(x\sum_{B\in\mathcal B}|B|)=\widetilde O(nx)$ space in total. Thus  we
can build these skeletons in $\widetilde O(x)$ passes. Altogether, with 
$R=O(h/\eps)$ and $h=(n/x)^{1/2}n^{o(1)}$, the algorithm uses
\[
 O(R\log x)+\widetilde O(x)=(n/x)^{1/2}n^{o(1)}/\eps+\widetilde O(x)
\]
passes. This is $n^{1/3+o(1)}/\eps$ passes for $x=n^{1/3}$. The full streaming algorithm and its analysis are
presented in \cref{sec:streaming-reductions}.

\paragraph{From approximate to exact distances.}
Our implicit hopsets only give $(1+\eps)$-approximate distances. To
compute exact distances, we adapt the reduction from exact to
approximate parallel SSSP in directed graphs with integer weights, due
to Rozho\v{n}, Haeupler, Martinsson, Grunau, and Žužić~\cite{Rozhon}.
It makes $O(\log^2n)$ calls to a $(1+\eps)$-approximate SSSP algorithm
with $\eps=\Theta(1/\log n)$. In
\cref{sec:streaming-exact-reduction}, we implement this reduction in the
streaming model using $\widetilde O(n)$ space.

\subsection{A new implicit hopset construction}
\label{sec:overview-hopset}

We now explain how we construct the implicit hopsets of
\cref{thm:implicit-hopset-intro}. Our starting point is the hopset of
Cao, Fineman, and Russell~\cite{CFR}, which we call the \emph{CFR
hopset}.
We first describe the CFR hopset and why it has small
hopbound. Then we
describe our implicit hopset construction, and explain how it
overcomes a technical barrier in~\cite{ABGS}.

\paragraph{The CFR hopset.}
The CFR hopset is built by an algorithm that recursively decomposes the
input graph $G$ into smaller induced subgraphs of $G$. The graph $G$
itself is at \emph{level}~$0$ of the recursion, and decomposing a 
subgraph at level $r$ gives subgraphs at level $r+1$. At each level of the recursion, 
randomly chosen vertices called \textit{pivots} add shortcut edges to and from
the vertices near them, and then the graph is partitioned according to
which pivots each vertex is near. Each level $r$ has a \emph{distance
scale} $D_r$, which decreases with $r$ and determines what ``near''
means at level $r$. Consider a subgraph $G'$ at level $r$. The
algorithm does the following in $G'$.
\begin{enumerate}[label=(\arabic*),itemsep=1pt]
 \item \emph{Sample pivots.} Each vertex of $G'$ becomes a pivot
 with probability about $k^r/n$, where $k=\Theta(\log n)$.
 \item \emph{Search from the pivots.} We compute single-source shortest paths (SSSP) in $G'$,  to and from each pivot $p$. These SSSP searches at $p$ are \emph{distance-limited} up to a random radius
 $\rho_p$ between about $\lambda D_r$ and $2\lambda D_r$, where $\lambda=\widetilde O(1)$.
 The \emph{in-ball} of $p$ is the set of
 vertices $v$ with $\dist_{G'}(v,p)\le\rho_p$, and the \emph{out-ball}
 of $p$ is the set of vertices $v$ with $\dist_{G'}(p,v)\le\rho_p$.
 \item \emph{Add a star.} The pivot $p$ adds shortcut edges from every
 vertex of its in-ball to $p$, and from $p$ to every vertex of its
 out-ball, weighted by the corresponding distances. These edges form
 the \emph{star} of $p$, and the hopset consists of the stars of all
 pivots.
 \item \emph{Partition and recurse.} Vertices that lie in both the
 in-ball and the out-ball of some pivot $p$ can be removed, since paths using these vertices can be rerouted through the star of $p$.\footnote{Special care is needed here to ensure that this detour through $p$ does not increase the length of the path too much. This is the main added source of complexity for the CFR hopset.}
 Each remaining vertex gets a
 label that records, for each pivot, whether the vertex is in the pivot's 
 in-ball, its out-ball, or neither.  We then partition the vertices based on their label sets and recurse.  Vertices with the same label configurations form an induced 
 subgraph at level $r+1$.\footnote{The CFR hopset also
 creates an extra subgraph around the \emph{boundary} of each pivot ball,
 because a path can leave the search radius of a pivot and come back
 later. We ignore this detail here.}

\end{enumerate}
The sampling probability grows by a factor of $k$ per level, so after
about $\log_kn=O(\log n/\log\log n)$ levels, every vertex is a pivot,
and the recursion stops. 
Additionally, the partition scheme makes neighborhoods shrink, so that at level $r$,  every vertex has at most  $n/k^r$ vertices within
distance $O(\lambda D_r)$ in each direction~\cite[Lemma 3]{CFR}.

\paragraph{Following a path through the recursion.}
To see why the stars shortcut every shortest path in $G$ to few edges, fix a shortest
path $P$ in $G$. The analysis of~\cite{CFR} studies how the recursive partitioning of the CFR algorithm 
breaks $P$ into smaller and smaller subpaths. At each level $r$, we
keep track of a set of disjoint subpaths of $P$ that still need to be
shortcut, which we call the \emph{pieces} of $P$ at level $r$. Each
piece of $P$ lies inside a single subgraph at level $r$ and has length
at most $D_r$, so that it is short compared to the search radii at
level $r$. Now consider a piece $Q$ of $P$ at level $r$, in a subgraph
$G'$. There are two cases.
\begin{itemize}[itemsep=1pt]
 \item \emph{A pivot shortcuts $Q$.} If the first vertex of $Q$ is in
 the in-ball of some pivot $p$, and the last vertex of $Q$ is in the 
 out-ball of $p$, then the star of $p$ shortcuts subpath $Q$ to two hops. 
 \item \emph{The pivots split $Q$.} Otherwise, consider a single pivot
 $p$. If a vertex of $Q$ is in the in-ball of $p$, then every earlier
 vertex of $Q$ can also reach $p$, by following $Q$. Thus the
 vertices of $Q$ in the in-ball of $p$ form a prefix of $Q$. Similarly,
 the vertices of $Q$ in the out-ball of $p$ form a suffix of $Q$.\footnote{This is not completely true, since an earlier
 vertex reaches $p$ at a slightly larger distance, which may exceed
 $\rho_p$. This is why the CFR hopset needs the extra subgraphs around
 the boundaries of the pivot balls.} Each pivot changes the labels along $Q$ in at most two places.
 Thus the pivots split $Q$ into contiguous subpaths, each of which lies
 in a single subgraph at level $r+1$. Finally, we split each of these subpaths further into subpaths
 of length at most $D_{r+1}$. The resulting subpaths are the pieces of
 $P$ at level $r+1$.
\end{itemize}
At the last level of the recursion, every vertex is a pivot, so the
pivots shortcut all remaining pieces of $P$. As a result, the pivot stars shortcut 
$P$ to a path with roughly as many edges as there are pieces of $P$ 
over all levels of the recursion. It remains to bound the number of pieces of $P$.

There are two ways a subpath of $P$ can be cut into multiple pieces. A \emph{type~1 split} happens when a
pivot splits a subpath of $P$. A \emph{type~2 split} happens when we
split a subpath of $P$ into subpaths of length at most $D_{r+1}$. Cao,
Fineman, and Russell show that in expectation, type~1 splits increase
the number of pieces of $P$ by a factor of only about $\sqrt k$ per
level. On the other hand, type~2 splits create about
$w(P)/D_r$ additional pieces of $P$ at level $r$. Type~2 splits
create fewer pieces if $D_r$ decreases slowly. Nevertheless, the CFR
hopset shrinks the distance scale $D_r$ quickly and sets
$D_{r+1}=D_r/\Theta(\sqrt k)$. We will explain why the CFR hopset shrinks
$D_r$ so quickly when we describe our construction. This fast decrease
causes the number of pieces $w(P)/D_r$ created by type~2 splits to grow
by a factor of $\Theta(\sqrt k)$ per level, the same rate as for
type~1 splits. After the roughly $\log_kn$ levels of the recursion,
$P$ has about $(\sqrt k)^{\log_kn}=n^{1/2}$ pieces, up to $n^{o(1)}$
factors. This gives an $(n^{1/2+o(1)},\eps)$-hopset.

\paragraph{A first attempt: block closures in the later levels.}
Our goal is to construct an implicit hopset with hopbound
$h=(n/x)^{1/2}n^{o(1)}$ whose blocks have at most $x$ vertices. 
Our approach begins with the following observation. At recursion level $r_0\approx\log_k(n/x)$ in the CFR hopset, the in-ball and the out-ball of every pivot have at most about $x$ vertices. Thus we can choose these balls to be the blocks of our implicit hopset.\footnote{This approach roughly corresponds to the reachability algorithm in~\cite{ABGS}.}

We call $r_0$ the threshold level, the levels $r<r_0$ the \emph{early} levels,
and the levels $r\ge r_0$ the \emph{late} levels. The \emph{ball} of a pivot is the union of its
in-ball and its out-ball. A natural first attempt changes only the late
levels of the CFR hopset. At a late level, each pivot adds the weighted
transitive closure $\TC_w(G[B])$ of its ball $B$ to the hopset, instead
of building a star. Then the balls $B$ become the blocks of our implicit hopset, and the block closures $\TC_w(G[B])$ are the implicit edges. 

We now explain how the block closures $\TC_w(G[B])$ could help us
obtain a better hopbound. Recall that the CFR hopset shortcuts $P$ to
a path with roughly as many edges as there are pieces of $P$. Type~1
splits increase the number of pieces of $P$ by a factor of about
$\sqrt k$ per level. We show that in the late levels, the
block closures stop type~1 splits from creating new pieces of $P$.
Consider a piece $Q$ of $P$ at a late level $r$, and a pivot $p$. As
in the CFR hopset, the vertices of $Q$ in the in-ball of $p$ form a
prefix of $Q$. This whole prefix lies in the ball
of $p$, so a single edge of the block closure of this ball shortcuts
the prefix. Similarly, a single block-closure edge shortcuts the suffix
of $Q$ in the out-ball of $p$. Now let $Q=\langle v_0,\dots,v_\ell\rangle$.
Let $v_i$ be the last vertex of $Q$ that lies in an in-ball, and let
$v_j$ be the first vertex of $Q$ that lies in an out-ball. Then two
block-closure edges shortcut the prefix of $Q$ up to $v_i$ and the
suffix of $Q$ from $v_j$. If the prefix and the suffix overlap, these
two edges shortcut all of $Q$. Otherwise, $Q$ looks as follows.
\begin{equation}
 \label{eq:overview-middle}
 \underbrace{v_0\to\cdots\to v_i}_{\text{prefix: shortcut by one edge}}
 \to
 \underbrace{v_{i+1}\to\cdots\to v_{j-1}}_{\text{middle: untouched by any pivot}}
 \to
 \underbrace{v_j\to\cdots\to v_\ell}_{\text{suffix: shortcut by one edge}}
\end{equation}
No vertex in the middle part of $Q$ lies in an in-ball or an
out-ball. So all of them have the same (empty) label, and the middle
part of $Q$ lies in a single subgraph at level $r+1$. Thus, in the late
levels of the recursion, type~1 splits no longer create new pieces of $P$. At most one
subpath of each piece of $P$, its middle part, continues to the next
level.

We can now see how the block closures could give us the hopbound
$h=(n/x)^{1/2}n^{o(1)}$ that we want. In the early levels, the number
of pieces of $P$ grows by a factor of about $\sqrt k$ per level. Thus
only $(\sqrt k)^{r_0}\approx(n/x)^{1/2}$ pieces of $P$ reach the
threshold level $r_0$, up to $n^{o(1)}$ factors. Suppose that in the
late levels, type~2 splits did not create new pieces of $P$ either.
Then the block closures would shortcut each piece of $P$ at level $r_0$
to a path with a few edges per late level. As a result, $P$ would be
shortcut to a path with $(n/x)^{1/2}n^{o(1)}$ edges.

\paragraph{Why this approach fails.}
 Unfortunately, this approach fails to give a hopset with hopbound $h=(n/x)^{1/2}n^{o(1)}$.
As noted in~\cite[Remark~6.8]{ABGS}, the number of pieces of $P$ may still grow too quickly. The
main obstacle is the shrinking distance scale $D_r$. The middle part of
a piece of $P$ at a late level $r$, as in \eqref{eq:overview-middle},
could still have length close to $D_r$. Then type~2 splits would divide this middle part into
$D_r / D_{r+1} = \Theta(\sqrt k)$ pieces of $P$ at level $r+1$, since
the first attempt keeps the rule $D_{r+1}=D_r/\Theta(\sqrt k)$ of the CFR hopset. So in the late levels, the number of
pieces of $P$ could still grow by a factor of $\Theta(\sqrt k)$ per
level, just as in the early levels. As a result, $P$ would be shortcut
to a path with about $n^{1/2}$ edges, no better than with the CFR
hopset itself.

The hopset construction of~\cite{ABGS} takes a different approach that avoids block closures. Their hopset adds shortcut edges
between every vertex and its $x$ nearest vertices in each direction.
This gives hopbound $(n/x)^{1/2}n^{o(1)}$. However, the natural blocks for these edges are the sets of $x$ nearest vertices of all $n$
vertices, which have total size $nx$. Thus their hopset construction
does not give an implicit hopset that can be stored in $\widetilde O(n)$ space.

\paragraph{Our construction: a two-speed recursion.}
To overcome the obstacle caused by type~2 splits, we would like the
distance scale $D_r$ to shrink more slowly in the late levels. However,
the CFR hopset needs $D_r$ to shrink quickly in order to obtain
$(1+\eps)$-approximate shortest paths. 
When a pivot $p$ shortcuts a piece $Q$ of $P$, the two star
edges go through $p$. Their total weight can be as large as about twice
the search radius $\rho_p$, much more than the length of $Q$. We call
the extra weight beyond the length of $Q$ the \emph{detour} through
$p$. The analysis of~\cite{CFR} bounds the total detour by
$\eps\,w(P)$, but this bound requires $D_r$ to shrink quickly.

Our key insight is that the switch from stars to block closures in the
late levels also removes the reason for shrinking the distance scale
$D_r$ quickly. The late pivots do not build stars. Every shortcut
edge that a late pivot adds is a block-closure edge, whose weight is at
most the length of the subpath of $P$ that it shortcuts. Thus the late
levels have no detours to pay for, and we can run the recursion at two
speeds. In the early levels $r<r_0$, we keep the CFR hopset unchanged,
including the rule $D_{r+1}=D_r/\Theta(\sqrt k)$. In the late levels $r\ge r_0$, where
pivots add block closures, we let $D_r$ shrink slowly. To make this work,
we change the first attempt in two ways.

\emph{Change 1: slow down the distance scale.} In the late levels, we
let the distance scale shrink by only a constant factor per level, say
$D_{r+1}=D_r/4$. The middle part of a piece of $P$ at level $r$ has
length at most $D_r$, so type~2 splits divide it into only $O(1)$
pieces of $P$ at level $r+1$.

\emph{Change 2: slightly enlarge the balls.} To obtain
\eqref{eq:overview-middle}, we assumed that if a vertex of a piece $Q$ of $P$ is in
the in-ball of a pivot $p$, then so is every earlier vertex of $Q$. With
bounded search radii, this is not completely true. An earlier vertex of
$Q$ can reach $p$ by first following $Q$, but this only shows that its
distance to $p$ is at most $\rho_p+D_r$. We define the \emph{enlarged
ball} of $p$ as the set of vertices within distance $\rho_p+D_r$ of $p$
in either direction. In the late levels, each pivot $p$ adds the block
closure of its enlarged ball instead of its ball.
Now if some vertex $v$ of $Q$ is in the in-ball of $p$, then the whole
prefix of $Q$ up to $v$ lies in the enlarged ball of $p$. So the block
closure of the enlarged ball contains an edge from the first vertex of
$Q$ to $v$ whose weight is at most the length of this prefix.

\paragraph{The hopbound of our construction.}
We now sketch why our construction has hopbound
$h=(n/x)^{1/2}n^{o(1)}$. The early levels are unchanged, so the
analysis of the CFR hopset still shows that only $(n/x)^{1/2}n^{o(1)}$
pieces of $P$ reach the threshold level $r_0$. Now consider a piece $Q$
of $P$ at a late level $r$. By Change~2, two block-closure edges
shortcut the prefix and the suffix of $Q$ in \eqref{eq:overview-middle},
and only the middle part of $Q$ continues to level $r+1$. By Change~1,
type~2 splits divide this middle part into only $O(1)$ pieces of $P$ at
level $r+1$. Since there are $O(\log n/\log\log n)$ levels, each piece
of $P$ that reaches level $r_0$ is shortcut to a path with
$O(1)^{O(\log n/\log\log n)}=n^{o(1)}$ edges. Therefore $P$ is shortcut
to a path with $(n/x)^{1/2}n^{o(1)}$ edges.
The weight of our shortcutted path increases only through the detours
of the early stars, which
the analysis of the CFR hopset bounds by $\eps\,w(P)$. The full
construction and its analysis are presented in \cref{sec:sequential}.

\paragraph{The blocks fit in memory.}
Our implicit hopset $(H_0,\mathcal B)$ stores the early stars
explicitly in $H_0$, and the enlarged balls of the late pivots as
blocks in $\mathcal B$. We choose the threshold level $r_0$ between the
early and the late levels so that every block has at most $x$
vertices. The blocks also have total size
$\sum_{B\in\mathcal B}|B|=\widetilde O(n)$. As in the CFR hopset, there
are $|H_0|=\widetilde O(n/\eps^2)$ star edges, which is
$\widetilde O(n)$ since $\eps\ge\log^{-c}n$.

\paragraph{Building the implicit hopset in the streaming model.}
The main computational cost in our implicit hopset construction is
computing single-source shortest paths from the pivots. In
\cref{sec:overview-sssp}, we described an algorithm for computing
shortest paths in few passes. However, this algorithm assumed that we
had already computed an implicit hopset and stored it in memory.
Cao, Fineman, and Russell
resolve this chicken-and-egg problem by building the hopset in
$O(\log n)$ stages, and we do the same. Each stage runs our two-speed
recursion on $G$ together with the implicit hopset built in the earlier
stages. With high probability, the new stars and block closures of a stage shortcut
any fixed path with at most $2h$ edges to a path with at most $h$
edges. This increases the length of the path by only a factor of
$1+\eps/O(\log n)$. Thus the
number of edges needed by a near-shortest path halves in each stage.
After $O(\log n)$ stages, every pair of vertices has a near-shortest
path with at most $h$ edges.

A stage only needs to shortcut paths with at most $2h$ edges. So after
scaling and rounding, each of its searches runs up to distance
$\widetilde O(h/\eps)$. Many searches run at
the same time and share passes. They fit in $\widetilde O(n)$ space, since each vertex is reached by
few searches and the blocks have total size $\sum_{B\in\mathcal B}|B|=\widetilde O(n)$. So each stage takes $\widetilde O(h/\eps+x)$ passes, which
is $n^{1/3+o(1)}/\eps$ for $x=n^{1/3}$.

\paragraph{A complication specific to implicit hopsets.}
The streaming algorithm never stores the edges in the block closures
$\TC_w(G[B])$. It only stores the vertex set of each block. This causes
a complication in later stages. Consider an edge $(u,v)$ in the block
closure of a block $B$ from an earlier stage. This edge corresponds to a
shortest path from $u$ to $v$ in $G[B]$, which we call the
\emph{underlying path} of $(u,v)$. When we recurse on a smaller induced
subgraph $G'$, it may be possible that $G'$ does not contain the
underlying path of $(u,v)$. Then a search inside $G'$ cannot use the
edge $(u,v)$, since the algorithm of \cref{sec:overview-sssp} can only recover this edge from its
underlying path. This is a problem because in later stages, the paths
that we shortcut use such edges. Losing $(u,v)$ inside $G'$ can create
additional pieces of such a path, and the analysis of the CFR hopset
does not account for them. \Cref{sec:streaming-closures} handles this
issue by choosing the search radii with a small random shift, so that
underlying paths rarely cross subgraph boundaries.

\paragraph{The parallel algorithm.}
We now quickly sketch our improved parallel algorithm for SSSP.
In the parallel model, we explicitly store in memory the entire
hopset $H=H_0\cup\bigcup_{B\in\mathcal B}\TC_w\bigl((G\cup H_0)[B]\bigr)$.
We use fast matrix multiplication to compute the
block closures. For a block with $b\le x$ vertices, Zwick's
algorithm~\cite{Zwick} computes $(1+\eps)$-approximate distances
between all pairs of its vertices using about $b^\omega$ work and
polylogarithmic depth. This additional $(1+\eps)$-approximation is
harmless, since each block-closure edge that shortcuts a subpath of $P$
has weight at most $1+\eps$ times the length of that subpath. Since the blocks
have total size $\widetilde O(n)$, all block closures together take
$\sum_B|B|^\omega\le x^{\omega-1}\sum_B|B|=\widetilde O(nx^{\omega-1})$
work. In contrast, the hopset construction of~\cite{ABGS} needs about
$nx^2$ work. Finding the $x$ nearest vertices of a vertex takes about
$x^2$ work, since the subgraph induced by these vertices can have $x^2$
edges. This is why the parallel shortest-path tradeoff of~\cite{ABGS}
is worse than ours. As in the CFR algorithm, we build the hopset in $O(\log n)$
stages, so every search has radius $\widetilde O(h/\eps)$ after scaling
and rounding. A search of radius $R$ takes $\widetilde O(R)$ depth, by
processing the distances $0,1,\dots,R$ in order. So the construction
takes $(n/x)^{1/2}n^{o(1)}$ depth and $\widetilde O(m+nx^{\omega-1})$
work. Then one more search in $G\cup H$ computes approximate distances
from $s$. We choose $x\approx(m/n)^{1/(\omega-1)}$, so that
$nx^{\omega-1}\approx m$. Applying the reduction of~\cite{Rozhon} from
exact to approximate SSSP then gives \cref{thm:exact-tradeoff}. The
full parallel algorithm and its analysis are presented in
\cref{sec:parallel}.

\section{Preliminaries}
\label{sec:prelim}

The notation $\widetilde O(\cdot)$ hides $\polylog n$ factors. We say
that a guarantee holds \emph{with high probability} if it can be made to hold with probability at least $1-n^{-c}$ for any fixed constant $c>0$.

\paragraph{Graphs and distances.}
We assume that the input graph $G=(V,E,w)$ has $n$ vertices and
$m\ge n$ edges, with positive edge weights of polynomial aspect ratio
$W=w_{\max}/w_{\min}$, where $w_{\max}$ and $w_{\min}$ denote the
largest and smallest edge weights.  Let $G[U]$ denote the 
subgraph of $G$ induced by $U\subseteq V$. The \emph{length} $w(P)$
of a path $P$ is the total weight of the edges of $P$. The number of
these edges is called the \emph{hop length} of $P$. Let $\dist_G(u,v)$
denote the length of a shortest path from $u$ to $v$ in $G$. The
minimum length of a $(u, v)$-path with at most $h$ edges is denoted by
$\dist_G^{(h)}(u,v)$.

Given a source $s\in V$, the single-source shortest paths problem
(SSSP) asks for $\dist_G(s,v)$ for all $v\in V$ and a shortest-path
tree rooted at $s$. In $(1+\eps)$-approximate SSSP, we instead compute
\emph{$(1+\eps)$-approximate distances} from $s$, i.e., estimates
$\widehat d(v)$ with
$\dist_G(s,v)\le\widehat d(v)\le(1+\eps)\dist_G(s,v)$ for every
vertex $v$. We define an approximate hopset as follows.

\begin{definition}
\label{def:hopset}
Given a weighted directed graph $G=(V,E,w)$, an integer $h\ge1$, and
$\eps\in(0,1)$, a set $H$ of weighted edges is an
\emph{$(h,\eps)$-hopset} of $G$ if
\[
 \dist_G(u,v)\le\dist_{G\cup H}^{(h)}(u,v)
 \le(1+\eps)\dist_G(u,v)
 \qquad\text{for all }u,v\in V.
\]
Here $h$ is the \emph{hopbound} of $H$.
\end{definition}

An edge $(u,v)$ is \emph{noncontracting} in $G$ if its weight is at
least $\dist_G(u,v)$. The first inequality in \cref{def:hopset} holds
exactly when all edges of $H$ are noncontracting. Throughout the paper,
we assume that $\eps\le1/2$ and that
$1/\eps=O(\log^{c_0}n)$ for an arbitrary fixed constant $c_0\ge1$.

The central object of this paper is the $(h,\eps,x)$-implicit hopset
$(H_0,\mathcal B)$ of \cref{def:implicit-hopset-intro}, which
represents the hopset
$H=H_0\cup\bigcup_{B\in\mathcal B}\TC_w\bigl((G\cup H_0)[B]\bigr)$. The sets $B \in \mathcal{B}$ are the \textit{blocks} of the hopset. 
We refer to the set  $\TC_w((G\cup H_0)[B])$ as the \emph{block closure} of $B$.

\paragraph{Computational models.}
Our algorithms work in the streaming and parallel models.
In the streaming model, the algorithm can make multiple passes over a
stream in which the edges of the input graph arrive one at a time in an
arbitrary order. We measure space in words of $O(\log n)$ bits. In the
parallel model, the \emph{work} of an algorithm is the total number of
operations it performs. The \emph{depth} is defined as the length of
the longest chain of dependent operations~\cite{JaJa92}.

\subsection{Scaling and rounding}
\label{sec:prelim-scaling}

The scaling and rounding technique of Klein and Subramanian~\cite{KS}
allows us to use a hopset to compute $(1+\eps)$-approximate SSSP in 
few rounds, by reducing it to exact SSSP up to distance $O(h/\eps)$ in
$O(\log n)$ reweighted versions of $G\cup H$. This reduction is often 
applied implicitly in the graph algorithms literature. We state it explicitly because the details of this reduction affect our streaming shortest paths algorithm in \Cref{sec:core-algorithm}.

For a scaling parameter $\alpha>0$, let $G_\alpha$ and $H_\alpha$ be
obtained from $G$ and $H$ by giving each edge $e$ the integer weight
$w_\alpha(e)=\lceil w(e)/\alpha\rceil$. These are the rounded graphs
$\bar G$ and $\bar H$ of \cref{sec:overview-sssp}. As
$w(e)\le\alpha\,w_\alpha(e)<w(e)+\alpha$ for every edge $e$, any path
$P$ in $G\cup H$ with at most $h$ edges satisfies
\begin{equation}
 w(P)\le\alpha\,w_\alpha(P)<w(P)+h\alpha.
 \label{eq:path-rounding}
\end{equation}

The scaling and rounding reduction can be summarized with the following lemma.

\begin{lemma}[Scaling and rounding~\cite{KS}]
\label{lem:scale-round}
Let $H$ be an $(h,\eps)$-hopset of $G$, let $s\in V$, and let
$\Delta>0$ be a distance guess. For $\alpha=\eps\Delta/h$ and
$R=\lceil4h/\eps\rceil$, each vertex $v\in V$ satisfies
$\alpha\cdot\dist_{G_\alpha\cup H_\alpha}(s,v)\ge\dist_G(s,v)$. If
moreover $\Delta\le\dist_G(s,v)<2\Delta$, then
$\dist_{G_\alpha\cup H_\alpha}(s,v)\le R$ and
\[
 \alpha\cdot\dist_{G_\alpha\cup H_\alpha}(s,v)
 \le(1+2\eps)\dist_G(s,v).
\]
\end{lemma}

The lemma follows easily from \eqref{eq:path-rounding}, applied to a
path from $s$ to $v$ in $G\cup H$ with at most $h$ edges and length at
most $(1+\eps)\dist_G(s,v)$. Thus we only need to compute distance-bounded
shortest paths, up to distance $R=O(h/\eps)$.
Consider the $O(\log n)$ distance guesses $\Delta=2^jw_{\min}$ with
$0\le j\le\lceil\log_2(nW)\rceil$. For each guess $\Delta$, we set
$\alpha=\eps\Delta/h$ and define
$d_\Delta(v)=\dist_{G_\alpha\cup H_\alpha}(s,v)$ if this distance is at
most $R$ and $d_\Delta(v)=\infty$ otherwise. Any vertex $v\ne s$
reachable from $s$ has $\Delta\le\dist_G(s,v)<2\Delta$ for one of these
guesses. \Cref{lem:scale-round} then implies that
$\min_\Delta(\eps\Delta/h)\,d_\Delta(v)$ is a $(1+2\eps)$-approximation
of $\dist_G(s,v)$ for all $v\in V$.

\section{Streaming Shortest Paths}
\label{sec:streaming-reductions}

In this section, we give the proof of our main result.

\streamingthm*

Our algorithm first computes an implicit hopset $(H_0,\mathcal B)$ of
$G$, which is stored in memory. Then the implicit hopset is used to compute
distances from~$s$. Both of these steps rely on an algorithm that computes distance-bounded shortest paths in $G\cup H$ while storing only the pair $(H_0,\mathcal B)$. We define this problem below and solve it in \cref{sec:core-algorithm}. 

\subsection{The \core problem}
\label{sec:core-problem}

In this subsection, we define the distance-bounded shortest paths
problem that our algorithm solves both when it builds the implicit
hopset and when it uses the implicit hopset to compute distances
from~$s$.

To compute distances from $s$ with an implicit hopset
$(H_0,\mathcal B)$, we apply the scaling and rounding reduction of
\cref{sec:prelim-scaling} to the hopset
$H=H_0\cup\bigcup_{B\in\mathcal B}\TC_w((G\cup H_0)[B])$. For a
scaling parameter $\alpha>0$, the rounded graphs $G_\alpha$ and
$H_\alpha$ of \cref{sec:prelim-scaling} give each edge $e$ of $G$ and
$H$ the integer weight $\lceil w(e)/\alpha\rceil$. Inside a block $B$,
let $\dist_B(u,v)=\dist_{(G\cup H_0)[B]}(u,v)$ denote the distance from
$u$ to $v$ in the \emph{block subgraph} $(G\cup H_0)[B]$. This distance
is the weight of the block-closure edge $(u,v)$ of $B$, so this edge
has the weight
\begin{equation}
 \left\lceil\frac{\dist_B(u,v)}{\alpha}\right\rceil
 \label{eq:core-query-weights}
\end{equation}
in $H_\alpha$.

Recall from \cref{sec:prelim-scaling} that after scaling and rounding,
computing approximate distances with an $(h,\eps)$-hopset only requires exact
distances up to a radius $R=O(h/\eps)$ in $G_\alpha\cup H_\alpha$. The
following problem asks for these distances when only the pair
$(H_0,\mathcal B)$ is stored in memory. It is the main subroutine of
our streaming algorithm. Our hopset construction also solves this
problem while it is still building $(H_0,\mathcal B)$, so the problem
does not assume that $(H_0,\mathcal B)$ is an implicit hopset.

\begin{problem}[\core]
\label{prob:core-distance}
We are given a directed graph $G$ with positive edge weights as a
stream. We store in memory a set $H_0$ of edges with positive weights
and a collection $\mathcal B$ of blocks, with $|B|\le x$ for all
$B\in\mathcal B$ and $|H_0|+\sum_{B\in\mathcal B}|B|=\widetilde O(n)$.
Let $H=H_0\cup\bigcup_{B\in\mathcal B}\TC_w\bigl((G\cup H_0)[B]\bigr)$.
Given a scaling parameter $\alpha>0$, a source $s\in V$, and an
integer radius $R\ge0$, the goal is to compute, for each vertex
$v\in V$, the distance $\dist_{G_\alpha\cup H_\alpha}(s,v)$ if it is
at most $R$, and $\infty$ otherwise.
\end{problem}

The following theorem formalizes the batched Dijkstra's algorithm
sketched in \cref{sec:overview-sssp}. It solves \core in few passes
when the radius $R$ and the block size $x$ are small. We prove it in
\cref{sec:core-algorithm}.

\begin{restatable}{theorem}{boundedssspthm}
\label{thm:core-distance}
There is a deterministic streaming algorithm that solves \core using
$\widetilde O(n)$ space and $\widetilde O(R+x)$ passes.
\end{restatable}

\subsection{Proof of \texorpdfstring{\cref{cor:streaming-exact-sssp}}{the main theorem}}
\label{sec:streaming-reduction-chain}

We first combine \cref{thm:core-distance} with the scaling and
rounding reduction of \cref{sec:prelim-scaling}. This gives an
algorithm that computes approximate distances from $s$, assuming that
an implicit hopset is already stored in memory.

\begin{lemma}
\label{lem:approx-from-hopset}
Let $G$ be a directed graph with positive edge weights of polynomial
aspect ratio, given as a stream, and suppose that an
$(h,\eps,x)$-implicit hopset of $G$ is stored in memory. Then for any
source $s$, we can deterministically compute $(1+2\eps)$-approximate
distances from $s$ using $\widetilde O(n)$ space and
$\widetilde O(h/\eps+x)$ passes.
\end{lemma}
\begin{proof}
Let $(H_0,\mathcal B)$ be the implicit hopset stored in memory. We
treat $(H_0,\mathcal B)$ as the input to \core. We apply the scaling
and rounding reduction of \cref{sec:prelim-scaling} to compute
approximate distances from $s$. For each of the $O(\log n)$ distance
guesses $\Delta$, we compute the values $d_\Delta(v)$ with the
algorithm of \cref{thm:core-distance}, using $\alpha=\eps\Delta/h$ and
$R=\lceil4h/\eps\rceil$. Each run takes $\widetilde O(n)$ space and
$\widetilde O(h/\eps+x)$ passes. By \cref{lem:scale-round},
$\min_\Delta(\eps\Delta/h)\,d_\Delta(v)$ is a
$(1+2\eps)$-approximation of $\dist_G(s,v)$ for all $v\in V$.
\end{proof}

Next, we explain how to compute the implicit hopset that
\cref{lem:approx-from-hopset} assumes is stored in memory. We use the
hopbound
\begin{equation}
 h_x=\left\lceil\sqrt{n/x}\,n^{C_h/\log\log n}\right\rceil,
 \label{eq:streaming-hopbound}
\end{equation}
where $C_h>0$ is a sufficiently large constant and $x\in[n]$. The
following lemma gives a streaming construction of implicit hopsets with
this hopbound. It is the streaming version of the construction
sketched in \cref{sec:overview-hopset}, and we prove it in
\cref{sec:streaming-closures}.

\begin{restatable}[Streaming implicit hopsets]{lemma}{implicithopsetlemma}
\label{lem:implicit-hopset-reduction}
Given a directed graph $G$ with positive edge weights of polynomial
aspect ratio as a stream, an integer $x\in[n]$, and an accuracy
parameter $\eps$, there is a randomized streaming algorithm that, with
high probability, computes an $(h_x,\eps,x)$-implicit hopset of $G$
using $\widetilde O(n)$ space and $\widetilde O(h_x/\eps+x)$ passes.
\end{restatable}

Since $h_x=(n/x)^{1/2}n^{o(1)}$, \cref{lem:implicit-hopset-reduction}
implies \cref{thm:implicit-hopset-intro}.

By computing an implicit hopset with
\cref{lem:implicit-hopset-reduction} and then using it in
\cref{lem:approx-from-hopset}, we obtain the following streaming
algorithm for approximate SSSP. Together with an exact-to-approximate
reduction, it will give \cref{cor:streaming-exact-sssp}.

\begin{theorem}
\label{thm:streaming-sssp}
There is a randomized streaming algorithm that, given a directed graph
$G$ with positive edge weights of polynomial aspect ratio as a stream,
a source $s$, an integer $x\in[n]$, and an accuracy parameter $\eps$,
computes $(1+\eps)$-approximate distances from $s$ with high
probability, using $\widetilde O(n)$ space and
$\widetilde O(h_x/\eps+x)$ passes. For $x=\lceil n^{1/3}\rceil$, this
is $n^{1/3+o(1)}$ passes.
\end{theorem}
\begin{proof}
We compute an $(h_x,\eps/2,x)$-implicit hopset of $G$ with
\cref{lem:implicit-hopset-reduction} and then apply
\cref{lem:approx-from-hopset} with accuracy $\eps/2$. For
$x=\lceil n^{1/3}\rceil$, we have $h_x\le n^{1/3+O(1/\log\log n)}$.
\end{proof}

Finally, in \cref{sec:streaming-exact-reduction}, we will implement
the reduction of~\cite{Rozhon} from exact to approximate SSSP in the
streaming model and prove the following lemma.

\begin{restatable}[Exact from approximate SSSP~\cite{Rozhon}]{lemma}{exactreduction}
\label{thm:streaming-exact-reduction}
For any constant $a\ge1$, there is a constant $c=c(a)>0$ such that
the following holds. Suppose that a streaming algorithm computes
$(1+c/\log n)$-approximate distances from a source in directed graphs
with $O(n)$ vertices and positive, polynomially bounded integer edge
weights, using $\widetilde O(n)$ space and $T_{\rm apx}$ passes. Then
there is a streaming algorithm that computes an exact shortest-path tree in $n$-vertex directed graphs with
integer edge weights in $\{0,1,\dots,n^a\}$, using $\widetilde O(n)$
space and $O(T_{\rm apx}\log^2n)$ passes. If the approximate algorithm
succeeds with high probability, then so does the exact algorithm.
\end{restatable}

\begin{proof}[Proof of \cref{cor:streaming-exact-sssp}]
Let the edge weights of $G$ lie in $\{0,1,\dots,n^a\}$, and let
$c=c(a)$ be the constant of \cref{thm:streaming-exact-reduction}. In
graphs with $n'=O(n)$ vertices and positive, polynomially bounded
integer weights, \cref{thm:streaming-sssp} with $\eps=c/\log n$ and
$x=\lceil n'^{1/3}\rceil$ computes $(1+c/\log n)$-approximate
distances with high probability, using $\widetilde O(n)$ space and
$T_{\rm apx}=n^{1/3+o(1)}$ passes. \Cref{thm:streaming-exact-reduction}
then gives exact distances from $s$ and a shortest-path tree with high
probability, using $\widetilde O(n)$ space and
$O(T_{\rm apx}\log^2n)=n^{1/3+o(1)}$ passes.
\end{proof}

\subsection{An algorithm for \core}
\label{sec:core-algorithm}

In this subsection, we prove \cref{thm:core-distance}, which solves
\core deterministically using $\widetilde O(n)$ space and
$\widetilde O(R+x)$ passes. Recall that an instance
$(H_0,\mathcal B,\alpha,s,R)$ of \cref{prob:core-distance} consists of
the stored pair $(H_0,\mathcal B)$, a scaling parameter $\alpha$, a
source $s$ and a radius $R$. The goal is to compute the distances
from $s$ in the rounded graph $G_\alpha\cup H_\alpha$ of
\cref{sec:prelim-scaling} that are at most $R$. From now on, we fix
such an instance.

Recall that $G_\alpha\cup H_\alpha$ gives each edge $e$ of $G\cup H_0$
the integer weight $w_\alpha(e)=\lceil w(e)/\alpha\rceil$. By
\eqref{eq:core-query-weights}, it gives each block-closure edge
$(u,v)\in\TC_w((G\cup H_0)[B])$ of a block $B$ the weight
$\lceil\dist_B(u,v)/\alpha\rceil$. Our algorithm computes distances in
$G_\alpha\cup H_\alpha$, so all estimates $d(v)$ and round numbers $r$
below are distances in $G_\alpha\cup H_\alpha$. The only lengths taken
with the original weights are the distances $\dist_B(u,v)$ in the block
subgraphs $(G\cup H_0)[B]$, which our searches compute. We scale and
round these distances as needed to recover distances in
$G_\alpha\cup H_\alpha$.

\paragraph{Dijkstra's algorithm in rounds.}
Our algorithm is the procedure \batchdijkstra of
\Cref{alg:core-dijkstra}. Like the algorithm sketched in
\cref{sec:overview-sssp}, it is a variant of Dijkstra's algorithm on
$G_\alpha\cup H_\alpha$ that runs in rounds. Each vertex $v$ maintains
a distance estimate $d(v)$, with $d(s)=0$ and $d(v)=\infty$ for
$v\ne s$ initially. In round $r=0,1,\dots,R$, we \emph{settle} every
vertex $u$ with $d(u)=r$, i.e., we declare its estimate final. We will
show that round $r$ settles exactly the vertices at distance $r$ from $s$ in
$G_\alpha\cup H_\alpha$. This invariant is the main similarity between
our algorithm and Dijkstra's algorithm.

In round $r$, after settling all vertices $u$ with $d(u)=r$, we relax
the edges of $G\cup H_0$ out of these vertices.\footnote{Relaxing an
edge $(u,v)$ of $G_\alpha\cup H_\alpha$ means setting
$d(v)\gets\min\{d(v),d(u)+w_\alpha(u,v)\}$.} This takes one pass over
the stream. Unlike Dijkstra's algorithm, we do not relax the
block-closure edges, i.e., the edges of $\TC_w((G\cup H_0)[B])$ for the
blocks $B\in\mathcal B$, which are not stored. Relaxing the
block-closure edges of a block $B$ out of a settled vertex $u\in B$
would require the distances $\dist_B(u,v)$ for all $v\in B$, and
computing them requires loading the whole block subgraph
$(G\cup H_0)[B]$, which can have $x^2$ edges. Instead, in round $r$, we
only look for the vertices to which the block-closure edges give the
estimate~$r$. To make this precise, let
\[
 U=\{v\in V:\dist_{G_\alpha\cup H_\alpha}(s,v)=r\}
\]
denote the set of \emph{newly settled vertices} in round $r$. For a
block $B$, let $U_B$ be the set of vertices $v\in B$ not settled before
round $r$ for which some block-closure edge
$(u,v)\in\TC_w((G\cup H_0)[B])$ with $u$ settled before round $r$
satisfies $d(u)+\lceil\dist_B(u,v)/\alpha\rceil\le r$. Intuitively,
$U_B$ consists of the newly settled vertices of round $r$ that we would
find by relaxing the block-closure edges of $B$ out of the settled
vertices.

We will show that the following two invariants hold at the start of
each round $r$. The second one says that the edges of $G\cup H_0$ out
of the settled vertices have been relaxed, as in Dijkstra's algorithm.
\begin{enumerate}[label=(I\arabic*),itemsep=1pt]
 \item Each vertex $v$ has $d(v)\ge\dist_{G_\alpha\cup H_\alpha}(s,v)$.
 The vertices settled before round $r$ are exactly the vertices $v$
 with $\dist_{G_\alpha\cup H_\alpha}(s,v)<r$, and each of them has the
 estimate $d(v)=\dist_{G_\alpha\cup H_\alpha}(s,v)$.
 \item For each vertex $u$ settled before round $r$ and each edge
 $(u,v)$ of $G\cup H_0$, we have $d(v)\le d(u)+w_\alpha(u,v)$.
\end{enumerate}
A vertex settles in round $r$ either through an edge of $G\cup H_0$,
which an earlier round has relaxed, or through a block-closure edge.
More precisely, (I1) and (I2) imply that
\begin{equation}
 U=\{v\in V:v\text{ is not settled before round }r\text{ and }d(v)=r\}
 \cup\bigcup_{B\in\mathcal B}U_B.
 \label{eq:core-round}
\end{equation}
To see this, let $v$ be a vertex not settled before round $r$, which
has distance at least $r$ from $s$ by (I1). If $d(v)=r$, then this
distance is at most $r$ by (I1). If $v\in U_B$, then some vertex
$u\in B$ settled before round $r$ has
$d(u)+\lceil\dist_B(u,v)/\alpha\rceil\le r$. A shortest path from $s$
to $u$, which has length $d(u)$ by (I1), followed by the block-closure
edge $(u,v)$, is then a path from $s$ to $v$ of length at most $r$ in
$G_\alpha\cup H_\alpha$. In both cases, $v\in U$, so the right side of
\eqref{eq:core-round} is contained in $U$. Conversely, let $v\in U$. If $v=s$, then $r=0$ and $d(s)=0$.
Otherwise, let $(u,v)$ be the last edge of a shortest path from $s$ to
$v$ in $G_\alpha\cup H_\alpha$. Its weight is at least $1$, so $u$ has
distance less than $r$ from $s$ and was settled before round $r$ with
$d(u)=\dist_{G_\alpha\cup H_\alpha}(s,u)$. If $(u,v)$ is an edge of
$G\cup H_0$, then $d(v)\le d(u)+w_\alpha(u,v)=r$ by (I2), and thus
$d(v)=r$ by (I1). If $(u,v)$ is a block-closure edge of a block $B$, then
$v\in U_B$.

By \eqref{eq:core-round}, round $r$ settles exactly the set $U$ if it
first finds the sets $U_B$ and gives their vertices the estimate $r$,
and then settles the vertices $u$ with $d(u)=r$. The rest of this
subsection explains how we find the sets $U_B$. For this purpose, the
algorithm stores two things for each block $B$. The first is a set
$\mathsf{Known}_B\subseteq B$, which will be the set of vertices of $B$
settled so far. The second is a round $\mathsf{next}_B\ge1$, which is
the next round in which the algorithm searches $B$ for newly settled
vertices, and which is $\infty$ while no such search is scheduled.

\begin{algorithm}[t]
\caption{Our batched Dijkstra's algorithm for \core.}
\label{alg:core-dijkstra}
\begin{algorithmic}[1]
\Procedure{BatchedDijkstra}{$s,R$}
 \State $d(s)\gets0$, and $d(v)\gets\infty$ for $v\ne s$
 \State $\mathsf{Known}_B\gets\varnothing$ for all $B\in\mathcal B$
 \Comment{settled vertices of $B$}
 \State $\mathsf{next}_B\gets\infty$ for all $B\in\mathcal B$
 \Comment{next round in which we search $B$}
 \For{$r=0,1,\dots,R$}
  \For{each block $B$ with $\mathsf{next}_B=r$}
  \Comment{search $B$ for newly settled vertices}
   \State $(U_B,e_B)\gets{}$\Call{SearchBlock}{$B,r$}
   \State $d(v)\gets r$ for all $v\in U_B$, and
   $\mathsf{next}_B\gets\lceil e_B\rceil$
  \EndFor
  \For{each vertex $u$ with $d(u)=r$} \Comment{settle $u$}
   \For{each block $B\ni u$ with $u\notin\mathsf{Known}_B$}
    \State Insert $u$ into $\mathsf{Known}_B$ with $d_B(u)\gets r$, and
    set $\mathsf{next}_B\gets r+1$
    \label{line:core-join}
   \EndFor
  \EndFor
  \State Relax all edges $(u,v)$ of $G\cup H_0$ with $d(u)=r$
  \Comment{one pass}
 \EndFor
 \State \Return $d$
\EndProcedure
\Statex
\Procedure{SearchBlock}{$B,r$}
 \For{$i=0,1,\dots,\lceil\log n\rceil$}
  \State $\tau\gets2^i$
  \State Build a $\tau$-skeleton $F$ of $(G\cup H_0)[B]$ without the
  edges into $\mathsf{Known}_B$ \Comment{one pass}
  \State Compute $d_F(v)=\min_{u\in\mathsf{Known}_B}\bigl(d_B(u)
  +\dist_F(u,v)/\alpha\bigr)$ for $v\in B$
  \Comment{Dijkstra in $F$}
  \State $U_F\gets\{v\in B\setminus\mathsf{Known}_B:d_F(v)\le r\}$
  \If{$|U_F|<\tau$} \Comment{then $U_F=U_B$}
   \State Insert each $v\in U_F$ into $\mathsf{Known}_B$ with
   $d_B(v)\gets d_F(v)$
   \State \Return $U_F$ and
   $e_B=\min\{d_F(v):v\in B\setminus\mathsf{Known}_B\}$
  \EndIf
 \EndFor
\EndProcedure
\end{algorithmic}
\end{algorithm}

\paragraph{Block distances.}
To find $U_B$, we use the estimate that the block-closure edges of $B$
give each vertex, before rounding. For a block $B$ and a round $r$, the
\emph{block distance} of a vertex $v\in B$ is
\[
 d_B(v)=\min_{\substack{u\in B\text{ settled}\\\text{before round }r}}
 \bigl(d(u)+\dist_B(u,v)/\alpha\bigr).
\]
By (I3) below, the vertices $u$ in this minimum are the vertices of
$\mathsf{Known}_B$ at the start of round $r$. As each $d(u)$ is an
integer and the ceiling function is nondecreasing,
minimizing over the same vertices $u$ gives
\begin{equation}
 \left\lceil d_B(v)\right\rceil
 =\min_u\bigl(d(u)+\left\lceil\dist_B(u,v)/\alpha\right\rceil\bigr).
 \label{eq:core-ceiling}
\end{equation}
The right side is the estimate that $v$ would get from relaxing the
block-closure edges of $B$ out of the settled vertices, so $d_B(v)$ is
this estimate before rounding. Hence $U_B$ can be written as
\[
 U_B=\{v\in B:v\text{ is not settled before round }r\text{ and }
 d_B(v)\le r\}.
\]
Our goal is now to compute $U_B$, and by this characterization, it
suffices to compute the block distances $d_B(v)$ for $v\in B$. The
value $d_B(v)$ is the distance to $v$ in $(G\cup H_0)[B]$, with lengths
divided by $\alpha$, from the settled vertices of $B$, where each
settled vertex $u$ starts at $d(u)$. We can therefore compute all block
distances of $B$ with a single run of Dijkstra's algorithm in
$(G\cup H_0)[B]$.

Along with each vertex $u\in\mathsf{Known}_B$, our algorithm stores the
block distance $d_B(u)$. The proof of \cref{lem:core-global-search}
shows that these sets satisfy the following invariant.
\begin{enumerate}[label=(I\arabic*),itemsep=1pt,start=3]
 \item For each block $B$, the set $\mathsf{Known}_B$ consists of the
 vertices of $B$ settled before round $r$, and each
 $u\in\mathsf{Known}_B$ is stored with its block distance $d_B(u)$.
\end{enumerate}

\paragraph{Finding $U_B$ with $\tau$-skeletons.}
However, as in \cref{sec:overview-sssp}, we do not load the whole
block subgraph $(G\cup H_0)[B]$, which can have $x^2$ edges. Instead, we load a skeleton subgraph of
it, whose size grows with $|U_B|$. For an integer $\tau\ge1$, a
\emph{$\tau$-skeleton} of a weighted graph is a subgraph that keeps the
$\tau$ lightest outgoing edges at each vertex and drops the rest. To
search $B$ in round $r$, we build in one pass a $\tau$-skeleton $F$ of
the graph obtained from $(G\cup H_0)[B]$ by deleting every edge $(a,b)$
with $b\in\mathsf{Known}_B$. Explicitly, for each vertex $a\in B$, the
skeleton $F$ contains the $\tau$ lightest edges $(a,b)$ of $G\cup H_0$
with $b\in B\setminus\mathsf{Known}_B$, or all of them if there are
fewer than $\tau$.\footnote{If $G\cup H_0$ has several edges from $a$
to $b$, we keep only the lightest one in $(G\cup H_0)[B]$.} We then
compute the values
\[
 d_F(v)=\min_{u\in\mathsf{Known}_B}\bigl(d_B(u)+\dist_F(u,v)/\alpha\bigr)
 \quad\text{for all }v\in B
\]
by running Dijkstra's algorithm in $F$. This uses no additional passes,
since $F$ is stored in memory. The search then reports the set
\[
 U_F=\{v\in B\setminus\mathsf{Known}_B:d_F(v)\le r\}.
\]

The value $d_F(v)$ is the block distance of $v$ computed in $F$ instead
of the whole block subgraph. Its definition mirrors that of $d_B(v)$,
with two differences. First, the distances are taken in $F$, which a
search can afford to load. Second, each $u\in\mathsf{Known}_B$ starts
at its block distance $d_B(u)$ instead of $d(u)$. This is needed
because $F$ has no edges into $\mathsf{Known}_B$. A shortest path in
$(G\cup H_0)[B]$ to $v$ may pass through $u$, and $d_B(u)$ accounts for
its part up to $u$. (Because of rounding, $d_B(u)$ can be smaller than
$d(u)$. In \cref{sec:overview-sssp}, which ignores rounding, each
settled vertex $u$ starts at $d(u)$.) We will show that the values
$d_F(v)$ are never smaller than the block distances, and that they are
equal to them on $U_B$ if the search finds fewer than $\tau$ vertices.
As a consequence, the search of $B$ reports only vertices of $U_B$, and
it finds all of them if it finds fewer than $\tau$, as the next lemma
shows.

\begin{lemma}[Skeletons find all new vertices]
\label{lem:core-skeleton-certificate}
Suppose that $\mathsf{Known}_B$ consists of the vertices of $B$ settled
before round $r$, stored with their block distances $d_B(u)$. Then
$U_F\subseteq U_B$. If moreover $|U_F|<\tau$, then $U_F=U_B$.
\end{lemma}
\begin{proof}
We first show that $d_F(v)\ge d_B(v)$ for all $v\in B$. For
$u\in\mathsf{Known}_B$, the triangle inequality for $\dist_B$ gives
$d_B(u)+\dist_B(u,v)/\alpha\ge d_B(v)$. As $F$ is a subgraph of
$(G\cup H_0)[B]$, we have $\dist_F(u,v)\ge\dist_B(u,v)$, and the claim
follows.

Next, we show that if $d_F(v)>d_B(v)$ for some
$v\in B\setminus\mathsf{Known}_B$, then at least $\tau$ vertices
$z\in B\setminus\mathsf{Known}_B$ have $d_F(z)\le d_B(v)$. Let $P$ be a
shortest path in $(G\cup H_0)[B]$ from a settled vertex $y$ to $v$ with
$d(y)+w(P)/\alpha=d_B(v)$. Let $u$ be the last vertex of $P$ in
$\mathsf{Known}_B$, and let $P'$ be the suffix of $P$ from $u$ to $v$.
No edge of $P'$ goes into $\mathsf{Known}_B$, and as
$d_B(u)\le d(y)+w(P[y,\dots,u])/\alpha$, we have
$d_B(u)+w(P')/\alpha\le d_B(v)$. If $F$ contained all edges of $P'$,
then we would have $d_F(v)\le d_B(v)$. Hence $F$ omits an edge of $P'$;
let $(a,b)$ be the first one. As $(a,b)$ does not go into
$\mathsf{Known}_B$, the skeleton $F$ keeps instead $\tau$ edges $(a,z)$
with $w(a,z)\le w(a,b)$ that do not go into $\mathsf{Known}_B$. These
edges go to $\tau$ distinct vertices $z\in B\setminus\mathsf{Known}_B$.
Since $F$ contains the prefix $P'[u,a]$ and the edge $(a,z)$, each such
$z$ satisfies
\begin{align*}
 d_F(z)&\le d_B(u)+\bigl(w(P'[u,a])+w(a,z)\bigr)/\alpha\\
 &\le d_B(u)+w(P')/\alpha\le d_B(v).
\end{align*}

Finally, each $v\in U_F$ is not settled before round $r$ and satisfies
$d_B(v)\le d_F(v)\le r$, which gives $U_F\subseteq U_B$. Now let
$|U_F|<\tau$, and suppose that some $v\in U_B$ is not in $U_F$. Then
$d_B(v)\le r<d_F(v)$, so at least $\tau$ vertices
$z\in B\setminus\mathsf{Known}_B$ have $d_F(z)\le d_B(v)\le r$. All of
them lie in $U_F$, which contradicts $|U_F|<\tau$.
\end{proof}

By \cref{lem:core-skeleton-certificate}, the procedure
\textsc{SearchBlock} of \Cref{alg:core-dijkstra} finds $U_B$ with the
standard doubling trick: it builds a $\tau$-skeleton for
$\tau=2^0,2^1,2^2,\dots$ until $|U_F|<\tau$. Since $\tau$ is doubled only if
$|U_B|\ge|U_F|\ge\tau$, a search of $B$ builds $O(1+\log x)$ skeletons
with $O(|B|(|U_B|+1))$ edges in total, each with at most $2x^2$ edges.
We assume that $x\le\sqrt{n/2}$, so that no skeleton has more than $n$
edges. Larger $x$ only needs routine changes, which we omit. They use
that the $\tau$ vertices $v\in B\setminus\mathsf{Known}_B$ with the
smallest values $d_F(v)$ satisfy $d_F(v)=d_B(v)$, by the argument of
\cref{lem:core-skeleton-certificate}.

\paragraph{Scheduling the block searches.}
In \cref{sec:overview-sssp}, we search every block in every round.
Here we search a block $B$ only in rounds in which $U_B$ can be
nonempty, and $\mathsf{next}_B$ is the next such round. Besides $U_B$,
a search of $B$ returns the smallest value
\[
 e_B=\min\{d_F(v):v\in B\setminus\mathsf{Known}_B\},
\]
where $\mathsf{Known}_B$ already contains the vertices of $U_B$. By
\cref{lem:core-next-search} below, $e_B$ is the smallest block distance
of a vertex of $B$ outside $\mathsf{Known}_B$. As long as the block
distances of $B$ do not change, $U_B$ is empty before round
$\lceil e_B\rceil$, and we set $\mathsf{next}_B=\lceil e_B\rceil$.

A vertex $u\in B$ can also settle in a round $r$ without lying in
$U_B$, through an edge of $G\cup H_0$ or a block-closure edge of
another block. We then insert $u$ into $\mathsf{Known}_B$ and set
$d_B(u)=r$, which is the block distance of $u$ from round $r+1$ on
(Line~\ref{line:core-join} of \Cref{alg:core-dijkstra}). As $u$ can
lower the block distances of other vertices of $B$, we also set
$\mathsf{next}_B=r+1$, so that $B$ is searched again in the next round.
We will show that these choices maintain the following invariant.
\begin{enumerate}[label=(I\arabic*),itemsep=1pt,start=4]
 \item Each block $B$ satisfies
 \[
  r\le\mathsf{next}_B\le\lceil d_B(v)\rceil
  \quad\text{for all }v\in B\text{ not settled before round }r.
 \]
\end{enumerate}
The invariant (I4) says that $U_B$ is empty in each round
$r<\mathsf{next}_B$. Indeed, in such a round, (I4) gives
$\lceil d_B(v)\rceil\ge\mathsf{next}_B>r$, and thus $d_B(v)>r$, for each
vertex $v\in B$ not settled before round $r$. To maintain (I3) and
(I4), we need that a search of $B$ stores the correct block distances
$d_B(v)$, and that $e_B$ is the smallest block distance of a vertex of
$B$ outside the updated set $\mathsf{Known}_B$.

\begin{lemma}
\label{lem:core-next-search}
Suppose that $\mathsf{Known}_B$ consists of the vertices of $B$ settled
before round $r$, stored with their block distances $d_B(u)$, and
that
$|U_F|<\tau$. Then $d_F(v)=d_B(v)$ for all $v\in U_F$, and the smallest
value $d_F(v)$ of a vertex $v\in B\setminus(\mathsf{Known}_B\cup U_F)$
is the smallest block distance of such a vertex.
\end{lemma}
\begin{proof}
The proof of \cref{lem:core-skeleton-certificate} shows that
$d_F(v)\ge d_B(v)$ for all $v\in B$. It also shows that if
$d_F(v)>d_B(v)$ for some $v\in B\setminus\mathsf{Known}_B$, then at
least $\tau$ vertices $z\in B\setminus\mathsf{Known}_B$ have
$d_F(z)\le d_B(v)$. For $v\in U_F$, all these vertices $z$ would have
$d_F(z)\le d_B(v)\le r$ and lie in $U_F$, which contradicts
$|U_F|<\tau$. Hence $d_F(v)=d_B(v)$ for all $v\in U_F$.

Now let $v\in B\setminus(\mathsf{Known}_B\cup U_F)$ have the smallest
block distance. As $d_F\ge d_B$, it suffices to find a vertex
$z\in B\setminus(\mathsf{Known}_B\cup U_F)$ with $d_F(z)\le d_B(v)$. If
$d_F(v)=d_B(v)$, we take $z=v$. Otherwise, there are $\tau$ vertices
$z\in B\setminus\mathsf{Known}_B$ with $d_F(z)\le d_B(v)$, and as
$|U_F|<\tau$, one of them lies outside $U_F$.
\end{proof}

\paragraph{Correctness.}
We can now summarize \Cref{alg:core-dijkstra}. In each round $r$, its
procedure \batchdijkstra first searches each block $B$ with
$\mathsf{next}_B=r$ using \textsc{SearchBlock}, and it gives the
vertices of $U_B$ the estimate $r$. It then settles the vertices $u$
with $d(u)=r$ and inserts each of them into the sets $\mathsf{Known}_B$
that do not contain it yet. Finally, it relaxes the edges of
$G\cup H_0$ out of these vertices. We next show that \batchdijkstra
returns the correct distances, by proving the invariants (I1)--(I4).

\begin{lemma}
\label{lem:core-global-search}
For each vertex $v$, \batchdijkstra returns the estimate
$d(v)=\dist_{G_\alpha\cup H_\alpha}(s,v)$ if this distance is at most
$R$, and an estimate $d(v)>R$ otherwise.
\end{lemma}
\begin{proof}
We prove by induction on $r$ that (I1)--(I4) hold at the start of each
round $r=0,1,\dots,R+1$. At the start of round $0$, no vertex is
settled and all block distances are $\infty$, so the invariants hold
with $\mathsf{Known}_B=\varnothing$ and $\mathsf{next}_B=\infty$. For
$r=R+1$, (I1) is the claim of the lemma: the vertices at distance at
most $R$ from $s$ are settled with $d(v)=\dist_{G_\alpha\cup H_\alpha}(s,v)$,
and each other vertex $v$ has
$d(v)\ge\dist_{G_\alpha\cup H_\alpha}(s,v)>R$.

Now assume (I1)--(I4) at the start of a round $r\le R$.

\textit{Searches and settling.} By (I3) and
\cref{lem:core-skeleton-certificate}, a search of a block $B$ in round
$r$ returns $U_B$, and the vertices of $U_B$ get the estimate $r$. Each
block $B$ with $U_B\ne\varnothing$ is searched in round $r$, since a
vertex $v\in U_B$ gives $\mathsf{next}_B\le\lceil d_B(v)\rceil\le r$ by
(I4), and thus $\mathsf{next}_B=r$. By \eqref{eq:core-round}, the
vertices not settled before round $r$ that now have the estimate $r$
form the set $U$, and round $r$ settles them. Each estimate that round
$r$ sets is either the estimate $r$ of a vertex of $U$ or a value
$d(u)+w_\alpha(u,v)\ge\dist_{G_\alpha\cup H_\alpha}(s,v)$ with
$u\in U$. Hence (I1) holds at the start of round $r+1$. The relaxation
of the edges of $G\cup H_0$ out of the vertices of $U$ in round $r$
gives (I2) at the start of round $r+1$, since estimates never
increase.

\textit{Blocks.} Now fix a block $B$. Its block distances change only
through the vertices of $B$ that settle in round $r$. These are the
vertices of $U_B$, which the search of $B$ inserts into
$\mathsf{Known}_B$, and the vertices that Line~\ref{line:core-join}
inserts. A vertex $u\in U_B$ has $d(u)=r\ge d_B(u)$, so
$d(u)+\dist_B(u,z)/\alpha\ge d_B(u)+\dist_B(u,z)/\alpha\ge d_B(z)$ for
all $z\in B$. Hence the vertices of $U_B$ do not change the block
distances of $B$.

\textit{Case~1: Line~\ref{line:core-join} does not insert a vertex into 
$\mathsf{Known}_B$ in round $r$.} Then the block distances of $B$ do not
change. If $B$ was searched in round $r$, then the search stores the
block distances of the vertices of $U_B$ by \cref{lem:core-next-search},
which gives (I3). Each vertex $v$ of $B$ outside the updated set
$\mathsf{Known}_B$ has $d_F(v)>r$, since otherwise it would lie in
$U_F$. Hence $e_B>r$. Now \cref{lem:core-next-search} and the choice
$\mathsf{next}_B=\lceil e_B\rceil$ give (I4) at the start of round
$r+1$. If $B$ was not searched in round $r$, then $U_B=\varnothing$, and
$\mathsf{Known}_B$ and $\mathsf{next}_B$ do not change. As
$\mathsf{next}_B\ne r$, (I4) gives $\mathsf{next}_B\ge r+1$, and (I3)
and (I4) hold at the start of round $r+1$.

\textit{Case~2: Line~\ref{line:core-join} inserts at least one vertex
into $\mathsf{Known}_B$ in round $r$.} A vertex $u$ that
Line~\ref{line:core-join} inserts is not in $U_B$, since the search of
$B$ inserts the vertices of $U_B$. Thus $d_B(u)>r$ before round $r$,
and from round $r+1$ on, the block distance of $u$ is $d_B(u)=d(u)=r$,
which is the stored value. The
vertices $z\in B$ settled before round $r$ or in $U_B$ have
$d_B(z)\le r$. They keep their block distances, since each vertex $y$
settled in round $r$ has $d(y)+\dist_B(y,z)/\alpha\ge r$. Together with
\cref{lem:core-next-search} for the vertices of $U_B$, this gives (I3)
at the start of round $r+1$. Finally, Line~\ref{line:core-join} sets
$\mathsf{next}_B=r+1$, and each vertex $v\in B$ not settled by the end
of round $r$ has
$\lceil d_B(v)\rceil\ge\dist_{G_\alpha\cup H_\alpha}(s,v)\ge r+1$ by
\eqref{eq:core-ceiling} and (I1). This gives (I4) at the start of
round $r+1$.
\end{proof}

We now bound the passes and the space of \batchdijkstra.

\begin{proof}[Proof of \cref{thm:core-distance}]
By \cref{lem:core-global-search}, \batchdijkstra solves \core
deterministically if we replace each returned estimate above $R$ by
$\infty$. It uses $R+1$ passes for the relaxations. It remains to
bound the cost of the block searches. Let $\kappa$ be the total number
of insertions of vertices into the sets $\mathsf{Known}_B$ for
$B\in\mathcal B$. Each vertex $v$ is inserted into $\mathsf{Known}_B$ at
most once per block $B\ni v$, so
$\kappa\le\sum_{B\in\mathcal B}|B|=\widetilde O(n)$.

We charge each search of a block $B$ to a distinct insertion into
$\mathsf{Known}_B$. If Line~\ref{line:core-join} has inserted a vertex
into $\mathsf{Known}_B$ since the previous search of $B$, we charge the
search to one such insertion. This covers the first search of $B$, as
$\mathsf{next}_B=\infty$ until such an insertion. Otherwise, the search
runs in round $r=\lceil e_B\rceil$ for the value $e_B$ returned by the
previous search of $B$. By the proof of \cref{lem:core-global-search},
the block distances of $B$ have not changed since that search. By
\cref{lem:core-next-search}, some vertex outside $\mathsf{Known}_B$ has
the block distance $e_B\le r$, so it lies in $U_B$ and this search
inserts it into $\mathsf{Known}_B$. We charge the search to this
insertion. Hence there are at most $\kappa$ block searches in total.

Next, we bound the total number of skeleton edges. A search of a block
$B$ that returns $U_B$ builds skeletons with $O(|B|(|U_B|+1))$ edges in
total, which is $O(x(|U_B|+1))$ as $|B|\le x$. We sum this bound over
all block searches. The search of $B$ inserts the vertices of $U_B$
into $\mathsf{Known}_B$, so the sets $U_B$ of all searches have total
size at most $\kappa$. As there are also at most $\kappa$ searches, all
skeletons together have $O(x\kappa)$ edges.

In each round, we search several blocks. A search of a block $B$ reads
only $\mathsf{Known}_B$ and the stored block distances of $B$, so the
searches of different blocks in the same round do not affect each
other. We therefore run them side by side over the same passes, in
phases. In phase $i$, we build the $i$-th skeleton of each search of
the round that has not yet returned. As we have $O(n)$ space, each pass
loads skeleton edges with total size $O(n)$ into
memory.\footnote{This uses that no skeleton has more than $n$ edges.}
If the skeletons of a phase have $E$ edges in total, the phase
therefore takes $O(1+E/n)$ passes. With $O(1+\log x)$ phases per round,
the searches take
$O((R+1)(1+\log x)+x\kappa/n)=\widetilde O(R+x)$ passes. The remaining
memory, including the sets $\mathsf{Known}_B$, takes
$O(n+|H_0|+\sum_{B\in\mathcal B}|B|)=\widetilde O(n)$ space.
\end{proof}

\paragraph{Batches of searches.}
Our hopset construction in \cref{sec:streaming-closures} will run
\batchdijkstra from many pivots at once, inside the subgraphs of the
recursion. We therefore extend \cref{thm:core-distance} to
\emph{batches} of searches that share the scaling parameter $\alpha>0$
and the stored pair $(H_0,\mathcal B)$. Here $H_0$ is a set of edges
with positive weights, and $\mathcal B$ is a collection of blocks with
$|B|\le x$. Search $j$ of a batch has a stored vertex set
$S_j\subseteq V$, a direction (forward or backward), a source
$s_j\in S_j$ and an integer radius $R_j\ge0$. It solves the instance
$(H_0[S_j],\{B\cap S_j:B\in\mathcal B\},\alpha,s_j,R_j)$ of \core on the
graph $G[S_j]$, with all edges reversed if search $j$ is backward. That
is, it computes distances up to $R_j$ in the rounded version of the
graph
\[
 (G\cup H_0)[S_j]\cup\bigcup_{B\in\mathcal B}
 \TC_w\bigl((G\cup H_0)[B\cap S_j]\bigr).
\]
We restrict the blocks to $S_j$ because the proof of
\cref{thm:core-distance} charges each search of a block to a vertex of
the block that settles, and a vertex outside $S_j$ never settles in
search $j$.

We measure the cost of a batch of searches in \emph{distance
records}. When search
$j$ first gives a vertex $v$ an estimate $d(v)\le R_j$, it creates one
record $(j,v)$ and one record $(j,B,v)$ for each block $B\ni v$. Search
$j$ then settles $v$ by round $R_j$. Hence the total number of distance
records of the batch is
\[
 N_{\rm rec}=\sum_j\;\sum_{v\text{ settled by search }j}
 \Bigl(1+\bigl|\{B\in\mathcal B:v\in B\}\bigr|\Bigr).
\]
For a single search on all of $V$, this is at most
$n+\sum_{B\in\mathcal B}|B|$. The next lemma bounds the passes and the
space of a batch in terms of $N_{\rm rec}$.

\begin{lemma}
\label{lem:core-batch}
We can run a batch of searches with maximum radius $R=\max_jR_j$
deterministically in
\begin{equation}
 O\left((R+1)(1+\log x)+\frac{xN_{\rm rec}}{n}\right)\text{ passes}
 \label{eq:core-batch-cost}
\end{equation}
and $O(n+N_{\rm rec})$ space, in addition to the space that stores
$H_0$, $\mathcal B$, the distinct sets $S_j$, and the sources $s_j$. At
any time, this additional space is $O(n)$ plus the number of distance
records created so far.
\end{lemma}
\begin{proof}
We follow the proof of \cref{thm:core-distance}, with one instance of
\batchdijkstra for each search $j$, restricted to the stream edges with
both endpoints in $S_j$. The instances run their rounds in lockstep. In
each round, they share one relaxation pass and run their block searches
side by side in phases. In instance $j$, a vertex $v$ is inserted into
$\mathsf{Known}_{B\cap S_j}$ at most once, and only after search $j$
has given $v$ an estimate $d(v)\le R_j$, i.e., after the record $(j,v)$
has been created. Therefore, each insertion corresponds to a distinct
distance record $(j,B,v)$, which bounds the number of insertions by
$\kappa\le N_{\rm rec}$ and gives \eqref{eq:core-batch-cost}. The restricted
blocks are determined by the stored sets $B$ and $S_j$ and are never
stored themselves.

To bound the space, note that instance $j$ can discard each estimate
above $R_j$, which does not change its output. It then stores an
estimate $d(v)$ only after creating the record $(j,v)$. Each stored
block distance belongs to a vertex of some $\mathsf{Known}_{B\cap S_j}$.
The value $\mathsf{next}_{B\cap S_j}$ is set to a finite value only
after a vertex has been inserted into $\mathsf{Known}_{B\cap S_j}$.
Besides the $O(n)$ skeleton edges, each stored value thus belongs to an
existing distance record.
\end{proof}

When we apply \cref{lem:core-batch} in \cref{sec:streaming-closures},
each batch will create $N_{\rm rec}=\widetilde O(n)$ distance records.
A batch of searches then takes $\widetilde O(R+x)$ passes and
$\widetilde O(n)$ additional space, which matches the bounds of
\cref{thm:core-distance} for a single search up to polylogarithmic
factors.

\section{Sequential Hopset Construction}
\label{sec:sequential}

In this section, we present a sequential algorithm for our hopset
construction. This algorithm is the basis for our parallel and
streaming constructions in \cref{sec:parallel,sec:streaming-closures}.
Specifically, we prove the following theorem on sequential hopset
construction, where $\omega$ denotes the exponent of matrix
multiplication.

\begin{restatable}[Sequential hopset construction]{theorem}{hopsetthm}
\label{thm:hopset}
Given an $n$-vertex directed graph $G$ with $m$ edges and positive edge
weights of polynomial aspect ratio, a parameter $1\le x\le n$, and an
accuracy parameter $\eps$, there is a randomized algorithm that
computes, with high probability, an $(h,\eps)$-hopset $H$ of $G$ with
\[
 h=\sqrt{n/x}\,n^{O(1/\log\log n)}
 \qquad\text{and}\qquad
 |H|=\widetilde O(nx),
\]
using $\widetilde O(m+nx^{\omega-1})$ work.
\end{restatable}

Our construction modifies the hopset of Cao, Fineman, and
Russell~\cite{CFR}, the \emph{CFR hopset}, only at the late levels of
its recursion, as sketched in \cref{sec:overview-hopset}. We recall the
CFR hopset in \cref{sec:cfr} and describe our modification in
\cref{sec:two-speed}. The analysis then follows a fixed path through the
recursion. For the early levels, \cref{sec:early} imports facts
from~\cite{CFR}, and for the late levels, \cref{sec:late} gives a new
argument. \Cref{sec:one-execution} combines the two parts and
proves \cref{thm:hopset}. Finally, \cref{sec:shared-path-shortening}
shows how to build the hopset in $O(\log n)$ stages, which our parallel
and streaming algorithms will need.

\subsection{The CFR hopset}
\label{sec:cfr}

In this subsection, we recall the CFR hopset~\cite[Sections~3 and~5]{CFR}, which our construction keeps at the early levels. It is an $(n^{1/2+o(1)},\eps)$-hopset of size $\widetilde O(n/\eps^2)$, computed with $\widetilde O(m/\eps^2)$ work~\cite[Theorem~14]{CFR}.
For a subgraph $G'$ of $G$, a vertex $v$ of $G'$, and $d\ge0$, we let
\[
 \Rout_d(G',v)=\{u:\dist_{G'}(v,u)\le d\},
 \qquad
 \Rin_d(G',v)=\{u:\dist_{G'}(u,v)\le d\},
\]
and $\Rboth_d(G',v)=\Rin_d(G',v)\cup\Rout_d(G',v)$. These are the
vertices that $v$ reaches within distance $d$ in $G'$, the vertices
that reach $v$ within distance $d$, and their union.

\paragraph{Executions and trials.}
The CFR hopset handles paths of different lengths separately. Recall
that $w_{\min}$ denotes the smallest edge weight and
$W=w_{\max}/w_{\min}$ the aspect ratio. The construction has two
nested loops, as in~\cite[Algorithm~1]{CFR} and~\cite[Section~6.1]{ABGS}.
\begin{itemize}[itemsep=1pt]
 \item \emph{Inner loop over distance guesses.} For each guess
 $\Delta=2^jw_{\min}$ with $0\le j\le\lceil\log_2(2nW)\rceil$, a
 randomized recursion adds edges that shortcut paths of length about
 $\Delta$. We call one run of this recursion, for one fixed guess
 $\Delta$, the \emph{execution} for $\Delta$.
 \item \emph{Outer loop for boosting.} A \emph{trial} is one pass of
 the inner loop, that is, one execution for each of the $O(\log n)$
 guesses. A single trial shortcuts any fixed path only with constant
 probability, so the construction runs $O(\log n)$ independent trials
 and outputs the union of the edges that they add.
\end{itemize}
Thus a trial consists of $O(\log n)$ executions, and the whole
construction runs $O(\log^2n)$ executions.

\paragraph{Levels and ranks.}
An execution recursively decomposes $G$ into induced subgraphs. The
graph $G$ itself is at level~$0$, and decomposing a subgraph at level
$r$ gives subgraphs at level $r+1$. Let $k=\Theta(\log n)$ be an integer, with a sufficiently large constant factor, and let $\gamma=100$. Level $r$ has
the \emph{distance scale} $D_r=D_0/(\gamma\sqrt k)^r$, where
$D_0=\Delta/k^c$ for a polylogarithmic factor $k^c$ that is fixed
in~\eqref{eq:c-choice} below. At the start of an execution, each vertex
$v$ draws its \emph{rank}, the level at which $v$ will be a pivot. The
rank of $v$ is the first level $r$ at which an independent coin with
success probability
\[
 p_r=\frac{\gamma k^{r+1}\log n}{n}
\]
succeeds. For
$r_{\max}=\min\{r:\gamma k^{r+1}\log n\ge n\}=O(\log n/\log\log n)$, we have
$p_{r_{\max}}\ge1$, so each vertex has rank at most $r_{\max}$, and
$r_{\max}$ is the last level of the recursion.

\paragraph{Pivots and their balls.}
Consider a subgraph $G'$ at level $r$. Its \emph{pivots} are the
vertices of $G'$ of rank $r$. Let $\lambda=16\gamma^2k^2\lceil\log n\rceil^2$
be the radius factor, and let $\eta_{\max}=2\lambda$. The \emph{search} of a pivot $p$ computes
single-source shortest paths in $G'$ to and from $p$ up to distance
$\eta_{\max}D_r$. From its search, $p$ chooses a radius $\rho_p$ in
$[\lambda D_r,2\lambda D_r)$. The sets $\Rin_{\rho_p}(G',p)$ and
$\Rout_{\rho_p}(G',p)$ are the \emph{in-ball} and the \emph{out-ball} of
$p$, and their union is the \emph{ball} of $p$.

The radius is chosen by the following rule. Since $\lambda$ is a
multiple of $4k$, the integers in $[\lambda,2\lambda)$ are partitioned
into the $\lambda/(4k)$ groups
$I_\sigma=\{\lambda+4k(\sigma-1),\dots,\lambda+4k\sigma-1\}$ for
$\sigma=1,\dots,\lambda/(4k)$. The pivot $p$ samples a group
$\sigma_p\in\{1,\dots,\lambda/(4k)\}$ uniformly at random and picks as $\eta_p$ a
candidate $\eta\in I_{\sigma_p}$ that minimizes the size of the
\emph{shell}
\begin{equation}
 F_p(\eta)=\Rboth_{(\eta+1)D_r}(G',p)\setminus\Rboth_{(\eta-1)D_r}(G',p).
 \label{eq:fringe}
\end{equation}
Then $p$ gets the radius $\rho_p=\eta_pD_r$ and the \emph{fringe}
$F_p=F_p(\eta_p)$, which consists of the vertices near the boundary of
its ball.

\paragraph{Subgraphs at the next level.}
The subgraphs at level $r+1$ are of two kinds.
\begin{itemize}[itemsep=1pt]
 \item \emph{Core subgraphs.} The vertices that lie in both balls of
 some pivot are removed. The other vertices receive the label
 $p^{\rm in}$ (resp.\ $p^{\rm out}$) from each pivot $p$ whose in-ball
 (resp.\ out-ball) contains them. The core subgraphs are the subgraphs
 $G'[X]$ for the classes $X$ of remaining vertices with the same set of
 labels.
 \item \emph{Fringe subgraphs.} Each pivot $p$ also forms the fringe
 subgraph $G'[F_p]$.
\end{itemize}
A vertex can lie in several of these subgraphs, and we call each
occurrence a \emph{copy} of the vertex. Similarly, a \emph{copy} of an
edge is an occurrence of the edge in a subgraph. The fringe subgraphs
are needed because the searches are limited in distance. If a vertex of
a subpath $Q$ of length at most $D_r$ lies in the in-ball of $p$, then
the earlier vertices of $Q$ reach $p$ within distance $\rho_p+D_r$, but
not necessarily within $\rho_p$. Thus the vertices of $Q$ in the in-ball need not form a prefix of $Q$. They do form a prefix up to the last vertex of $Q$ within distance $\rho_p-D_r$ of $p$, since each earlier vertex reaches $p$ within distance $\rho_p$. The other vertices of $Q$ in the in-ball lie near the boundary of the ball, at distance more than $\rho_p-D_r$ from $p$, and such vertices are handled by the fringe subgraph of $p$. With the fringe subgraphs, the pivots divide $Q$ into a few contiguous subpaths, each in a single subgraph at level $r+1$~\cite[Lemma~8]{CFR}. The radius rule keeps the fringe subgraphs under control. A fixed vertex lies in the shells $F_p(\eta)$ of at most two integers $\eta$, so the random group makes it unlikely to lie in the fringe of $p$, and the minimization keeps each fringe small~\cite[Lemmas~3 and~4]{CFR}. The removed vertices are handled by the stars, which we describe next.

\paragraph{Stars built in advance.}
The edges of the CFR hopset form stars. The \emph{star} of a vertex $p$
with radius $d$ in a subgraph $G'$ consists of the edges $(u,p)$ for
$u\in\Rin_d(G',p)$ and $(p,u)$ for $u\in\Rout_d(G',p)$, weighted by
their distances in $G'$. A pivot does not build a star at its own
level. Instead, the vertices of rank $r\ge L$ build \emph{stars in
advance}, $L$ levels earlier. Such a vertex $p$ adds its star with
radius $\eta_{\max}D_{r-L}$ in each subgraph $G'$ at level $r-L$ that
contains $p$. Here $L=O(1)$ is fixed in~\eqref{eq:c-choice} below. In
addition, at the start of the execution, each vertex of rank at most $L$
adds its \emph{global star}, which is its star with radius $2\Delta$ in
the whole graph $G$.

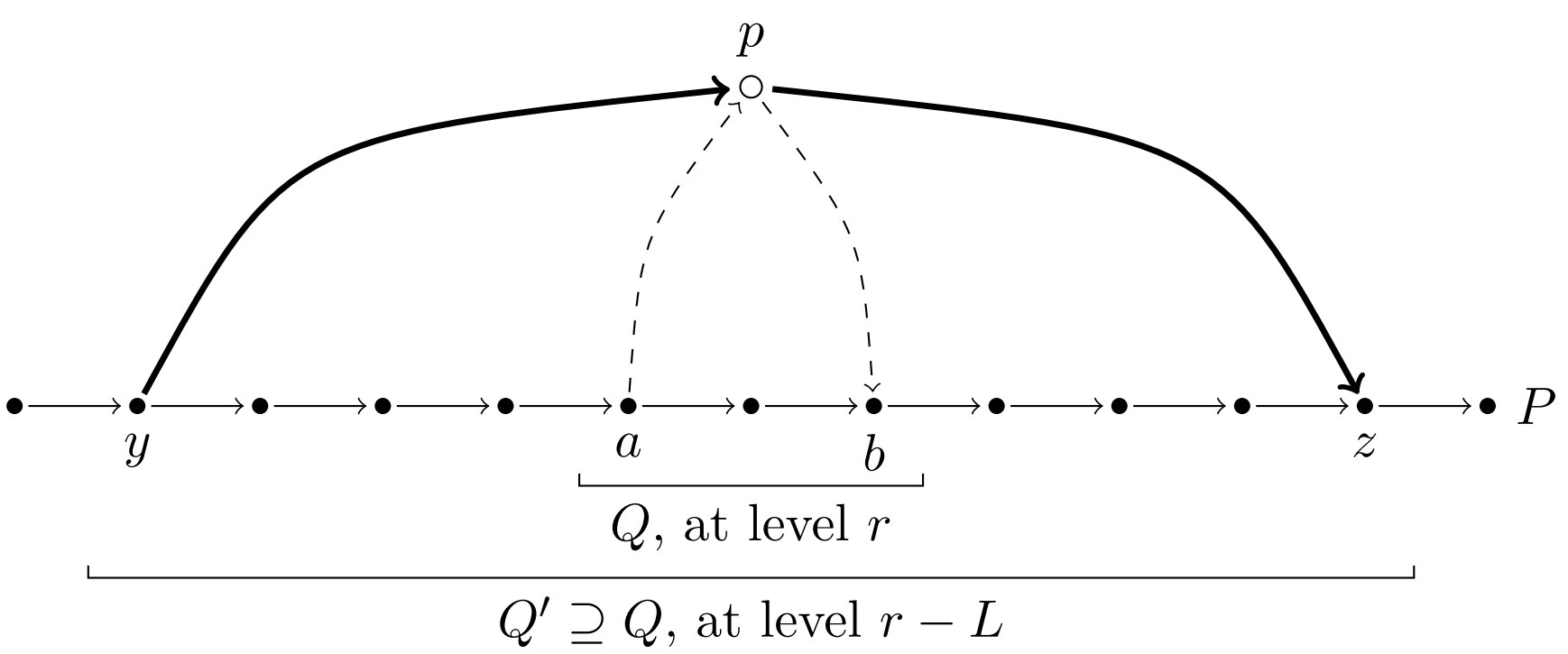
\begin{figure}[t]
\centering
\begin{tikzpicture}[x=0.9cm,y=0.9cm,
  vtx/.style={circle,fill,inner sep=1.2pt},
  piv/.style={circle,draw,fill=white,inner sep=1.6pt}]
 \foreach \i in {0,...,12} {\node[vtx] (u\i) at (\i,0) {};}
 \foreach \i [evaluate=\i as \j using int(\i+1)] in {0,...,11}
  {\draw[->,shorten <=1pt,shorten >=1.5pt] (u\i) -- (u\j);}
 \node[piv,label=above:{$p$}] (p) at (6,2.6) {};
 \draw[->,dashed,shorten <=1pt,shorten >=1.5pt] (u5) .. controls (5.1,1.4) .. (p);
 \draw[->,dashed,shorten <=1.5pt,shorten >=1pt] (p) .. controls (6.9,1.4) .. (u7);
 \draw[->,line width=1.3pt,shorten <=1pt,shorten >=2pt] (u1) .. controls (2.2,2.2) .. (p);
 \draw[->,line width=1.3pt,shorten <=2pt,shorten >=1pt] (p) .. controls (9.8,2.2) .. (u11);
 \path (u1) node[below=2pt] {$y$} (u11) node[below=2pt] {$z$}
  (u5) node[below=2pt] {$a$} (u7) node[below=2pt] {$b$}
  (u12) node[right=2pt] {$P$};
 \draw (4.6,-0.55) -- (4.6,-0.65) -- (7.4,-0.65) -- (7.4,-0.55);
 \node[below] at (6,-0.65) {$Q$, at level $r$};
 \draw (0.6,-1.3) -- (0.6,-1.4) -- (11.4,-1.4) -- (11.4,-1.3);
 \node[below] at (6,-1.4) {$Q'\supseteq Q$, at level $r-L$};
\end{tikzpicture}
\caption{A star built in advance. The vertex $a$ of a subpath $Q$ of $P$
at level $r$ reaches the pivot $p$ of rank $r$, and $p$ reaches the
vertex $b$ of $Q$, within distance $\eta_{\max}D_r$ (dashed). The star
that $p$ built at level $r-L$ has radius $\eta_{\max}D_{r-L}$, so it
contains the edges $(y,p)$ and $(p,z)$ (bold), which shortcut the longer subpath $Q'$ of $P$ from $y$ to $z$ with a detour of order
$\eta_{\max}D_r$.}
\label{fig:advance-star}
\end{figure}

Stars are built in advance to keep their detours small. When two star edges $(y,p)$ and $(p,z)$ shortcut the subpath of a path $P$ from $y$ to $z$, the \emph{detour} of the two edges is the amount by which their total weight exceeds the length of this subpath. Let $p$ be a pivot of rank $r$, and let $Q$ be a subpath of $P$ that has length at most $D_r$ and lies in a subgraph at level $r$. Suppose that
some vertex of $Q$ reaches $p$, and $p$ reaches some vertex of $Q$,
within distance $\eta_{\max}D_r$. Star edges through $p$ that shortcut $Q$ can have a detour of order $\eta_{\max}D_r$, much more than the
length of $Q$. The star in advance of $p$ has the larger radius
$\eta_{\max}D_{r-L}$. Its edges therefore shortcut a longer subpath
$Q'\supseteq Q$ of $P$ that has length at most $D_{r-L}$ and lies in a
subgraph at level $r-L$. The weights of these edges are at most the lengths of walks that follow $Q'$ to a vertex of $Q$, go through $p$, and return to $Q'$. Thus their detour is still of order $\eta_{\max}D_r$ (see \cref{fig:advance-star} for an illustration). Relative to the length of the subpath that they shortcut, which can be up to $D_{r-L}$ instead of $D_r$, the detour thus
shrinks by a factor of up to $D_{r-L}/D_r=(\gamma\sqrt k)^L$. The
analysis of~\cite{CFR} uses this to bound the total detour of all stars
by $\eps w(P)$ (see \cref{prop:cfr-detours} for the precise
statement). This bound also needs the
distance scale to shrink quickly. The number of subpaths of $P$ that
the analysis follows grows by a factor of about $\sqrt k$ per level,
and each of them can cause a detour of order $\eta_{\max}D_r$ at level
$r$. The total detour at level $r$ stays small only if $D_r$ decreases
at least as fast as this number grows. Since the CFR hopset shrinks the scale by the factor $\gamma\sqrt k$ at all levels, the number of subpaths of $P$ grows by a factor of about $\gamma\sqrt k$ per level throughout, and its hopbound $n^{1/2+o(1)}$ is of the order of the total number of these subpaths.

\paragraph{Parameters.}
Recall from \cref{sec:prelim} that $\eps\le1/2$ and that $1/\eps$ is
polylogarithmic. As in~\cite{CFR}, we let
\begin{equation}
 L=\lceil15-2\log_k\eps\rceil,
 \qquad
 k^c=\Theta\Bigl(\frac{\gamma^Lk^{(L-1)/2}}{\log^3n}\Bigr),
 \qquad\text{and}\qquad
 D_0=\Delta/k^c.
 \label{eq:c-choice}
\end{equation}
Thus $L=O(1)$, and $k^L$ and $k^c$ are polylogarithmic. The exact
values are chosen in~\cite[Section~4.3]{CFR} to bound the total detour.
We will compare the search radii with the guess $\Delta$ in the proof of
\cref{prop:cfr-detours} and in \cref{sec:shared-path-shortening}. For
this, note that $\eta_{\max}=O(\gamma^2k^2\log^2n)$, so the values above
give
\begin{equation}
 \eta_{\max}D_0\le5\Delta
 \label{eq:eta-max-D0}
\end{equation}
if the constant factor in $k=\Theta(\log n)$ is large enough.

\subsection{Our construction}
\label{sec:two-speed}

Our construction is given in \cref{alg:cfr-shell,alg:two-speed}. It
differs from the CFR hopset at the late levels, which use a new
recursion, and in three details: the ranks, the global stars, and a rule that aborts expensive executions. The \emph{threshold
level}
\[
 r_0=\min\{r_{\max},\lceil\log_k(n/x)\rceil\}
\]
separates the \emph{early} levels $r<r_0$ from the \emph{late} levels
$r\ge r_0$. It is chosen so that from level $r_0$ on, the balls of the
pivots have $\widetilde O(x)$ vertices. Then the block closures that the late pivots add, which are defined below, have $\widetilde
O(nx)$ edges in total and can be computed with
$\widetilde O(nx^{\omega-1})$ work.

\begin{algorithm}[t]
\caption{Our hopset construction \textsc{TwoSpeedHopset}$(G,x,\eps)$.}
\label{alg:cfr-shell}
\begin{algorithmic}[1]
\Procedure{TwoSpeedHopset}{$G,x,\eps$}
 \State $H\gets\varnothing$
 \For{$C_1\log n$ independent trials} \Comment{one trial per iteration}
  \For{each guess $\Delta=2^jw_{\min}$ with $0\le j\le\lceil\log_2(2nW)\rceil$} \Comment{one execution}
   \State Abort this execution and discard its edges if it becomes too expensive \label{line:abort}
   \State For each vertex $v$, draw $\operatorname{rank}(v)\in\{0,\dots,r_0-1\}\cup\{\infty\}$
   \State Add the global star of each $v$ with $\operatorname{rank}(v)<L$, with radius $8\Delta$ in $G$, to $H$
   \State \Call{Recurse}{$G,0$}, which adds stars to $H$ and records blocks $B_p$
   \State Add the $(1+\eps)$-approximate block closures of the recorded blocks $B_p$ to $H$
  \EndFor
 \EndFor
 \State \Return $H$
\EndProcedure
\end{algorithmic}
\end{algorithm}

\begin{algorithm}[t]
\caption{The recursion of our hopset construction.}
\label{alg:two-speed}
\begin{algorithmic}[1]
\Procedure{Recurse}{$G',r$}
 \State Run \Call{EarlyLevel}{$G',r$} if $r<r_0$, and \Call{LateLevel}{$G',r$} otherwise
\EndProcedure
\Statex
\Procedure{EarlyLevel}{$G',r$} \Comment{$r<r_0$, recursive step of the CFR hopset}
 \For{each $p\in V(G')$ with $\operatorname{rank}(p)=r+L$}
  \State Add the star of $p$ with radius $\eta_{\max}D_r$ in $G'$ to $H$ \Comment{star in advance}
 \EndFor
 \State $\Pi\gets\{v\in V(G'):\operatorname{rank}(v)=r\}$ \Comment{pivots}
 \For{each pivot $p\in\Pi$}
  \State Compute single-source shortest paths in $G'$ to and from $p$ up to $\eta_{\max}D_r$
  \State Sample $\sigma_p\in\{1,\dots,\lambda/(4k)\}$, and let $\eta_p\in I_{\sigma_p}$ minimize $|F_p(\eta)|$ \Comment{radius rule} \label{line:radius-rule}
  \State $\rho_p\gets\eta_pD_r$ and $F_p\gets F_p(\eta_p)$
  \State \Call{Recurse}{$G'[F_p],r+1$} \Comment{fringe subgraph}
 \EndFor
 \State Remove the vertices in both balls of some pivot, and label the other vertices
 \For{each class $X$ of remaining vertices with the same labels}
  \State \Call{Recurse}{$G'[X],r+1$} \Comment{core subgraphs}
 \EndFor
\EndProcedure
\Statex
\Procedure{LateLevel}{$G',r$} \Comment{$r\ge r_0$, our new recursive step}
 \If{$r<r_{\max}$}
  \State Let $\Pi$ contain each $v\in V(G')$ independently with probability $p_r$ \Comment{fresh pivots}
 \Else
  \State $\Pi\gets V(G')$
 \EndIf
 \For{each pivot $p\in\Pi$}
  \State Compute single-source shortest paths in $G'$ to and from $p$ up to $\eta_{\max}D_r$
  \State Choose $\eta_p$ as in Line~\ref{line:radius-rule}, and let $\rho_p\gets\eta_pD_r$
  \State Record the block $B_p=\Rboth_{\rho_p+D_r}(G',p)$
 \EndFor
 \If{$r<r_{\max}$}
  \State $X\gets$ the vertices of $G'$ that do not lie in any pivot's ball
  \State \Call{Recurse}{$G'[X],r+1$} \Comment{core subgraph with the empty label}
 \EndIf
\EndProcedure
\end{algorithmic}
\end{algorithm}

\paragraph{The late levels.}
A subgraph $G'$ at a late level $r$ is handled by the procedure
\textsc{LateLevel} of \cref{alg:two-speed}, which differs from the CFR
recursion in three ways.
\begin{enumerate}[label=(\arabic*),itemsep=2pt]
 \item \emph{Block closures instead of stars.} The pivots of $G'$ do
 not build stars, either at their own level or in advance. Instead, each
 pivot $p$ records its \emph{block}
 $B_p=\Rboth_{\rho_p+D_r}(G',p)$, which is the ball of $p$ with the
 radius enlarged from $\rho_p$ to $\rho_p+D_r$. At the end of the
 execution, we add the $(1+\eps)$-approximate block closure of $B_p$
 (defined below) to the hopset. An edge of this block closure that shortcuts a subpath inside $B_p$ weighs at most $1+\eps$ times the length of the
 subpath, so the late levels cause no detours. The margin $D_r$ in the
 radius of $B_p$ ensures that if a subpath of length at most $D_r$ has a
 vertex in the in-ball of $p$, then its part up to that vertex lies in
 $B_p$. We prove this in \cref{clm:prefix-containment}.
 \item \emph{A slowly shrinking distance scale.} Without detours at the
 late levels, the distance scale no longer has to shrink quickly there,
 and we let it shrink by a factor of $4$ per level:
 \begin{equation}
  D_r=
  \begin{cases}
   D_0/(\gamma\sqrt k)^r, & 0\le r\le r_0,\\[1mm]
   D_{r_0}/4^{r-r_0}, & r_0<r\le r_{\max}.
  \end{cases}
  \label{eq:two-speed-scales}
 \end{equation}
 As noted in~\cite[Remark~6.8]{ABGS}, the fast decrease of the scale
 is the obstacle to adding block closures to the CFR hopset. With
 $D_{r+1}=D_r/\Theta(\sqrt k)$, a subpath of length close to $D_r$
 splits into $\Theta(\sqrt k)$ subpaths of length at most $D_{r+1}$,
 even if no pivot divides it.
 \item \emph{Fresh pivots and a single core subgraph.} The ranks of the CFR hopset fix
 the level of each pivot at the start of the execution, so that the
 pivot can build its star in advance. The late pivots do not build
 stars, so they are freshly sampled in each subgraph, independently of
 the earlier levels. If $r<r_{\max}$, each vertex of $G'$ is a pivot independently
 with probability $p_r$, and at the last level $r_{\max}$, all vertices
 of $G'$ are pivots. They choose their radii in the same way as the
 pivots of the CFR hopset in \cref{sec:cfr}. If $r<r_{\max}$, the
 recursion continues only in the \emph{core subgraph with the empty
 label}, which is induced by the vertices of $G'$ that do not lie in any pivot's ball. This suffices because, as we prove in \cref{lem:late-peeling}, the block closures shortcut the parts of a short subpath that lie in balls. No fringe subgraphs are formed, so the minimization in the radius rule plays no role at the late levels. We keep it so that all pivots choose their radii in the same way.
\end{enumerate}

\paragraph{Ranks and global stars.}
Recall from \cref{sec:cfr} that the rank of a vertex $v$ is the level at
which $v$ will be a pivot. In our construction, the late pivots are
freshly sampled, so we need ranks only for the early pivots. We let $\operatorname{rank}(v)$ be the first level
$r<r_0$ at which the coin with success probability $p_r$ succeeds, and
$\operatorname{rank}(v)=\infty$ if there is none. The early pivots are
then those of the CFR hopset, and so are the stars in advance, except that the vertices whose CFR rank would be at least $r_0$ do not build stars in advance. Only the vertices of rank less than $L$ build global stars, since a vertex of rank $L$ already builds a star in advance at level~$0$. Our global stars have radius $8\Delta$ instead of $2\Delta$.

\paragraph{Approximate block closures.}
For a block $B_p$ of a subgraph $G'$, the
\emph{$(1+\eps)$-approximate block closure} of $B_p$ has an edge
$(u,v)$ for each pair $(u,v)\in\TC(G'[B_p])$, with a weight between
$\dist_{G'[B_p]}(u,v)$ and $(1+\eps)\dist_{G'[B_p]}(u,v)$. Since
the subgraphs of the recursion are induced subgraphs of $G$, the block
subgraph $G'[B_p]$ is $G[B_p]$. The $(1+\eps)$-approximate block
closure of $B_p$ can be computed with Zwick's algorithm.

\begin{proposition}[\cite{Zwick}]
\label{prop:apsp}
Given a directed graph $G'$ with $b$ vertices and positive edge weights
of polynomial aspect ratio and $0<\eps\le1/2$, the algorithm
$\apsp(G',\eps)$ computes estimates $\widehat d(u,v)$ with
$\dist_{G'}(u,v)\le\widehat d(u,v)\le(1+\eps)\dist_{G'}(u,v)$ for all
$u,v\in V(G')$, using $\widetilde O(b^\omega/\eps)$ work.
\end{proposition}

\paragraph{Aborting expensive executions.}
To bound the size and the work of each execution of our hopset
construction deterministically, we use the following \emph{abort
rule}. We abort an execution as soon as it has more than $Cn\log^Cn$ copies of
vertices and vertex visits of searches, more than $Cm\log^Cn$ copies of
edges and edge scans (including the scans that form the subgraphs
$G'[B_p]$), or a block $B_p$ with more than $Cx\log^Cn$ vertices.
Here $m$ is the number of edges of the input graph, and $C$ is a
sufficiently large constant. Since the bounds of
\cref{prop:cfr-copies} on these counts grow with $k^L$, the constant $C$ may depend on the constant
$c_0$ with $1/\eps=O(\log^{c_0}n)$ of \cref{sec:prelim}. The edges of an aborted
execution are discarded. We show in \cref{lem:one-execution} that an
execution is aborted with probability $o(1)$, and the independent
trials make up for these failures.

\subsection{Following a path through the early levels}
\label{sec:early}

We now analyze one execution of our hopset construction for a fixed
guess $\Delta$. For the rest of this section, we fix $\Delta$ and a
simple path $P$ in $G$ with $\Delta\le w(P)\le3\Delta$. Let $H_\Delta$
be the set of edges that the execution for $\Delta$ adds to the
hopset. We want to show that with probability at least $1/2$, the
graph $G\cup H_\Delta$ contains a path from the first vertex of $P$ to
its last vertex with few edges and length at most $(1+2\eps)w(P)$. We
call this path our \emph{shortcut path}.
\Cref{thm:hopset} only needs paths with $w(P)<2\Delta$, since the guesses are powers of $2$ times $w_{\min}$. The slack up to
$3\Delta$ leaves room for the rounding in
\cref{sec:shared-path-shortening}, which can make a path longer.

As in~\cite{CFR} and \cref{sec:overview-hopset}, we construct our
shortcut path by following $P$ through the recursion. At each level
$r$, we keep track of disjoint subpaths of $P$ that remain to be shortcut, which we call the \emph{pieces} of $P$ at level $r$. Each
piece of $P$ at level $r$ lies in a single subgraph at level $r$ and has
length at most $D_r$. At the early levels, star edges through
\emph{bridges} shortcut subpaths of $P$. At the late levels,
block-closure edges shortcut subpaths of $P$, as we show in
\cref{sec:late}. In this subsection, we define the pieces of $P$ at the
early levels and import their analysis from~\cite{CFR}.

\paragraph{Pieces of $P$ at the early levels.}
Our shortcut path will consist of edges that shortcut pieces of $P$
and of boundary edges between them, so to bound its number of edges,
we need to count the pieces of $P$. The pieces of $P$ at the early
levels are defined by the following three notions.
\begin{itemize}[itemsep=2pt]
 \item \emph{Type~2 splits.} At each level $r$, we split $P$ into
 pieces of length at most $D_r$, and we call these splits type~2
 splits. Formally, the \emph{type~2 split} of a subpath $Q$ of $P$ at
 scale $D$ greedily divides $Q$ into longest possible subpaths of
 length at most $D$, with one \emph{boundary edge} of $Q$ between
 consecutive ones. Each new subpath except the last, together with the
 next boundary edge, has length more than $D$, so there are at most
 $w(Q)/D+1$ new subpaths. The pieces of $P$ at level~$0$ are the
 subpaths obtained by the type~2 split of $P$ at scale $D_0$.
 \item \emph{Bridges.} Let $Q$ be a piece of $P$ at an early level
 $r<r_0$, in a subgraph $G'$. Following~\cite{JLS19,CFR}, we say that a
 pivot $p$ of $G'$ is a \emph{bridge} of $Q$ if $Q$ has a vertex in
 $\Rin_{\eta_{\max}D_r}(G',p)$ and a vertex in
 $\Rout_{\eta_{\max}D_r}(G',p)$ (recall that $\eta_{\max}D_r$ is the
 radius of the search of $p$).\footnote{The analysis of~\cite{CFR}
 defines bridges with the radius $\rho_p$ of the ball of $p$ in place
 of $\eta_{\max}D_r$. We use the larger search radius because then a
 pivot $p$ that is not a bridge of $Q$ has no vertex of $Q$ within
 distance $\eta_{\max}D_r$ in one of the two directions. Our analysis
 of the streaming construction in \cref{sec:streaming-underlying}
 relies on this.} This is the precise form of the case ``a pivot
 shortcuts $Q$'' of \cref{sec:overview-hopset}. If some pivot $p$ of
 $G'$ is a bridge of $Q$, then two star edges through $p$ shortcut a
 subpath of $P$ that contains $Q$, as we show in
 \cref{prop:cfr-detours}. Then $Q$ contributes $O(1)$ edges to our
 shortcut path, so we stop counting $Q$, and $Q$ has no pieces at level
 $r+1$.
 \item \emph{Type~1 splits.} If no pivot of $G'$ is a bridge of $Q$,
 then no vertex of $Q$ is removed, since a vertex of $Q$ in both balls
 of a pivot $p$ would make $p$ a bridge of $Q$. In this case,
 by~\cite[Lemma~8]{CFR}, the pivots divide $Q$ into contiguous
 subpaths, each in a single core or fringe subgraph of $G'$, with
 boundary edges between them. This is the \emph{type~1 split} of $Q$.
 The pieces of $P$ at level $r+1$ inside $Q$ are obtained by type~2
 splits of these subpaths at scale $D_{r+1}$. Thus a pivot that is not
 a bridge of $Q$ does not shortcut $Q$, but it can divide $Q$ into more
 pieces of $P$.
\end{itemize}

The next proposition bounds the number of pieces of $P$ at the early
levels. Recall from~\eqref{eq:c-choice} that $k^c=\Delta/D_0$ is
polylogarithmic. At the threshold level $r_0$, the bound is
$\sqrt{n/x}\,n^{O(1/\log\log n)}$, the hopbound of \cref{thm:hopset}.

\begin{proposition}[{\cite[Lemmas~8--10]{CFR}}]
\label{prop:cfr-pieces}
With probability $1-o(1)$, for all levels $r\le r_0$, there are
$O(\gamma^rk^{c+(r+1)/2}\log^2n)$ pieces of $P$ at level $r$.
\end{proposition}

The expected number of pieces of $P$ is
$O(\gamma^rk^{c+(r+1)/2}\log n)$ by~\cite[Lemma~10]{CFR}. There, the
path has length at most $k^cD_0$, while our path $P$ has length up to
$3\Delta=3k^cD_0$, which changes the bounds only by a constant factor.
The analysis of~\cite{CFR} defines the pieces of a path in the same way
as we do, except that it defines bridges with the radius
$\rho_p<\eta_{\max}D_r$ of the ball of $p$ in place of
$\eta_{\max}D_r$. Thus a bridge in the sense of~\cite{CFR} is also a
bridge in our sense, and whenever the analysis of~\cite{CFR} stops
counting a piece of $P$, we stop counting it too. Our pieces of $P$
therefore form a subset of the subpaths counted
in~\cite[Lemma~10]{CFR}.

Next, we describe the subpaths of $P$ that the star edges through the
bridges shortcut. Let $Q$ be a piece of $P$ at an early level $r$ that
has a bridge, and fix a pivot $p$ of rank $r$ that is a bridge of $Q$.
We let $Q'$ be the piece of $P$ at level $r-L$ that contains $Q$ if
$r\ge L$, and $Q'=P$ if $r<L$. These are the subpaths that the star in
advance of $p$ and the global star of $p$, respectively, will shortcut
(see \cref{fig:advance-star} for an illustration of the first case). We
call the maximal subpaths among all such subpaths $Q'$ the
\emph{bridge subpaths} of $P$. Any two pieces of $P$ are nested or
disjoint, and so are the subpaths $Q'$, which makes the bridge
subpaths disjoint.

Recall that the detour of two star edges $(y,p)$ and $(p,z)$ that
shortcut a subpath of $P$ from $y$ to $z$ is the amount by which their
total weight exceeds the length of the subpath. The next proposition,
which is also from~\cite{CFR}, shows that two star edges shortcut each
bridge subpath, and it bounds the total detour of these edges. These
detours are the only error that the early levels add to the length of
our shortcut path. We include a proof, because our global stars have
radius $8\Delta$ and $P$ can have length up to $3\Delta$. The second
half of the proof, which adds up the detours over all levels, follows
the calculation in~\cite[Section~4.3]{CFR}.

\begin{proposition}[Lemma~11 and Section~4.3 of~\cite{CFR}]
\label{prop:cfr-detours}
Suppose that the execution of our hopset construction for $\Delta$ is
not aborted and that the event of \cref{prop:cfr-pieces} holds. Then
for each bridge subpath $Q'$ of $P$, the star edges contain an edge
from the first vertex of $Q'$ to a vertex $p$ and an edge from $p$ to
the last vertex of $Q'$. The total weight of these edges exceeds the
total length of the bridge subpaths by at most $\eps w(P)$.
\end{proposition}
\begin{proof}
Let $Q'$ be a bridge subpath from $y$ to $z$, obtained from a piece
$Q\subseteq Q'$ at level $r$ and a pivot $p$ of rank $r$ that is a
bridge of $Q$. In the subgraph at level $r$ that contains $Q$, some
vertex $a$ of $Q$ reaches $p$, and $p$ reaches some vertex $b$ of $Q$,
within distance $\eta_{\max}D_r$. Suppose first that $r\ge L$, and let
$G'$ be the subgraph at level $r-L$ that contains $Q'$. The subgraph at
level $r$ that contains $Q$ is obtained from $G'$ by the recursion, so
it is an induced subgraph of $G'$, and it contains $p$. Following $Q'$
from $y$ to $a$ gives
$\dist_{G'}(y,p)\le D_{r-L}+\eta_{\max}D_r\le\eta_{\max}D_{r-L}$, as
$D_r\le D_{r-L}/2$. Hence the star that $p$ built in advance in $G'$
contains the edge $(y,p)$, and symmetrically the edge $(p,z)$. The
two edges weigh at most $\eta_{\max}D_r$ plus the lengths of the
subpaths of $Q'$ from $y$ to $a$ and from $b$ to $z$, respectively.
These two subpaths overlap at most in a subpath of $Q$, so their lengths
add up to at most $w(Q')+w(Q)$. As $w(Q)\le D_r$, the
detour of the two edges is at most $3\eta_{\max}D_r$. If $r<L$, then
$Q'=P$, and in the same way,
$\dist_G(y,p)\le w(P)+\eta_{\max}D_0\le3\Delta+5\Delta$ by~\eqref{eq:eta-max-D0}, so the global star of $p$ contains the edges
$(y,p)$ and $(p,z)$, with detour at most $3\eta_{\max}D_0$.

It remains to add up the detours. For $r\ge L$, the bridge subpaths
obtained from pivots of rank $r$ are pieces of $P$ at level $r-L$. Since $D_r=D_0/(\gamma\sqrt k)^r$ at the early levels, \cref{prop:cfr-pieces} shows that their detours sum to
\[
 O\bigl(\gamma^{r-L}k^{c+(r-L+1)/2}\log^2n\bigr)\cdot
 \frac{3\eta_{\max}D_0}{(\gamma\sqrt k)^{r}}
 =O\bigl(\eta_{\max}\gamma^{-L}k^{(1-L)/2}\log^2n\bigr)\cdot\Delta.
\]
If $P$ is a bridge subpath obtained from a pivot of rank less than
$L$, its detour is at most
$3\eta_{\max}D_0=O(\eta_{\max}\gamma^{-L}k^{(1-L)/2}\log^3n)\cdot\Delta$
by the choice of $k^c$ in~\eqref{eq:c-choice}, which is of the same order up to a factor of $\log n$. This bound is much stronger than~\eqref{eq:eta-max-D0}, which
we only use to compare distances with search radii. As $\eta_{\max}=O(\gamma^2k^2\log^2n)$, $\gamma^{2-L}\le1$, and $k^{-L/2}\le\eps k^{-15/2}$ by the choice of $L$, the detours over
the at most $\log n$ levels sum to $O(\eps\log^5n/k^5)\cdot\Delta$.
This is at most $\eps\Delta\le\eps w(P)$ if the constant factor in
$k=\Theta(\log n)$ is large enough.
\end{proof}

\subsection{Following a path through the late levels}
\label{sec:late}

At the late levels, we will show in \cref{lem:late-branching} that the block closures shortcut each piece of $P$ at a late level $r$ to a path with at most
$13^{r_{\max}-r+1}=n^{o(1)}$ edges. This is
the new part of our analysis. At the early levels, type~1 and type~2
splits together make the number of pieces of $P$ grow by a factor of
about $\gamma\sqrt k$ per level, which gives the bound of
\cref{prop:cfr-pieces}. At the late
levels, the block closures stop type~1 splits from creating new pieces,
as we show in \cref{lem:late-peeling}. Since $D_{r+1}=D_r/4$, the type~2 split of
what remains of a piece creates at most five pieces at the next level.
Recall from \cref{sec:two-speed} that at a
late level $r$, each pivot $p$ of a subgraph $G'$ has the radius
$\rho_p=\eta_pD_r$, where $\eta_p$ is an integer in $[\lambda,2\lambda)$,
and records its block $B_p=\Rboth_{\rho_p+D_r}(G',p)$. The
recursion continues only in the core subgraph of $G'$ with the empty
label, which is induced by the vertices of $G'$ that do not lie in any
pivot's ball.

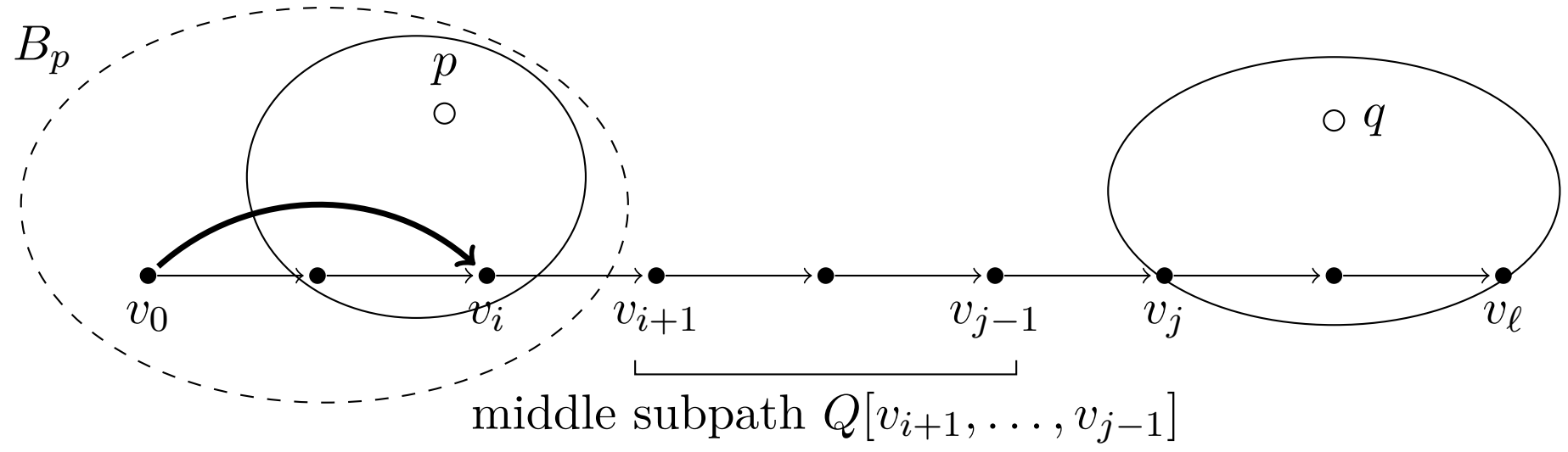
\begin{figure}[t]
\centering
\begin{tikzpicture}[x=1.1cm,y=1.1cm,
  vtx/.style={circle,fill,inner sep=1.3pt},
  piv/.style={circle,draw,fill=white,inner sep=1.6pt}]
 \draw[dashed] (1.25,0.5) ellipse (2.15 and 1.4);
 \draw (1.9,0.7) ellipse (1.2 and 1.0)
  (8.4,0.6) ellipse (1.6 and 0.95)
  (3.45,-0.6) -- (3.45,-0.7) -- (6.15,-0.7) -- (6.15,-0.6);
 \path (-0.75,1.6) node {$B_p$}
  (2.1,1.15) node[piv,label=above:{$p$}] {}
  (8.4,1.1) node[piv,label=right:{$q$}] {}
  (4.8,-0.7) node[below] {middle subpath $Q[v_{i+1},\dots,v_{j-1}]$}
  foreach \i/\x in {0/0,1/1.2,2/2.4,3/3.6,4/4.8,5/6.0,6/7.2,7/8.4,8/9.6}
   {(\x,0) node[vtx] (v\i) {}}
  foreach \i/\j in {0/1,1/2,2/3,3/4,4/5,5/6,6/7,7/8}
   {(v\i) edge[->,shorten >=1pt] (v\j)}
  (v0) edge[->,line width=1.3pt,bend left=42,shorten <=1pt,shorten >=1.5pt] (v2)
  (v0) node[below=2pt] {$v_0$} (v2) node[below=2pt] {$v_i$}
  (v3) node[below=2pt] {$v_{i+1}$} (v5) node[below=2pt] {$v_{j-1}$}
  (v6) node[below=2pt] {$v_j$} (v8) node[below=2pt] {$v_\ell$};
\end{tikzpicture}
\caption{The prefix, middle subpath, and suffix of a piece $Q$ of $P$
at a late level, with the in-ball of a pivot $p$ and the out-ball of a
pivot $q$ (solid curves). The in-ball of $p$ contains $v_i$ but not
$v_0$, whereas the block $B_p$ (dashed) contains the whole prefix
$Q[v_0,\dots,v_i]$, so the block closure of $B_p$ has the edge
$(v_0,v_i)$ (bold). Only the middle subpath survives to the next
level.}
\label{fig:late-level}
\end{figure}

\paragraph{Pieces of $P$ at the late levels.}
Consider a piece $Q=\langle v_0,\dots,v_\ell\rangle$ of $P$ at a late
level $r\ge r_0$, in a subgraph $G'$. For $a\le b$, we write
$Q[v_a,\dots,v_b]$ for the subpath of $Q$ from $v_a$ to $v_b$. Let $i$
be the largest index such that $v_i$ lies in the in-ball of some pivot
of $G'$. Similarly, let $j$ be the smallest index such that $v_j$ lies
in the out-ball of some pivot of $G'$.\footnote{We let $i=-1$ or
$j=\ell+1$ if there is no such vertex.} If $j\le i$, then two
block-closure edges shortcut all of $Q$, as we show in the proof of
\cref{lem:late-branching}. Thus $Q$ contributes only two edges to our
shortcut path, so we stop counting $Q$, and $Q$ has no pieces at level
$r+1$. Otherwise, $Q$ consists of the prefix $Q[v_0,\dots,v_i]$, the
middle subpath $Q[v_{i+1},\dots,v_{j-1}]$, and the suffix
$Q[v_j,\dots,v_\ell]$, joined by the boundary edges $(v_i,v_{i+1})$ and
$(v_{j-1},v_j)$. The pieces of $P$ at level $r+1$ inside $Q$ are the
subpaths obtained by the type~2 split of the middle subpath at scale
$D_{r+1}$. \Cref{fig:late-level} shows an example.

The block closures have to shortcut the prefix $Q[v_0,\dots,v_i]$ with
a single edge $(v_0,v_i)$, and symmetrically the suffix. The in-ball of
a pivot $p$ that contains $v_i$ would not suffice in place of the block
$B_p$, because the vertices before $v_i$ need not lie in it. For $a<i$,
following $Q$ from $v_a$ to $v_i$ gives no better bound than
$\dist_{G'}(v_a,p)\le w(Q[v_a,\dots,v_i])+\rho_p$. If the prefix leaves
the in-ball, the block closure of the in-ball can miss the edge
$(v_0,v_i)$ or give it a much larger weight than
$w(Q[v_0,\dots,v_i])$. The margin $D_r$ in
the radius of $B_p$ is chosen to match the length bound $w(Q)\le D_r$,
so that all vertices of the prefix lie in $B_p$.

\begin{claim}
\label{clm:prefix-containment}
Let $p$ be any pivot of $G'$. If $v_a$ lies in the in-ball of $p$, then
all vertices of the prefix $Q[v_0,\dots,v_a]$ lie in the block $B_p$.
Similarly, if $v_a$ lies in the out-ball of $p$, then the suffix
$Q[v_a,\dots,v_\ell]$ lies in $B_p$.
\end{claim}
\begin{proof}
Let $0\le a'\le a$. Following $Q$ from $v_{a'}$ to $v_a$ and then a
shortest path from $v_a$ to $p$ gives
\[
 \dist_{G'}(v_{a'},p)\le w(Q[v_{a'},\dots,v_a])+\dist_{G'}(v_a,p)\le D_r+\rho_p,
\]
where the first inequality holds since $Q$ is a path in $G'$ and the
second since $w(Q)\le D_r$ and $v_a\in\Rin_{\rho_p}(G',p)$. Hence
$v_{a'}\in\Rin_{\rho_p+D_r}(G',p)\subseteq B_p$. The suffix is
handled symmetrically.
\end{proof}

The next lemma shows that only the middle subpath of $Q$ has to be
followed to the next level. The block closures shortcut the prefix and the suffix of $Q$ with single edges, and the middle subpath lies in a
single subgraph at level $r+1$.

\begin{lemma}[Prefix, middle, and suffix]
\label{lem:late-peeling}
Let $Q=\langle v_0,\dots,v_\ell\rangle$ be a piece of $P$ at a late
level $r$, in a subgraph $G'$, with $i$ and $j$ as above.
\begin{enumerate}[label=(\arabic*),itemsep=1pt]
 \item For all $0<a\le i$, the $(1+\eps)$-approximate block closures
 of the blocks of the pivots of $G'$ contain an edge $(v_0,v_a)$
 of weight at most $(1+\eps)w(Q[v_0,\dots,v_a])$. Similarly, for all
 $j\le a<\ell$, they contain an edge $(v_a,v_\ell)$ of weight at most
 $(1+\eps)w(Q[v_a,\dots,v_\ell])$.
 \item If $i<j$, then the middle subpath $Q[v_{i+1},\dots,v_{j-1}]$
 lies in the core subgraph of $G'$ with the empty label.
 \item If $r=r_{\max}$, then $j\le i$.
\end{enumerate}
\end{lemma}
\begin{proof}
For $0<a\le i$, let $p$ be any pivot whose in-ball contains $v_i$. The
prefix $Q[v_0,\dots,v_i]$ lies in $B_p$ by
\cref{clm:prefix-containment}, so $G'[B_p]$ contains the path
$Q[v_0,\dots,v_a]$. The $(1+\eps)$-approximate block closure of $B_p$
then contains the edge $(v_0,v_a)$, of weight at most
$(1+\eps)\dist_{G'[B_p]}(v_0,v_a)\le(1+\eps)w(Q[v_0,\dots,v_a])$.
Symmetrically, a pivot whose out-ball contains $v_j$ gives the edges
$(v_a,v_\ell)$. If $i<a<j$, then by the choice of $i$ and $j$, the
vertex $v_a$ does not lie in any in-ball or out-ball of a pivot. This
proves Item~2.
Finally, at the last level every vertex is a pivot. In particular, $v_0$
lies in the in-ball and in the out-ball of the pivot $v_0$, so
$j=0\le i$.
\end{proof}

\Cref{lem:late-peeling} is what stops type~1 splits from creating new pieces of $P$ at the late levels. At an early level, a pivot that is not a bridge of
$Q$ can still divide $Q$ at the boundary of its ball. A type~1 split
can then divide $Q$ into many subpaths, all of which are followed to the
next level. At a late level, only the middle subpath is followed,
however many pivots lie near $Q$, since single block-closure edges shortcut the prefix and the suffix. For the same reason, the recursion
can continue only in the core subgraph with the empty label. Every
vertex of $Q$ outside this subgraph lies in an in-ball or an out-ball,
and hence in the prefix or the suffix.
The shortcut set construction
of~\cite[Section~4.2]{ABGS} uses the same argument for reachability.
There, the searches have no distance limit, so the prefix lies in the
ball itself, and no enlargement is needed. With
$D_{r+1}=D_r/4$, the middle subpath splits into at most five pieces at
level $r+1$, which gives the following bound.

\begin{lemma}
\label{lem:late-branching}
Let $H_\Delta$ be the set of edges added to our hopset by an execution
of our algorithm for $\Delta$ that is not aborted. For any
piece $Q$ of $P$ at a late level $r$, the graph $G\cup H_\Delta$
contains a path from the first vertex of $Q$ to its last vertex with
at most $13^{r_{\max}-r+1}$ edges and length at most $(1+\eps)w(Q)$.
\end{lemma}
\begin{proof}
We use induction on $r$, from $r_{\max}$ down to $r_0$. Let
$Q=\langle v_0,\dots,v_\ell\rangle$, with $i$ and $j$ as in
\cref{lem:late-peeling}. If $j\le i$, which by Item~3 always happens
at the last level, we can choose an index $a$ with $j\le a\le i$. By
Item~1, the edges $(v_0,v_a)$ and $(v_a,v_\ell)$ then shortcut $Q$ to a path with two edges and length at most $(1+\eps)w(Q)$.

Otherwise, $i<j$. The edges $(v_0,v_i)$ and $(v_j,v_\ell)$ of Item~1 shortcut the prefix and the suffix. The boundary edges
$(v_i,v_{i+1})$ and $(v_{j-1},v_j)$ connect them to the middle subpath,
which lies in a single subgraph at level $r+1$ by Item~2. The middle
subpath has length at most $w(Q)\le D_r=4D_{r+1}$, so the type~2 split
at scale $D_{r+1}$ divides it into at most $4+1=5$ pieces of $P$ at
level $r+1$, with at most $4$ boundary edges between them. By the
induction hypothesis, each of these pieces $Q'$ has a path in
$G\cup H_\Delta$ between the endpoints of $Q'$ with at most
$13^{r_{\max}-r}$ edges and length at most $(1+\eps)w(Q')$. Together
with the two block-closure edges, the two boundary edges around the
middle subpath, and the at most $4$ boundary edges inside it, these
paths form a path for $Q$ with at most
\[
 2+2+4+5\cdot13^{r_{\max}-r}
 \le8\cdot13^{r_{\max}-r}+5\cdot13^{r_{\max}-r}
 =13^{r_{\max}-r+1}
\]
edges.

For the length, note that the block-closure edges and the paths for
the pieces are at most $1+\eps$ times as long as the subpaths of $Q$
that they replace, and that these subpaths and the boundary edges
partition $Q$. Hence the path has length at most $(1+\eps)w(Q)$.

For $r=r_0$, this gives at most
$13^{r_{\max}-r_0+1}=n^{O(1/\log\log n)}$ edges per piece, since
$r_{\max}-r_0\le\log_kn=O(\log n/\log\log n)$.
\end{proof}

\subsection{Analysis of one execution of our construction}
\label{sec:one-execution}
\label{sec:cfr-facts}

We now bound the size and the work of an execution of our hopset
construction for a guess $\Delta$, and we combine the analyses of the
early and the late levels. Our parallel and streaming algorithms will
run these executions on graphs that contain $G$ and the hopset edges
of earlier stages, as described in \cref{sec:shared-path-shortening}.
We therefore analyze an execution on any directed graph $K$ on the
vertex set of $G$, with $m_K$ edges and positive edge weights of
polynomial aspect ratio. For \cref{thm:hopset}, $K$ is $G$ itself, and
in \cref{sec:shared-path-shortening}, $K$ is obtained from a graph
that contains $G$ by rounding its edge weights. The definitions and
facts of \cref{sec:cfr,sec:two-speed,sec:early,sec:late} hold for $K$
in place of $G$.

\paragraph{Neighborhoods and copies.}
We first import bounds on the number of vertices near a vertex and on
the number of copies of a vertex. These bounds will control the sizes
of the blocks and the cost of the searches. When we condition on the levels before $r$, we fix all random choices of the levels before $r$, which determine the subgraphs at the levels up to $r$. In this section, these are the coins of the levels before $r$ and the groups $\sigma_p$ of their pivots. We do not fix the ranks of at least $r$, even though they determine which vertices build stars in advance at the earlier levels. Given the levels before $r$, each vertex of a subgraph at level $r$ is a pivot with probability at most $p_r$, independently of the other vertices. At an early level, a vertex whose rank is not below $r$ has rank $r$ with probability $p_r$. At a late level, the pivots are sampled with probability $p_r$.

Recall that $\eta_{\max}D_r$ is the radius of the searches at level
$r$, and that $\Rin_d(G',v)$ and $\Rout_d(G',v)$ contain the vertices
within distance $d$ of $v$ in the two directions. The next proposition
shows that at each level $r$, few vertices lie within the search
radius of any vertex.

\begin{proposition}[Paraphrasing Lemma~3 of~\cite{CFR}]
\label{prop:cfr-balls}
Let $b_r=\lceil n/k^r\rceil$ for each level $r$. With probability at
least $1-n^{-10}$, for all levels $r$ of an execution of our hopset
construction, all subgraphs $G'$ at level $r$, and all vertices $v$ of
$G'$, we have
\[
 |\Rin_{\eta_{\max}D_r}(G',v)|\le b_r
 \qquad\text{and}\qquad
 |\Rout_{\eta_{\max}D_r}(G',v)|\le b_r.
\]
\end{proposition}
\begin{proof}[Proof sketch]
For $r\le r_0$, this is~\cite[Lemma~3]{CFR}. The induction there
extends to the late levels. At a late level $r<r_{\max}$, given the
levels before $r$, each vertex of a subgraph at level $r$ is a pivot
independently with probability $p_r$, where $p_rb_{r+1}\ge\gamma\log n$.
Moreover, the scales stay separated, in the sense that the search
radius $\eta_{\max}D_{r+1}$ of level $r+1$ is smaller than
$(\lambda-1)D_r$. Recall that the fringe $F_p=F_p(\eta_p)$ of a pivot
$p$ excludes the vertices within distance $(\eta_p-1)D_r\ge(\lambda-1)D_r$
of $p$. Thus each vertex within distance $\eta_{\max}D_{r+1}$ of a pivot
$p$ at level $r$ lies in the ball of $p$ and outside its fringe. The
separation follows from $\eta_{\max}=2\lambda$ and $D_r\ge4D_{r+1}$. The
sampling argument of~\cite{JLS19,CFR} uses these two facts and the fact
that distances do not decrease in subgraphs. Its idea is the following.
Suppose that more than $b_{r+1}$ vertices reach a vertex $v$ within
distance $\eta_{\max}D_{r+1}$. Then with high probability, a pivot $p$
at level $r$ is among the $b_{r+1}$ of them that come last in a
topological order, with extra care for cycles~\cite{CFR}. By the
separation, $v$ lies deep inside the out-ball of $p$, and the labels of
$p$ separate $v$ from the vertices that come earlier in this order. The
argument fails with probability $n^{-\Omega(\gamma)}$ per core
subgraph. Fringe subgraphs are formed
only at the early levels and are handled as in~\cite{CFR}.
\end{proof}

Next, we bound how often a vertex is copied into subgraphs and visited
by searches. These bounds will give the size and the work of an
execution. Recall that the radius rule of \cref{sec:cfr} samples the
group $\sigma_p$ of each pivot $p$ uniformly from $\lambda/(4k)$
groups.

\begin{proposition}[Adapted from Lemmas~4--7 of~\cite{CFR}]
\label{prop:cfr-copies}
Condition on the levels before a level $r$, and suppose that the bounds
of \cref{prop:cfr-balls} hold at level $r$. Then for each subgraph $G'$
at level $r$, in expectation, each copy of a vertex in $G'$
\begin{enumerate}[label=(\arabic*),itemsep=1pt]
 \item lies in at most $16kb_rp_r/\lambda\le2/(\gamma\log n)$ fringe
 subgraphs of $G'$,
 \item is visited by the searches of $O(\gamma k\log n)$ pivots and of
 $O(\gamma k^{L+1}\log n)$ vertices that build their stars in advance in
 $G'$, and
 \item lies in the blocks of $O(\gamma k\log n)$ pivots.
\end{enumerate}
Consequently, an execution of our hopset construction has
$\widetilde O(n)$ copies of vertices and $\widetilde O(m_K)$ copies of
edges in expectation.\footnote{As in~\cite{CFR,ABGS}, we bound such
expectations assuming the bounds of \cref{prop:cfr-balls}.}
\end{proposition}

Below the threshold level, these bounds follow from the proofs
of~\cite[Lemmas~4--7]{CFR}. At a late level, a copy of a vertex $v$ of
$G'$ lies in at most one subgraph at the next level. The pivots whose
searches or blocks contain this copy lie in
$\Rboth_{\eta_{\max}D_r}(G',v)$, which has at most $2b_r$ vertices and
hence $2b_rp_r\le4\gamma k\log n$ pivots in expectation. At the last
level $r_{\max}$, where all vertices are pivots, there are at most
$2b_{r_{\max}}=O(\gamma k\log n)$ of them by the choice of $r_{\max}$.
A subgraph that contains an edge also contains the first vertex of the
edge, which gives the bound for edges.

The late pivots add the block closures of their blocks. Next,
we bound the size $|B_p|$ of each block and the expected total
size of the blocks. These are the two quantities that
\cref{def:implicit-hopset-intro} bounds for the blocks of an implicit
hopset.

\begin{corollary}
\label{cor:small-balls}
If the bounds of \cref{prop:cfr-balls} hold, then each block has
$|B_p|=O(x+k\log n)=\widetilde O(x)$ vertices. Moreover, the blocks of
an execution of our hopset construction have expected total size
$\E\sum_p|B_p|=\widetilde O(n)$.
\end{corollary}
\begin{proof}
Since $\eta_p$ is an integer in $[\lambda,2\lambda)$, we have
$\rho_p+D_r\le2\lambda D_r=\eta_{\max}D_r$. Thus $B_p$ lies in the
region $\Rboth_{\eta_{\max}D_r}(G',p)$ explored by the search of $p$,
which has at most $2b_r\le2b_{r_0}$ vertices by
\cref{prop:cfr-balls}.
Here $b_{r_0}\le\max\{\lceil x\rceil,\lceil\gamma k\log n\rceil\}$, since $r_0=\min\{r_{\max},\lceil\log_k(n/x)\rceil\}$ and $\gamma k^{r_{\max}+1}\log n\ge n$. The second bound follows from
\cref{prop:cfr-copies}.
\end{proof}

We can now analyze an execution of our hopset construction on $K$.
Recall the hopbound
\[
 h_x=\bigl\lceil\sqrt{n/x}\,n^{C_h/\log\log n}\bigr\rceil
\]
of \eqref{eq:streaming-hopbound}, where $C_h$ is a sufficiently large
constant, and that an edge $(u,v)$ is noncontracting in $K$ if its
weight is at least $\dist_K(u,v)$.

\begin{lemma}
\label{lem:one-execution}
\label{lem:shared-path-shortening}
\label{lem:hopset-correct}
Let $K$ be a directed graph on $n$ vertices with $m_K$ edges and
positive edge weights of polynomial aspect ratio, let $1\le x\le n$, and
let $0<\eps\le1/2$ with $1/\eps=O(\log^{c_0+1}n)$. Consider the execution
for a guess $\Delta>0$ of
\textnormal{\textsc{TwoSpeedHopset}}$(K,x,\eps)$ in \cref{alg:cfr-shell}.
This execution adds a set $H_\Delta$ of $\widetilde O(nx)$ edges that
are noncontracting in $K$, using $\widetilde O(m_K+nx^{\omega-1})$
work. Moreover, let $P$ be any simple path in $K$ from $u$ to $v$ with
$\Delta\le w_K(P)\le3\Delta$, chosen independently of the random
choices of the execution. Then with probability at least $1/2$,
\[
 \dist_{K\cup H_\Delta}^{(h_x)}(u,v)\le(1+2\eps)\,w_K(P).
\]
\end{lemma}
\begin{proof}
\underline{Size and work.} The abort rule of \cref{sec:two-speed}
(Line~\ref{line:abort} of \Cref{alg:cfr-shell}) caps the numbers of
copies, vertex visits, and edge scans at $Cn\log^Cn$ and $Cm_K\log^Cn$,
and the sizes of the blocks at $Cx\log^Cn$. It is checked
throughout, so the following bounds also hold for aborted executions.
The recursion takes $O(\log n)$ work per copy, visit, and scan, or
$\widetilde O(n+m_K)$ in total. There are $\widetilde O(n)$ star edges,
at most one per vertex visit. Each block $B_p$ lies in the
region explored by the search of $p$ (see the proof of
\cref{cor:small-balls}), so the abort rule ensures that
$\sum_p|B_p|\le Cn\log^Cn$ and $\max_p|B_p|\le Cx\log^Cn$. As
$1/\eps$ is polylogarithmic, \cref{prop:apsp} bounds the work of the
block closures by
\[
 \sum_p\widetilde O\bigl(|B_p|^\omega/\eps\bigr)
 \le\widetilde O\Bigl(\max_p|B_p|^{\omega-1}\sum_p|B_p|\Bigr)
 =\widetilde O(nx^{\omega-1}).
\]
They add at most
$\sum_p|B_p|^2\le\max_p|B_p|\sum_p|B_p|=\widetilde O(nx)$ edges. Each
added edge $(u,v)$ weighs at least the distance from $u$ to $v$ in an
induced subgraph of $K$, and hence at least $\dist_K(u,v)$. Thus all
added edges are noncontracting in $K$.

\underline{Failure probability.} We show that with probability
$1-o(1)$, the execution is not aborted and the event of
\cref{prop:cfr-pieces} holds. By \cref{prop:cfr-copies}, the expected
numbers of copies of vertices and of vertex visits in the recursion are
$\widetilde O(n)$, and those of copies of edges and of edge scans are
$\widetilde O(m_K)$. These scans include those that form the subgraphs
$G'[B_p]$, since each copy of a vertex lies in $O(\gamma k\log n)$
blocks in expectation. A vertex builds a global star with
probability at most $\sum_{r<L}p_r\le2\gamma k^L\log n/n$, so there are
$O(\gamma k^L\log n)$ global stars in expectation. Each of them is one
search in $K$, which adds at most $n$ visits and $m_K$ scans. By
Markov's inequality, each count of the abort rule exceeds its cap with
probability at most $1/\log n$ if $C$ is large enough. By
\cref{prop:cfr-balls,cor:small-balls}, some block has more than
$Cx\log^Cn$ vertices with probability at most $n^{-10}$. Finally, the
event of \cref{prop:cfr-pieces} fails with probability $o(1)$.
Therefore, with probability $1-o(1)\ge1/2$, the execution is not
aborted and the event of \cref{prop:cfr-pieces} holds. We assume this
from now on.

\underline{Hops.} We divide $P$ into three kinds of parts: (a)~the
bridge subpaths, (b)~the pieces of $P$ at the threshold level $r_0$
that do not lie in any bridge subpath, and (c)~the boundary edges that
do not lie in any bridge subpath. At each early level, a piece of $P$
is either contained in a bridge subpath or divided into pieces at the
next level and boundary edges. As any two pieces of $P$ are nested or
disjoint, $P$ is the concatenation of the subpaths of types (a) and
(b) and the edges of type (c). We replace each bridge subpath by the
two star edges of \cref{prop:cfr-detours} and each piece of type~(b)
by the path of \cref{lem:late-branching}, and we keep the boundary
edges of type~(c). Each bridge subpath other than $P$ is a piece of
$P$ at a level $r\le r_0$, and each boundary edge of type~(c) comes
right before a distinct such piece. Our shortcut path therefore has at
most $3+13^{r_{\max}-r_0+1}$ edges per piece of $P$ at the levels
$r\le r_0$, plus two. The bounds of \cref{prop:cfr-pieces} grow
geometrically with $r$, so there are
$O(\gamma^{r_0}k^{c+(r_0+1)/2}\log^2n)$ such pieces in total. The
factors $k^c$, $k^{1/2}\log^2n$, and $13^{r_{\max}-r_0+1}$ are all
$n^{O(1/\log\log n)}$, the last one by the end of the proof of
\cref{lem:late-branching}. Hence our shortcut path has at most
\begin{align*}
 (3+13^{r_{\max}-r_0+1})\cdot O\bigl(\gamma^{r_0}k^{c+(r_0+1)/2}\log^2n\bigr)+2
 &\le(k^{r_0})^{1/2+\log_k\gamma}\,n^{O(1/\log\log n)}\\
 &\le\sqrt{n/x}\,n^{O(1/\log\log n)}
 \le h_x
\end{align*}
edges.
The first inequality follows from
$\gamma^{r_0}=(k^{r_0})^{\log_k\gamma}$, the second from
$k^{r_0}\le kn/x$ and $\gamma=k^{O(1/\log\log n)}$, and the last holds
for $C_h$ large enough.

\underline{Error.} By \cref{prop:cfr-detours}, the star edges exceed
the total length of the bridge subpaths by at most $\eps w_K(P)$.
The path for a piece $Q$ of type~(b) has length at most
$(1+\eps)w_K(Q)$ by \cref{lem:late-branching}. Therefore, our
shortcut path has length at most $(1+2\eps)w_K(P)$.
\end{proof}

\begin{proof}[Proof of \cref{thm:hopset}]
We run \Cref{alg:cfr-shell} with $\eps/2$ in place of $\eps$. Its
executions are those of \cref{lem:one-execution} with $K=G$ and
accuracy $\eps/2$. In particular, the edges of $H$ are noncontracting
in $G$. Before the random choices, we
fix a shortest path $P_{u,v}$ in $G$ for each pair $(u,v)\in\TC(G)$,
together with the guess $\Delta$ with $\Delta\le w(P_{u,v})<2\Delta$.
In each of the $C_1\log n$ independent trials, where $C_1$ is a
sufficiently large constant, the execution for this guess adds, with
probability at least $1/2$, edges that give a path from $u$ to $v$ in
$G\cup H$ with at most $h_x$ edges and length at most
$(1+\eps)\dist_G(u,v)$. By a union bound over the pairs, $H$ is an
$(h_x,\eps)$-hopset with high probability. Each of the $O(\log^2n)$
executions adds $\widetilde O(nx)$ edges using
$\widetilde O(m+nx^{\omega-1})$ work.
\end{proof}

\subsection{Building the hopset in stages}
\label{sec:shared-path-shortening}

The proof of \cref{thm:hopset} does not need to build the hopset in
stages. Our parallel and streaming algorithms, however, need stages,
since they run the searches of our hopset construction on graphs with
rounded edge weights. Following~\cite[Section~6]{CFR}, they round the
edge weights up to multiples of $\alpha=\Theta(\eps\Delta/h)$, where
$h=h_x$. Then each search runs up to an integer distance $R=O(h/\eps)$,
which takes roughly $R$ rounds or passes. This rounding changes the
length of a path with $O(h)$ edges by a factor of at most $1+O(\eps)$,
but it can
make paths with many more edges much longer. The hopset is therefore
built in $O(\log n)$ stages. Each stage only has to shortcut paths with at most $2h$ edges, and it halves the number of edges that near-shortest
paths need. In this subsection, we fix the rounding and the stages, and
we prove the lemma that the trials of
\cref{sec:parallel,sec:streaming-closures} will rely on.

\paragraph{Rounding.}
In the execution of our hopset construction for a guess $\Delta$ on a
graph $K$, we let
\begin{equation}
 \alpha=\frac{\eps\Delta}{8h}
 \qquad\text{and}\qquad
 R=\left\lceil\frac{64h}{\eps}\right\rceil.
 \label{eq:stage-radius}
\end{equation}
Rounding each edge weight of $K$ up to a multiple of $\alpha$ gives the
graph $K^{(\Delta)}$. The searches of the pivots and of the stars in
advance at a level $r$ run up to distance
$\eta_{\max}D_r\le\eta_{\max}D_0$, which is at most $5\Delta$
by~\eqref{eq:eta-max-D0}. The global stars, which are built in the whole
graph, search up to distance $8\Delta$. Since $8\Delta\le R\alpha$, all
searches of an execution on $K^{(\Delta)}$ run up to distance at most
$R$ in units of $\alpha$. Rounding increases the weight of each edge by
less than $\alpha$. Hence, as in \eqref{eq:path-rounding} with $2h$ in
place of $h$, it increases the length of any path with at most $2h$
edges by less than $2h\alpha=\eps\Delta/4$. In particular, if $P$ is a
simple path in $K$ with at most $2h$ edges and $\Delta$ is the guess
with $\Delta\le w_K(P)<2\Delta$, then
\begin{equation}
 \Delta\le w_{K^{(\Delta)}}(P)<(1+\eps/4)\,w_K(P)<3\Delta.
 \label{eq:stage-path-rounding}
\end{equation}

\paragraph{Stages.}
\Cref{alg:stages} runs $T=\lceil\log_2n\rceil$ stages of
$C_2\log n$ independent trials each, for a sufficiently large constant
$C_2$, on the graphs $K_i=G\cup H_i$, where $H_i$ contains the edges of
the earlier stages. As in \Cref{alg:cfr-shell}, a trial runs one
execution of our hopset construction for each guess. We leave the trials abstract here, since
\cref{sec:parallel,sec:streaming-closures} implement them differently.
The next lemma states the condition that the trials must guarantee. It
is the argument of~\cite[Lemma~16]{CFR}, stated for any trial that
satisfies this condition.

\begin{algorithm}[ht]
\caption{Building the hopset in stages, \textsc{IteratedHopset}$(G,\eps)$}
\label{alg:stages}
\begin{algorithmic}[1]
\Procedure{IteratedHopset}{$G,\eps$}
 \State $T\gets\lceil\log_2n\rceil$ and $H_1\gets\varnothing$
 \For{$i=1,\dots,T$}
  \State $K_i\gets G\cup H_i$
  \State Run $C_2\log n$ independent trials on $K_i$ with accuracy $\eps$
  \State $H_{i+1}\gets H_i\cup{}$(the edges output by these trials)
 \EndFor
 \State \Return $H_{T+1}$
\EndProcedure
\end{algorithmic}
\end{algorithm}

\begin{lemma}[Variant of Lemma~16 of~\cite{CFR}]
\label{lem:shared-hopset-iteration}
Let $0<\eps\le1/(10\log n)$. Suppose that for all stages $i$ of
\Cref{alg:stages}, conditioned on the random choices of the earlier
stages, the following holds: for any simple path $P$ in $K_i$ from $u$
to $v$ with at most $2h$ edges and $w_{\min}\le w_{K_i}(P)\le2nw_{\max}$
that does not depend on the random choices of the trial, the output
$H'$ of each trial of stage $i$ satisfies
\begin{equation}
 \Pr\Bigl[\dist_{K_i\cup H'}^{(h)}(u,v)\le(1+3\eps)\,w_{K_i}(P)\Bigr]\ge1/2.
 \label{eq:trial-event}
\end{equation}
Suppose also that the edges of $H_{i+1}$ are noncontracting in $G$, and
that $K_{i+1}$ contains $K_i$ and the outputs of all trials of stage
$i$, with the same or smaller weights. Then, with high probability,
$H_{T+1}$ is an $(h,10\eps\log n)$-hopset of $G$.
\end{lemma}
\begin{proof}
Consider a pair $(u,v)\in\TC(G)$ and a shortest path $P^1$ from $u$ to
$v$ in $G$. For each stage $i$, we define a simple path $P^i$ from $u$
to $v$ in $K_i$ that is determined by the random choices of the earlier
stages and has length $w(P^i)\le(1+3\eps)^{i-1}\dist_G(u,v)$. Let $P^i$
have $\ell_i$ edges, and divide $P^i$ into $\lceil\ell_i/(2h)\rceil\le n$
chunks of at most $2h$ consecutive edges. Each chunk is a simple path
in $K_i$. Its length is at least $w_{\min}$, since the edges of $K_i$
are noncontracting in $G$, and at most
$(1+3\eps)^T\dist_G(u,v)\le2nw_{\max}$, since $3T\le5\log n$ for large
$n$ and hence $(1+3\eps)^T\le e^{3\eps T}\le e^{1/2}\le2$. Thus \eqref{eq:trial-event}
applies to each chunk. For a given chunk, the event of \eqref{eq:trial-event} fails in all $C_2\log n$ independent trials of stage $i$ with probability at most $2^{-C_2\log n}=n^{-C_2}$. By a
union bound over the chunks, with probability at least $1-n^{1-C_2}$,
the trials of stage $i$ shortcut each chunk to a path with at most $h$ edges and at most $1+3\eps$ times its length. In this case, we let
$P^{i+1}$ be the concatenation of these paths in $K_{i+1}$, with cycles
removed. If some chunk is not shortcut, we let $P^{i+1}=P^i$, which is
also a path in $K_{i+1}$. In both cases, $w(P^{i+1})\le(1+3\eps)w(P^i)$.
In the first case, each of the $\lceil\ell_i/(2h)\rceil$ chunks is
replaced by at most $h$ edges, so
$\lceil\ell_{i+1}/h\rceil\le\lceil\ell_i/(2h)\rceil$. If the first case
happens in all stages, then the number $\lceil\ell_i/h\rceil$, which is
at most $n$ for $i=1$, halves (rounded up) in each of the $T$ stages.
Then $P^{T+1}$ has at most $h$ edges and length
at most
$(1+3\eps)^T\dist_G(u,v)\le(1+6\eps T)\dist_G(u,v)\le(1+10\eps\log n)\dist_G(u,v)$,
where we use $e^y\le1+2y$ for $0\le y\le1$. A union
bound over the stages and the pairs shows that this happens for all
pairs with high probability. As the edges of $H_{T+1}$ are
noncontracting in $G$, the lower bound of \cref{def:hopset} holds as
well.
\end{proof}

We obtain an $(h,\eps)$-hopset by running \Cref{alg:stages} with
$\eps/(10\log n)$ in place of $\eps$. This
accuracy satisfies the assumption $1/\eps=O(\log^{c_0+1}n)$ of
\cref{lem:one-execution}.

\section{Parallel Shortest Paths}
\label{sec:parallel}

In this short section, we implement the hopset construction of
\cref{sec:sequential} in the parallel model and prove
\cref{thm:exact-tradeoff}. We can store the entire hopset explicitly
here; the obstacle is that our searches use Dijkstra's algorithm, which
is inherently sequential. We get around this with the scaling and
rounding technique of Klein and Subramanian~\cite{KS} from
\cref{sec:prelim-scaling}, which was used for the same purpose
in~\cite{CFR} and~\cite{ABGS}. We
let $\omega$ be the exponent of matrix multiplication for parallel
algorithms with polylogarithmic depth, following~\cite{ABGS}. The
following proposition will be proved in \cref{sec:parallel-approx}.

\begin{restatable}{proposition}{paralleloracle}
\label{prop:parallel-oracle}
Given an $n$-vertex directed graph $G$ with $m$ edges and positive edge
weights of polynomial aspect ratio, a source $s$, an accuracy parameter
$\eps$, and a parameter $1\le x\le n$, there is a randomized parallel
algorithm that computes, with high probability, $(1+\eps)$-approximate
distances from $s$ using $\widetilde O(m+nx^{\omega-1})$ work and
$\sqrt{n/x}\,n^{O(1/\log\log n)}$ depth.
\end{restatable}

\subsection{Approximate distances from \texorpdfstring{$s$}{s}}
\label{sec:parallel-approx}

Let $G$ be the input graph of \cref{prop:parallel-oracle}, with
smallest and largest edge weights $w_{\min}$ and $w_{\max}$, and let
$h=h_x$ be the hopbound of \eqref{eq:streaming-hopbound}. We build the
hopset in the stages of \Cref{alg:stages}. Recall from
\cref{sec:shared-path-shortening} that stage $i$ runs $C_2\log n$
independent trials on the graph $K_i=G\cup H_i$, where $H_i$ contains
the edges of the earlier stages. As in \Cref{alg:cfr-shell}, a trial
runs one execution of our hopset construction for each guess $\Delta$,
and \cref{lem:shared-hopset-iteration} states the condition that the
output of each trial must satisfy. \Cref{sec:shared-path-shortening}
also fixed how these executions round the edge weights. It remains to
define the trial of our parallel algorithm and to run it in small
depth.

Our \emph{parallel trial} on a graph $K=K_i$ runs, for each guess
$\Delta=2^jw_{\min}$ with $0\le j\le\lceil\log_2(2nW)\rceil$ in
parallel, the execution for $\Delta$ of
\textsc{TwoSpeedHopset}$(K^{(\Delta)},x,\eps)$ in \Cref{alg:cfr-shell},
where $K^{(\Delta)}$ is obtained from $K$ by rounding the edge weights
up to multiples of $\alpha=\eps\Delta/(8h)$. It outputs the edges added
by these executions, except those heavier than $4nw_{\max}$. This
weight cap keeps the aspect ratio of each graph $K_i$ polynomial, as
\cref{lem:one-execution} requires.

Next, we run each execution of our hopset construction in small depth.
Its searches are distance-bounded searches with integer weights, as in
the scaling and rounding reduction of \cref{sec:prelim-scaling}.
Indeed, by \cref{sec:shared-path-shortening}, the execution for
a guess $\Delta$ rounds the edge weights of $K$ up to multiples of $\alpha$,
and all of its searches run up to distance at most
$R=\lceil64h/\eps\rceil$ in units of $\alpha$. As in~\cite{KS}, such a
search settles the vertices at distance $0,1,\dots,R$ from its source
in $R+1$ rounds, and thus takes $\widetilde O(R)$ depth. The searches
of one recursion level form a batch and run their rounds side by side,
so that we pay for the $R+1$ rounds once per batch and not once per
search. We check the abort rule
of \cref{sec:two-speed} before every round of a batch. The other steps
take polylogarithmic depth. The radii and the subgraphs of
the recursion can be computed by parallel sorting and prefix sums
(see~\cite{Fineman18,JLS19}). Since $1/\eps$ is polylogarithmic, the
algorithm of Zwick~\cite{Zwick} behind \cref{prop:apsp} computes the
block closures with $O(\log n)$ rounds of independent matrix
multiplications.

The next lemma shows that \Cref{alg:stages} with the parallel trial
computes the hopset behind \cref{prop:parallel-oracle}.

\begin{lemma}
\label{lem:parallel-hopset}
Let $0<\eps\le1/(10\log n)$. With high probability, \Cref{alg:stages}
with the parallel trial computes an $(h,10\eps\log n)$-hopset $H$ of
$G$ with $|H|=\widetilde O(nx)$, using $\widetilde O(m+nx^{\omega-1})$
work and $\widetilde O(h/\eps)$ depth.
\end{lemma}

\begin{proof}
We check the conditions of \cref{lem:shared-hopset-iteration}. Using
\cref{lem:one-execution} and the weight cap, induction on $i$ shows
that every edge $(u,v)$ of $K_i$ weighs between
$\dist_G(u,v)\ge w_{\min}$ and $4nw_{\max}$, since rounding the weights
up does not decrease distances. Thus the edges of $K_i$ are noncontracting in $G$, and $K_i$
and the graphs $K_i^{(\Delta)}$ have polynomial aspect ratio, as
\cref{lem:one-execution} requires. As unions keep the lightest edge of
each pair, $K_{i+1}$ contains all edges of $K_i$ and of the trial
outputs, with the same or smaller weights.

Fix the earlier stages and a path $P$ from $u$ to $v$ in $K=K_i$ as in
\cref{lem:shared-hopset-iteration}, let $\Delta$ be the guess with
$\Delta\le w_K(P)<2\Delta$, and let $H_\Delta$ be the set of edges added
by the execution for $\Delta$. By \eqref{eq:stage-path-rounding},
\cref{lem:one-execution} applies to $P$ in $K^{(\Delta)}$, so with
probability at least $1/2$,
\[
 \dist^{(h)}_{K^{(\Delta)}\cup H_\Delta}(u,v)
 \le(1+2\eps)\,w_{K^{(\Delta)}}(P)
 \le(1+2\eps)(1+\eps/4)\,w_K(P)
 \le(1+3\eps)\,w_K(P).
\]
Each edge of a path that attains the left side weighs at most
$(1+3\eps)\cdot2nw_{\max}<4nw_{\max}$. The trial therefore keeps the
edges of this path from $H_\Delta$, and those from $K^{(\Delta)}$ are
no heavier in $K$.
Hence the output of the trial satisfies \eqref{eq:trial-event}, and
\cref{lem:shared-hopset-iteration} shows that $H$ is an
$(h,10\eps\log n)$-hopset of $G$ with high probability.

By \cref{lem:one-execution}, each of the $O(\log^3n)$ executions adds
$\widetilde O(nx)$ edges using $\widetilde O(|E(K_i)|+nx^{\omega-1})$
work, and the rounds of its $O(\log n)$ batches add
$\widetilde O(R)=\widetilde O(n)$ work. Hence $|H|=\widetilde O(nx)$,
each graph $K_i$ has $m+\widetilde O(nx)$ edges, and the total work is
$\widetilde O(m+nx^{\omega-1})$. Each stage runs its executions in
parallel, and an execution runs its batches one after another, each in
$\widetilde O(R)$ depth. As all other steps take polylogarithmic depth,
the $O(\log n)$ stages take $\widetilde O(R)=\widetilde O(h/\eps)$
depth.
\end{proof}

\begin{proof}[Proof of \cref{prop:parallel-oracle}]
We compute an $(h,\eps/2)$-hopset $H$ of $G$ by applying
\cref{lem:parallel-hopset} with $\eps/(20\log n)$ in place of $\eps$.
This takes $\widetilde O(m+nx^{\omega-1})$ work and
$\widetilde O(h\log n/\eps)=\widetilde O(h/\eps)$ depth. With $\eps/2$
in place of $\eps$,
\cref{lem:scale-round} shows that it suffices to
compute the distances from $s$ in $G_\alpha\cup H_\alpha$ up to radius
$O(h/\eps)$ for all $O(\log n)$ distance guesses $\Delta$. As
in~\cite{KS}, these searches take
$\widetilde O(m+nx)$ work and
$\widetilde O(h/\eps)=\sqrt{n/x}\,n^{O(1/\log\log n)}$ depth.
\end{proof}

\subsection{Exact distances}
\label{sec:parallel-exact}

To obtain exact distances, we use our parallel version of the
reduction of~\cite{Rozhon}, which we prove in
\cref{sec:streaming-exact-reduction}.

\begin{restatable}{lemma}{parallelreduction}
\label{lem:parallel-exact-reduction}
Let $a\ge1$ and $c=c(a)$ be as in \cref{thm:streaming-exact-reduction}.
If a parallel algorithm computes $(1+c/\log n)$-approximate distances
from a source in directed graphs with $O(n)$ vertices, $O(m)$ edges,
and positive, polynomially bounded integer weights using $W_{\rm apx}$
work and $D_{\rm apx}$ depth, then exact SSSP with a shortest-path tree
in $n$-vertex, $m$-edge directed graphs with integer weights in $\{0,1,\dots,n^a\}$
takes $\widetilde O(W_{\rm apx}+m)$ work and $\widetilde O(D_{\rm apx})$
depth, and it succeeds with high probability if the approximate
algorithm does.
\end{restatable}

\noindent Combining \cref{prop:parallel-oracle} with
\cref{lem:parallel-exact-reduction} gives exact distances.

\parallelthm*

\begin{proof}
The input graph has integer edge weights in $\{0,1,\dots,n^a\}$ for a
constant $a$. Let $c=c(a)$ be the constant of
\cref{lem:parallel-exact-reduction}. We let
$x=\lceil(m/n)^{1/(\omega-1)}\rceil\le n$, which balances the two terms
of the work bound of \cref{prop:parallel-oracle}:
\[
 nx^{\omega-1}=O(m)
 \qquad\text{and}\qquad
 \frac nx\le\left(\frac{n^\omega}{m}\right)^{1/(\omega-1)}.
\]
Each call of this reduction asks for $(1+c/\log n)$-approximate
distances in a graph with $O(n)$ vertices, $O(m)$ edges and positive,
polynomially bounded integer weights. It is answered by
\cref{prop:parallel-oracle} for the same $x$, in $\widetilde O(m)$ work
and $\sqrt{n/x}\,n^{o(1)}\le(n^\omega/m)^{1/(2\omega-2)}n^{o(1)}$
depth.
\Cref{lem:parallel-exact-reduction} therefore computes the distances
from $s$ and a shortest-path tree with high probability, using
$\widetilde O(m)$ work and $(n^\omega/m)^{1/(2\omega-2)}n^{o(1)}$
depth.
\end{proof}

Any other choice of $x\in[n]$ gives the more general tradeoff stated
after \cref{thm:exact-tradeoff}, with $\widetilde O(m+nx^{\omega-1})$
work and $(n/x)^{1/2}n^{o(1)}$ depth.

\section{Streaming Implicit Hopset Construction}
\label{sec:streaming-closures}

Our streaming algorithm of \cref{sec:streaming-reductions} first
computes an implicit hopset and then uses it to compute distances from
the source. \Cref{sec:streaming-reductions} showed how to use the
implicit hopset, assuming the following lemma for computing it. In this
section, we prove this lemma.

\implicithopsetlemma*

Like the CFR algorithm, we build our implicit hopset in the $O(\log n)$
stages of \Cref{alg:stages}. However, the hopset of the earlier stages
is stored only \emph{implicitly} as a pair $(H_0,\mathcal B)$. Each
trial outputs the stars of its early pivots and the blocks $B_p$ of its
late pivots, which are added to $H_0$ and to $\mathcal B$ after the
stage. The trials never compute the edges in the block closures
$\TC_w((G\cup H_0)[B])$. We run the searches of a trial with the
batches of searches of \cref{sec:core-algorithm}. A batch restricts
each block $B$ to the vertex set of a search, so that the cost of the
search stays proportional to the number of vertices that it settles.
Thus a search inside a subgraph of the recursion with vertex set $S$
uses only the block closures of the sets $B\cap S$. The \emph{underlying path} of an
edge $(u,v)$ in the block closure of $B$ is a shortest path from $u$ to
$v$ in $(G\cup H_0)[B]$. If $S$ misses a vertex of this path, then the
searches inside $S$ may see a heavier edge $(u,v)$, or none. A heavier
or missing edge is a problem because the paths that later stages
shortcut use block-closure edges. A subpath of such a path may then be
longer in the subgraph of the recursion that contains it than in the
whole graph, or not be a path there at all. This subpath then has to be split at this edge,
and the analysis of \cref{sec:sequential} does not count the extra
subpaths that these splits create. The vertex sets $S$ of the subgraphs
are cut out by the balls of the pivots. We will choose the radii
$\rho_p$ of these balls with a small random shift, which makes such
splits rare for the edges of any fixed path.

\subsection{One streaming trial}
\label{sec:streaming-trial}

In this subsection, we describe one trial of the streaming
construction. It modifies the construction of \cref{sec:sequential}.
The two main changes are that its searches run on the restricted graphs
defined next and that the pivots use shifted radii.

Consider a stage $i$ of \Cref{alg:stages}, and let $K=K_i=G\cup H_i$ be
the graph on which its trials run. The hopset $H_i$ of the earlier
stages is stored as a pair $(H_0,\mathcal B)$, so that
$H_i=H_0\cup\bigcup_{B\in\mathcal B}\TC_w((G\cup H_0)[B])$. Recall
from \cref{sec:core-problem} that $\dist_B$ denotes distances in the
block subgraph $(G\cup H_0)[B]$. For each block-closure edge $(u,v)$ of
$K$, we fix a block $B$ with $w_K(u,v)=\dist_B(u,v)$, together with an
underlying path in $(G\cup H_0)[B]$. For any $S\subseteq V$, a search
with vertex set $S$ in a batch of \cref{lem:core-batch} runs on the
\emph{restricted graph}
\[
 K_S=(G\cup H_0)[S]\cup\bigcup_{B\in\mathcal B}
 \TC_w\bigl((G\cup H_0)[B\cap S]\bigr),
\]
or on the reverse of $K_S$, with rounded edge weights. We will use
two simple properties of the restricted graphs. Item~1 below says that
distances do not decrease when the vertex set shrinks. Item~2 says that
a block-closure edge keeps its weight in $K_S$ when its underlying path
lies in $S$, and that each subpath of the underlying path that lies in
$S$ is shortcut in $K_S$ by an edge that is no heavier than the
block-closure edge.

\begin{observation}
\label{obs:restricted-blocks}
Let $S'\subseteq S\subseteq V$.
\begin{enumerate}[label=(\arabic*),itemsep=1pt]
 \item Each edge of $K_{S'}$ is also an edge of $K_S$, of at most the
 same weight.
 \item Let $\pi$ be the underlying path of a block-closure edge $(u,v)$
 of $K$, and let $\pi[y,\dots,z]$ be a subpath of $\pi$ that lies in $S$.
 Then $K_S$ has an edge $(y,z)$ of weight at most $w_K(u,v)$. If all of
 $\pi$ lies in $S$, then $(u,v)$ has the weight $w_K(u,v)$ in $K_S$.
\end{enumerate}
Both items also hold if the edge weights of $K$, $K_{S'}$ and $K_S$ are
all rounded up to multiples of some $\alpha>0$.
\end{observation}

\paragraph{Changes to the construction.}
In the execution of our construction for a guess $\Delta$, we run the construction of
\cref{sec:sequential} on the graph $K^{(\Delta)}$, which is obtained
from $K$ by rounding the edge weights up to multiples of
$\alpha=\eps\Delta/(8h)$ as in~\eqref{eq:stage-radius}, with $h=h_x$.
However, the subgraph of the recursion with vertex set $S$ is the graph
$K_S^{(\Delta)}$ obtained from $K_S$ by the same rounding, and not an
induced subgraph of $K^{(\Delta)}$. As $K_S^{(\Delta)}$ is, up to the
factor $\alpha$, the rounded graph on which a search of
\cref{lem:core-batch} with vertex set $S$ runs, the searches of one
level can be run as one batch. There are four further changes:
\begin{itemize}[itemsep=2pt]
 \item Recall from \cref{sec:cfr} that a pivot $p$ of a subgraph $G'$
 at a level $r$ samples one of the $\lambda/(4k)$ groups $I_\sigma$ of
 $4k$ consecutive integers in $[\lambda,2\lambda)$, say $I_{\sigma_p}$,
 uniformly at random. It then picks an integer $\eta\in I_{\sigma_p}$
 that minimizes the size of the shell
 $F_p(\eta)=\Rboth_{(\eta+1)D_r}(G',p)\setminus\Rboth_{(\eta-1)D_r}(G',p)$,
 where $\Rboth_d(G',p)$ is the union of the in-ball and the out-ball of
 radius $d$ around $p$.
 In our streaming construction, $p$ also samples a shift $\varphi_p$
 uniformly from $\{0,1/n,\dots,(n-1)/n\}$. It picks as $\eta_p$ an
 integer $\eta\in I_{\sigma_p}$ that minimizes the size of the shifted
 shell $F_p(\eta+\varphi_p)$, and its radius is
 $\rho_p=(\eta_p+\varphi_p)D_r$. As in \cref{sec:sequential}, the
 fringe and the block of $p$ are
 $F_p=\Rboth_{\rho_p+D_r}(G',p)\setminus\Rboth_{\rho_p-D_r}(G',p)$ and
 $B_p=\Rboth_{\rho_p+D_r}(G',p)$.
 \item We use the larger radius factor
 $\lambda=1536\gamma^2k^3\lceil\log n\rceil^2$ instead of
 $16\gamma^2k^2\lceil\log n\rceil^2$, which is still a multiple of
 $4k$. Together with the shifts, the larger radius factor makes the
 splits at block-closure edges described at the start of this section
 rare. We let $\eta_{\max}=2\lambda+2$, so that the search radius
 $\eta_{\max}D_r$ of a pivot $p$ exceeds the radius
 $\rho_p+D_r<(2\lambda+1)D_r$ of its block by more than $D_r$. We
 increase by two the number $L$ of levels by which the stars in advance
 are built early (see~\eqref{eq:c-choice}). The two extra levels of
 stars compensate for the larger $\lambda$ in the detours of the early
 stars and keep $\eta_{\max}D_0\le5\Delta$, so that the searches have
 radius at most $R=\lceil64h/\eps\rceil$ in units of $\alpha$. The
 threshold level becomes
 $r_0=\min\{r_{\max},\lceil\log_k(4n/x)\rceil\}$. For simplicity, we
 also assume that $x\ge4\lceil\gamma k\log n\rceil$, which is
 polylogarithmic. Then each block $B_p$ has at most $x$ vertices when
 the bounds of \cref{prop:cfr-balls} hold, as we show in the proof of
 \cref{lem:streaming-searches}.
 \item Each late pivot adds its block $B_p$ to $\mathcal B$, so that the
 hopset gains the block closure $\TC_w((G\cup H_0)[B_p])$. Unlike the
 approximate block closure of \cref{sec:sequential}, which is taken in
 the subgraph of the recursion, this block closure is taken in the
 block subgraph $(G\cup H_0)[B_p]$.
 \item For a sufficiently large constant $C$, the execution is aborted
 if some block $B_p$ has more than $x$ vertices, if the number of
 copies of vertices (their occurrences in the subgraphs of the
 recursion) or $\sum_p|B_p|$ exceeds $Cn\log^Cn$, or if the
 batches create more than $C\log^Cn\cdot(n+\sum_{B\in\mathcal B}|B|)$
 distance records, which measure the cost of the batches in
 \cref{lem:core-batch}.
\end{itemize}
The arguments of \cref{sec:cfr-facts} give
\cref{prop:cfr-balls,prop:cfr-copies} for these executions as well.
Recall that these propositions bound the number of vertices within the
search radius $\eta_{\max}D_r$ of any vertex by $b_r=\lceil n/k^r\rceil$,
and the numbers of copies and of search visits of a vertex. Their
arguments carry over, using $\lambda D_r\le\rho_p<2\lambda D_r$ and
$\eta_{\max}D_{r+1}\le\eta_{\max}D_r/4<(\lambda-1)D_r$ and the fact that
for a fixed shift, the shifted shells of one parity in a group are
disjoint subsets of $\Rboth_{\eta_{\max}D_r}(G',p)$. Item~1 of
\cref{obs:restricted-blocks} takes the place of the fact that distances
do not decrease in induced subgraphs.

One streaming trial, shown in \Cref{alg:implicit-trial}, differs from
one trial of our hopset construction in \Cref{alg:cfr-shell} in three
ways. The global stars run as one batch of searches in
$K^{(\Delta)}$, and the calls of each level of the recursion run as one
batch on the graphs $K_S^{(\Delta)}$. The pivots use the shifted radii.
Each execution that is not aborted outputs its star edges and its blocks
$B_p$ instead of their block closures.

\begin{algorithm}[t]
\caption{One trial of the streaming construction.}
\label{alg:implicit-trial}
\begin{algorithmic}[1]
\Procedure{ImplicitTrial}{$H_0,\mathcal B,x,\eps$}
 \For{every guess $\Delta=2^jw_{\min}$ with $0\le j\le\lceil\log_2(2nW)\rceil$}
  \State $\alpha\gets\eps\Delta/(8h_x)$
  \State Sample the ranks \Comment{levels of the early pivots}
  \State Build the global stars in $K^{(\Delta)}$ with one batch
  \For{$r=0,1,\dots,r_{\max}$}
   \State Run the calls of \Cref{alg:two-speed} at level $r$ on the
   graphs $K_S^{(\Delta)}$ with shifted radii, with one batch for all
   their searches
  \EndFor
  \If{the execution is not aborted}
   \State Output the star edges and the blocks $B_p$
  \EndIf
 \EndFor
\EndProcedure
\end{algorithmic}
\end{algorithm}

\subsection{Underlying paths in smaller subgraphs}
\label{sec:streaming-underlying}

In this subsection, we show that one streaming trial shortcuts any
fixed path of $K$ with at most $2h$ edges with constant probability,
which is the condition that \cref{lem:shared-hopset-iteration} requires
of the trials of \Cref{alg:stages}. Before the trial starts, we fix a
simple path $P$ in $K$ with at most $2h$ edges and
$w_{\min}\le w_K(P)\le2nw_{\max}$. Let $\Delta$ be the guess with
$\Delta\le w_K(P)<2\Delta$, so that
$\Delta\le w_{K^{(\Delta)}}(P)<(1+\eps/4)\,w_K(P)<3\Delta$ by
\eqref{eq:stage-path-rounding}. We follow $P$ through the execution for $\Delta$ with the
terminology of \cref{sec:sequential}, measuring the lengths of the
pieces of $P$ in $K^{(\Delta)}$. Recall that the pieces of $P$ at a
level $r$ are disjoint subpaths of $P$ of length at most $D_r$ that
remain to be shortcut, each in a single subgraph at level $r$.

The analysis of \cref{sec:sequential} relies on three properties of
induced subgraphs: (P1)~distances do not decrease from one level to the
next, (P2)~a piece of $P$ is also a path of the same length in the
subgraph that contains it, and (P3)~the block closure of a block $B_p$
contains each subpath of a piece of $P$ that lies inside $B_p$. Our
subgraphs of the recursion are the restricted graphs $K_S^{(\Delta)}$,
which are not induced subgraphs of $K^{(\Delta)}$. Property~(P1) still
holds by Item~1 of \cref{obs:restricted-blocks}. Properties~(P2)
and~(P3), however, can fail at the block-closure edges of $P$. By
Item~2 of \cref{obs:restricted-blocks}, a block-closure edge of a piece
of $P$ keeps its weight in the subgraph that contains the piece if this
subgraph contains the underlying path of the edge. Otherwise, the edge
can be heavier there, or missing.

We therefore keep track of the following invariant. We say that a piece $Q$ of $P$ in a subgraph with vertex
set $S$ satisfies the \emph{underlying-path invariant} if the underlying
path of each block-closure edge of $Q$ lies in $S$. In this case,
$Q$ has the same length in $K_S^{(\Delta)}$ as in $K^{(\Delta)}$ by
Item~2 of \cref{obs:restricted-blocks}, which gives~(P2). We show in
\cref{clm:streaming-prefix} that the invariant also gives~(P3).

\begin{figure}[t]
\centering
\begin{tikzpicture}[vtx/.style={circle,fill,inner sep=1.3pt}]
 \draw (-18:4) arc[start angle=-18,end angle=40,radius=4]
  node[above right] {$\Rout_\rho(G',p)$};
 \path (0,0) node[circle,draw,inner sep=1.6pt,label=left:{$p$}] {}
  (2.9,-0.8) node[vtx,label=below:{$u$}] (u) {}
  (3,0.9) node[vtx,label=above:{$v$}] (v) {}
  (5,0) node[vtx,label=right:{$z$}] (z) {};
 \draw[->,thick] (u) -- node[left] {$e$} (v);
 \draw[->,dashed] (u) .. controls (4.3,-0.9) .. (z);
 \draw[->,dashed] (z) .. controls (4.3,0.9) .. (v);
 \draw[->] (0.6,-2.35) -- (7.4,-2.35);
 \draw (2.9,-2.25) -- (2.9,-2.45)
  node[anchor=north east,xshift=4pt] {$\dist_{G'}(p,u)$};
 \draw (4,-2.25) -- (4,-2.45) node[below] {$\rho$};
 \draw (5,-2.25) -- (5,-2.45)
  node[anchor=north west,xshift=-4pt] {$\dist_{G'}(p,z)$};
 \draw (2.9,-1.95) -- (2.9,-1.85) -- (5.5,-1.85) -- (5.5,-1.95);
 \node[above] at (4.2,-1.85) {$w_{K^{(\Delta)}}(e)$};
\end{tikzpicture}
\caption{A type~3 split. The endpoints of the block-closure edge
$e=(u,v)$ lie in the out-ball $\Rout_\rho(G',p)$, but the underlying
path of $e$ (dashed) leaves the ball at $z$. Hence $\rho$ lies in the
window $[\dist_{G'}(p,u),\dist_{G'}(p,u)+w_{K^{(\Delta)}}(e))$, which a
randomly shifted radius hits with probability
$O(k\,w_{K^{(\Delta)}}(e)/(\lambda D_r))$.}
\label{fig:type3}
\end{figure}
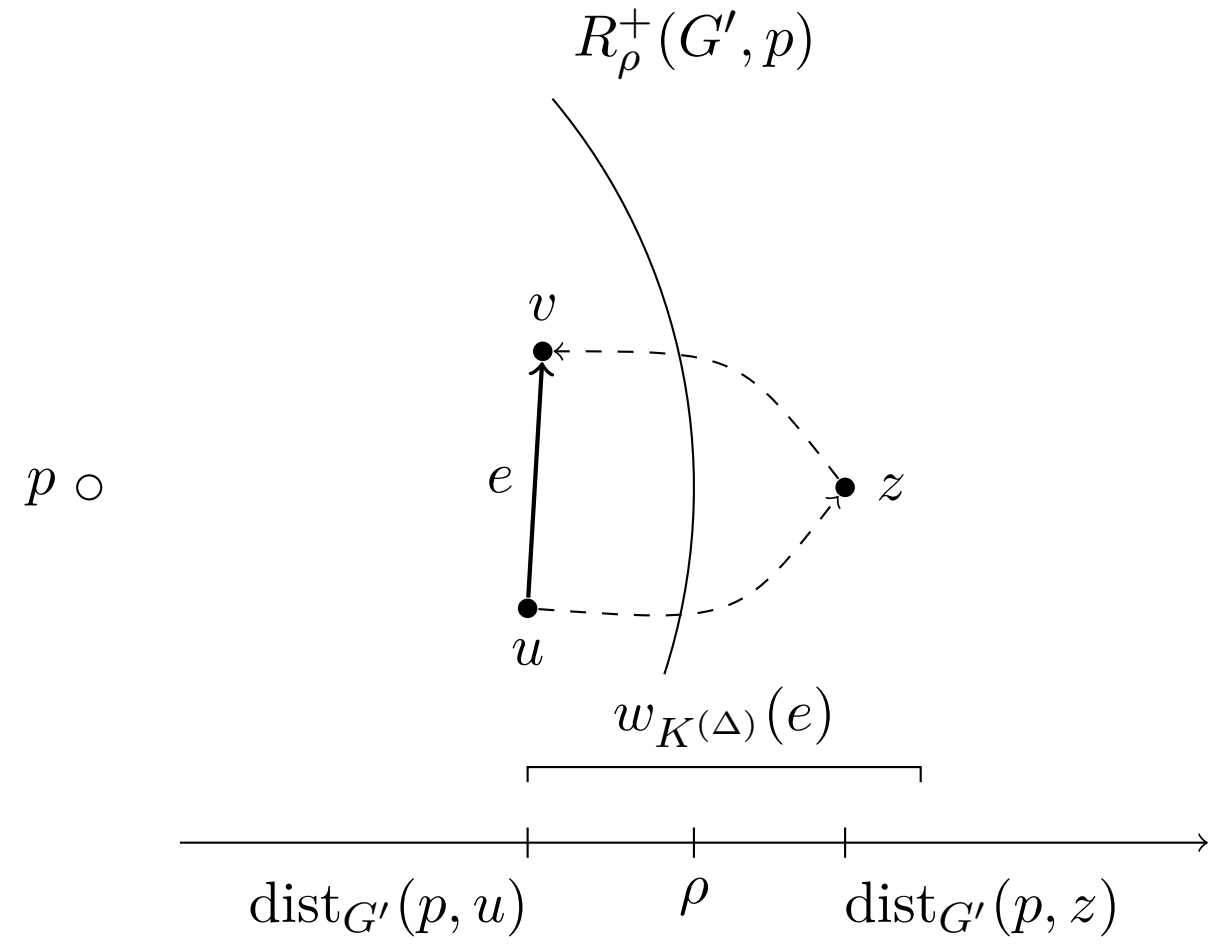

\paragraph{Type~3 splits.}
At level~$0$, the only subgraph is $K^{(\Delta)}$ itself, so the
underlying-path invariant holds. It can fail when we pass from a level
$r$ to level $r+1$. The subgraphs at level $r+1$ are cut out by the
balls of the pivots, and a ball can contain both endpoints of a
block-closure edge without containing all of its underlying path (see
\cref{fig:type3}). When this happens to a block-closure edge $(u,v)$ of
a subpath of $P$ that we follow to level $r+1$, we split the subpath at
$(u,v)$, so that the pieces of $P$ at level $r+1$ satisfy the invariant
again. We call this a type~3 split. The edge $(u,v)$ is an edge of $K$,
so our shortcut path can keep it as a boundary edge, just like the
boundary edges of type~1 and type~2 splits.

Formally, when we pass from level $r$ to level $r+1$, each subpath of a
piece $Q$ of $P$ that we follow lies in a single subgraph at level
$r+1$, say with vertex set $X$. At an early level, if no pivot is a
bridge of $Q$, these subpaths come from the type~1 split of $Q$, which
divides $Q$ into subpaths that each lie in a single core or fringe
subgraph. At a late level, the only such subpath is the middle subpath
of $Q$. By Item~2 of \cref{lem:late-peeling}, it does not lie in any
pivot's ball and lies in the core subgraph with the empty label. If
such a subpath contains a block-closure
edge $(u,v)$ whose underlying path does not lie in $X$, then a
\emph{type~3 split} divides the subpath at $(u,v)$, and $(u,v)$ becomes
a boundary edge. The type~2 split at scale $D_{r+1}$ then divides the
resulting subpaths into the pieces of $P$ at level $r+1$. As the
underlying path of each block-closure edge of these pieces lies in $X$,
the pieces of $P$ satisfy the underlying-path invariant at all levels.

Type~3 splits create pieces of $P$ that the analysis of
\cref{sec:sequential} does not count, so we must show that they are
rare. At level $r+1$, the core subgraphs are cut out by the in-balls
and out-balls of radius $\rho_p$ of the pivots $p$ at level $r$, and
the fringe subgraphs by the balls of radius $\rho_p-D_r$ and
$\rho_p+D_r$. For a type~3
split at a block-closure edge $e=(u,v)$, some ball of one of these
radii around a pivot $p$ must separate a vertex $z$ of the underlying
path of $e$ from $u$ and $v$. That is, the ball contains $z$ but neither
$u$ nor $v$, or it contains $u$ and $v$ but not $z$. In the subgraph at
level $r$ that contains $Q$, the vertex $u$
reaches $z$ and $z$ reaches $v$ within distance $w_{K^{(\Delta)}}(e)$.
The radius of the ball must therefore lie in one of four windows of
length $w_{K^{(\Delta)}}(e)$ that start or end at the distances between
$p$ and $u$ or $v$ (see \cref{fig:type3}).

The shift $\varphi_p$ makes such a radius unlikely. It spreads the
radii of $p$ over $[\lambda D_r,2\lambda D_r)$ at a finer resolution
than $w_{K^{(\Delta)}}(e)$. Whichever integer $\eta_p$ the
pivot $p$ picks in its group, a radius of $p$ thus lies in one of the
windows with probability $O(k\,w_{K^{(\Delta)}}(e)/(\lambda D_r))$.
The weights of the block-closure edges of $Q$ add up to at most $D_r$,
and only the few pivots near $Q$ matter, so the expected number of
type~3 splits of the subpaths of $Q$ is small for our choice of
$\lambda$.

Without the shift $\varphi_p$, an edge of small weight could cause a
type~3 split with a probability of order $k/\lambda$. Indeed, all radii
of $p$ would then be integer multiples of $D_r$. A window of length $w_{K^{(\Delta)}}(e)<D_r$ can contain such
a multiple $\eta D_r$, and whenever the group of $p$ contains $\eta$,
which happens with probability $4k/\lambda$, the pivot may pick
$\eta_p=\eta$ to minimize the size of its shell. Summed over the up to
$D_r/\alpha$ block-closure edges of $Q$, these probabilities would grow
with the number of edges of $Q$ instead of with its length
$w_{K^{(\Delta)}}(Q)\le D_r$.

The next lemma turns this intuition into a bound. In the proof of
\cref{lem:implicit-path-shortening} below, we use it to bound the extra
pieces of $P$ that type~3 splits create.

\begin{lemma}
\label{lem:streaming-boundary-charge}
Let $Q$ be a piece of $P$ at a level $r<r_{\max}$, and suppose that the
bounds of \cref{prop:cfr-balls} hold at level $r$. Given the levels
before $r$, the expected number of type~3 splits of the subpaths of $Q$
that we follow is at most $1/(4\gamma k\log n)$.
\end{lemma}

\begin{proof}
Let $G'$ be the subgraph at level $r$ that contains $Q$. For a pivot
$p$ of $G'$ and a block-closure edge $e=(u,v)$ of $Q$, the
\emph{windows} of $p$ and $e$ are the four half-open intervals of length
$w_{K^{(\Delta)}}(e)$ that start at $\dist_{G'}(p,u)$ or at
$\dist_{G'}(v,p)$, or that end at $\dist_{G'}(p,v)$ or at
$\dist_{G'}(u,p)$. They are determined by the levels before $r$.

\begin{claim}
\label{clm:streaming-cut}
If a type~3 split of a subpath of $Q$ happens at $e$, then for some
pivot $p$ of $G'$, one of the radii $\rho_p-D_r$, $\rho_p$ and
$\rho_p+D_r$ lies in a window of $p$ and $e$.
\end{claim}
\begin{proof}
Let $X\ni u,v$ be the vertex set of the subgraph at level $r+1$ that
contains the subpath. Some vertex $z$ of the underlying path of $e$
lies outside $X$. By the underlying-path invariant and Item~2 of
\cref{obs:restricted-blocks}, $z$ is a vertex of $G'$ with
$\dist_{G'}(u,z)\le w_{K^{(\Delta)}}(e)$ and
$\dist_{G'}(z,v)\le w_{K^{(\Delta)}}(e)$. Consider any out-ball
$\Rout_\rho(G',p)$. If it contains $u$ and $v$ but not $z$ (see
\cref{fig:type3}), then
$\dist_{G'}(p,u)\le\rho<\dist_{G'}(p,z)\le\dist_{G'}(p,u)+w_{K^{(\Delta)}}(e)$,
so $\rho$ lies in the window that starts at $\dist_{G'}(p,u)$. If it
contains $z$ but neither $u$ nor $v$, then
$\dist_{G'}(p,v)-w_{K^{(\Delta)}}(e)\le\dist_{G'}(p,z)\le\rho<\dist_{G'}(p,v)$,
so $\rho$ lies in the window that ends at $\dist_{G'}(p,v)$. Arguing in
the same way for in-balls, we conclude that $\rho$ lies in a window of
$p$ and $e$ whenever $z$ differs from both $u$ and $v$ in membership in
$\Rout_\rho(G',p)$ or in $\Rin_\rho(G',p)$.

\textit{Case~1: $X$ is the vertex set of a core subgraph.} Recall that
the vertices of a core subgraph are not removed and have the same
labels. That is, $u$ and $v$ lie in the same in-balls and out-balls of
radius $\rho_p$ of the pivots $p$, and not in both balls of any pivot.
The vertex $z$ is
removed or has a different label, since $z\notin X$. Hence for some
pivot $p$, the vertex $z$ differs from $u$ and $v$ in membership in
$\Rin_{\rho_p}(G',p)$ or in $\Rout_{\rho_p}(G',p)$, so that $\rho_p$
lies in a window of $p$ and $e$.

\textit{Case~2: $X=F_p$ for a pivot $p$.} Here $r$ is an early level
and $p$ is not a bridge of $Q$, so $Q$ has no vertex in
$\Rin_{\eta_{\max}D_r}(G',p)$ or no vertex in
$\Rout_{\eta_{\max}D_r}(G',p)$. By symmetry, we may assume the former. From
$u,v\in F_p$ and $\rho_p+D_r<\eta_{\max}D_r$, we get that both $u$ and
$v$ lie in $\Rout_{\rho_p+D_r}(G',p)$ but not in
$\Rboth_{\rho_p-D_r}(G',p)$. If $z\in\Rboth_{\rho_p-D_r}(G',p)$, then
$\rho_p-D_r$ lies in a window of $p$ and $e$. Otherwise, $z\notin F_p$
implies $z\notin\Rout_{\rho_p+D_r}(G',p)$, so that $\rho_p+D_r$ lies in
a window of $p$ and $e$.
\end{proof}

Now fix a pivot $p$ of $G'$, one of its three radii $\rho_p-D_r$,
$\rho_p$ and $\rho_p+D_r$, and one window of $p$ and $e$. As the radius
is $(\eta_p+\varphi_p)D_r$ plus $-D_r$, $0$ or $D_r$, it lies in the
window if and only if $\eta_p+\varphi_p$ lies in a certain interval of
length $w_{K^{(\Delta)}}(e)/D_r$. As the groups partition the integers
in $[\lambda,2\lambda)$, each multiple of $1/n$ in $[\lambda,2\lambda)$
is $\eta+\varphi$ for exactly one pair $(\sigma,\varphi)$ with
$\eta\in I_\sigma$, since $\varphi$ must be its fractional part and
$\eta$ its integer part. For large $n$, we
have $D_0/\alpha\le8h/\eps\le n$, since $h=n^{1/2+o(1)}$ and $1/\eps$ is
polylogarithmic. Hence
$1/n\le\alpha/D_0\le w_{K^{(\Delta)}}(e)/D_r$, since
$w_{K^{(\Delta)}}(e)\ge\alpha$ and $D_r\le D_0$. The interval then
contains at most
$n\,w_{K^{(\Delta)}}(e)/D_r+1\le2n\,w_{K^{(\Delta)}}(e)/D_r$ multiples
of $1/n$. For all but at most $2n\,w_{K^{(\Delta)}}(e)/D_r$ of the
$\lambda n/(4k)$ pairs $(\sigma,\varphi)$, the radius thus lies outside
the window, whichever integer $\eta\in I_\sigma$ the pivot $p$ picks. The
pair $(\sigma_p,\varphi_p)$ is uniform and independent of the levels
before $r$ and of the pivots. Hence the radius lies in the window with
probability at most
\[
 \frac{2n\,w_{K^{(\Delta)}}(e)/D_r}{\lambda n/(4k)}
 =\frac{8k\,w_{K^{(\Delta)}}(e)}{\lambda D_r}.
\]
Over the three radii and the four windows, one of the radii of $p$
lies in a window of $p$ and $e$ with probability at most
$96k\,w_{K^{(\Delta)}}(e)/(\lambda D_r)$.

Next, write $Q=\langle v_0,\dots,v_\ell\rangle$, and observe that only
pivots near $Q$ can have a radius $\rho$ in a window of $p$ and $e$,
where $\rho\le\rho_p+D_r<(2\lambda+1)D_r$. If $\rho$ lies in the window
that starts at $\dist_{G'}(p,u)$, then $p$ reaches $u$ within distance
$\rho$. If it lies in the window that ends at $\dist_{G'}(p,v)$, then
$p$ reaches $v$ within distance $\rho+w_{K^{(\Delta)}}(e)$, and the part
of $Q$ after $v$ has length at most $D_r-w_{K^{(\Delta)}}(e)$. In both
cases, following $Q$ shows that
$\dist_{G'}(p,v_\ell)\le\rho+D_r<\eta_{\max}D_r$ by our choice of
$\eta_{\max}=2\lambda+2$. Similarly, $\dist_{G'}(v_0,p)<\eta_{\max}D_r$
for the other two windows. By \cref{prop:cfr-balls}, at most $2b_r$
vertices of $G'$ are this close to $Q$. Recall that, given the levels
before $r$, each of them is a pivot with probability at most $p_r$ (see
\cref{sec:cfr-facts}). By \cref{clm:streaming-cut} and a union bound
over the block-closure edges $e$ of $Q$ and these vertices, the expected
number of type~3 splits of the subpaths of $Q$ is at most
\[
 \sum_e2b_r\,p_r\cdot\frac{96k\,w_{K^{(\Delta)}}(e)}{\lambda D_r}
 \le\frac{192k\,b_rp_r}{\lambda}
 \le\frac{1}{4\gamma k\log n},
\]
where the first step uses $\sum_ew_{K^{(\Delta)}}(e)\le w_{K^{(\Delta)}}(Q)\le D_r$
and the second uses $b_rp_r\le2\gamma k\log n$ and
$\lambda\ge1536\gamma^2k^3\log^2n$.
\end{proof}

We can now prove the main lemma of this subsection, which says that one
streaming trial shortcuts the path $P$ with constant probability.

\begin{lemma}
\label{lem:implicit-path-shortening}
For the stage and the path $P$ fixed above, let $H^+$ consist of the
star edges output by \textnormal{\textsc{ImplicitTrial}}$(H_0,\mathcal B,x,\eps)$
and of the block closures $\TC_w((G\cup H_0)[B])$ of the blocks $B$
that it outputs. Then with probability at least $1/2$, the graph
$K\cup H^+$ contains a path from the first vertex of $P$ to its last
vertex with at most $h$ edges and length at most $(1+2\eps)w_K(P)$.
\end{lemma}
\begin{proof}
We follow the proof of \cref{lem:one-execution}, but must take type~3
splits and underlying paths into account. Let $H_\Delta$ consist of the
star edges of the execution of our construction for the guess $\Delta$,
run without aborting,
and of the block closures $\TC_w((G\cup H_0)[B_p])$ of the
blocks $B_p$ of this execution. We have
$H_\Delta\subseteq H^+$ whenever the execution is not aborted.

\underline{Early levels.} At the early levels, type~3 splits only
change the constants in the analysis of~\cite{CFR} behind two imported
results. These are \cref{prop:cfr-pieces}, which bounds the number of
pieces of $P$ at the early levels, and \cref{prop:cfr-detours}, which
bounds the total detour of the star edges that shortcut the bridge
subpaths. To see this, let $Z_r$ denote the number of subpaths of
$P$ created by the splits at level $r-1$ and $Y_r$ the number of
vertices within distance $\eta_{\max}D_r$ of them. We let $Z'_r$ and
$Y'_r$ be these numbers after the type~2 split at scale $D_r$. The
bound of \cref{prop:cfr-pieces} is derived from two one-level
recurrences~\cite[Lemmas~8 and~9]{CFR},
\[
 \E Y_{r+1}\le\frac{4\,\E Z'_r}{p_r}
 \qquad\text{and}\qquad
 \E Z_{r+1}\le2p_r\,\E Y'_r+\E Z'_r.
\]
Both proofs use (P1) and~(P2). Each type~3 split at level $r$ divides
one subpath into two and adds one subpath of length at most $D_r$ to
$Z_{r+1}$. By \cref{prop:cfr-balls} at level $r$ and~(P1), at most $2b_r$
vertices lie within distance $\eta_{\max}D_{r+1}$ of this subpath, so
the split adds at most $2b_r$ to $Y_{r+1}$. By
\cref{lem:streaming-boundary-charge}, the expected number of type~3
splits at level $r$ is at most $\E Z'_r/(4\gamma k\log n)$. Using also
$b_rp_r\le2\gamma k\log n$, we find that type~3 splits add at
most $\E Z'_r$ to $\E Z_{r+1}$ and at most
$2b_r\E Z'_r/(4\gamma k\log n)\le\E Z'_r/p_r$ to $\E Y_{r+1}$. The
constant $4$ in the first recurrence thus becomes $5$, while the
coefficient $1$ of $\E Z'_r$ in the second becomes $2$. With these
constants, the
induction of~\cite[Lemma~10]{CFR} can be carried out as in~\cite{CFR}. It
gives $\E Z'_r=O(\gamma^rk^{c+(r+1)/2}\log n)$ for all $r\le r_0$.
Markov's inequality then shows that with probability $1-o(1)$, the
bound of \cref{prop:cfr-pieces} holds with a larger constant. In this
case, the argument behind \cref{prop:cfr-detours}, which also relies on
(P1) and~(P2), bounds the total detour of the stars by
$\eps\,w_{K^{(\Delta)}}(P)$, where the two extra levels of stars and
the constant in $k=\Theta(\log n)$ absorb the larger constants.

\underline{Late levels.} Recall that each late pivot adds its block
$B_p$ to $\mathcal B$, so that the block closure of $B_p$ is taken in
the block subgraph $(G\cup H_0)[B_p]$ and not in the subgraph of the
recursion. The block-closure edges of a subpath of a piece of $P$
inside $B_p$ are not edges of $(G\cup H_0)[B_p]$, so this block closure
shortcuts the subpath if $B_p$ also contains their underlying paths.
Recall that
\cref{clm:prefix-containment} shows that the block $B_p$ contains the
prefix of a piece of $P$ up to any vertex in the in-ball of $p$. The
following stronger form of \cref{clm:prefix-containment} shows that
$B_p$ also contains these underlying paths, which gives~(P3).

\begin{claim}
\label{clm:streaming-prefix}
Let $Q=\langle v_0,\dots,v_\ell\rangle$ be a piece of $P$ at a late
level, in a subgraph $G'$, and let $p$ be a pivot of $G'$ whose in-ball
contains $v_b$. Then $B_p$ contains all vertices of $Q[v_0,\dots,v_b]$
and of the underlying paths of the block-closure edges of
$Q[v_0,\dots,v_b]$. In particular, for all $0<a\le b$, the block
closure $\TC_w((G\cup H_0)[B_p])$ has an edge $(v_0,v_a)$ of weight at
most $w_K(Q[v_0,\dots,v_a])$. The case of an out-ball and a suffix of
$Q$ is symmetric.
\end{claim}
\begin{proof}
By the underlying-path invariant, $Q$ is a path of length at most $D_r$
in $G'$. The argument of \cref{clm:prefix-containment} then shows that
all vertices of $Q[v_0,\dots,v_b]$ lie in $B_p$. Consider any vertex $z$ of the
underlying path of a block-closure edge $(v_q,v_{q+1})$ with $q<b$. The
invariant and Item~2 of \cref{obs:restricted-blocks} bound the distance
from $z$ to $v_{q+1}$ in $G'$ by the weight of $(v_q,v_{q+1})$
in $K^{(\Delta)}$. Hence
$\dist_{G'}(z,p)\le w_{K^{(\Delta)}}(Q[v_q,\dots,v_b])+\dist_{G'}(v_b,p)\le
D_r+\rho_p$, i.e., $z\in\Rin_{\rho_p+D_r}(G',p)\subseteq B_p$.
Replacing all block-closure edges of $Q[v_0,\dots,v_a]$ by their
underlying paths gives a walk from $v_0$ to $v_a$ in
$(G\cup H_0)[B_p]$ of length $w_K(Q[v_0,\dots,v_a])$.
\end{proof}

Let $Q=\langle v_0,\dots,v_\ell\rangle$ be a piece of $P$ at a late
level $r$. Recall from \cref{sec:late} that $v_i$ is the last vertex of
$Q$ in the in-ball of a pivot, that $v_j$ is the first vertex of $Q$ in
the out-ball of a pivot, and that the middle subpath
$Q[v_{i+1},\dots,v_{j-1}]$ lies between them if $i<j$. We show by
induction from $r=r_{\max}$ down to $r_0$ that, given the levels before $r$,
the graph $K\cup H_\Delta$ contains a path $\widehat Q$ from $v_0$ to
$v_\ell$ of length at most $w_K(Q)$ with
$\E|\widehat Q|\le14^{r_{\max}-r+1}$ edges. The base $14$ instead of
the $13$ of \cref{lem:late-branching} leaves room for the type~3
splits. The argument is that of \cref{lem:late-branching}, with the
edges of \cref{clm:streaming-prefix} in place of those of
\cref{lem:late-peeling}, except that type~3 splits also divide the
middle subpath. Without type~3 splits, the type~2 split at scale
$D_{r+1}=D_r/4$ divides the middle subpath into at most five pieces
$Q'$ of $P$ at level $r+1$. Then $\widehat Q$ consists of the paths
$\widehat{Q'}$ and at most eight other edges, namely two edges of
\cref{clm:streaming-prefix}, the two boundary edges around the middle
subpath, and at most four boundary edges inside it. Each type~3 split
of the middle subpath adds one boundary edge and at most one piece
$Q'$. The edges of \cref{clm:streaming-prefix} are no heavier
than the subpaths of $Q$ that they shortcut. Boundary edges keep their
weights in $K$. It follows by induction that $\widehat Q$ has length at
most $w_K(Q)$. The pieces $Q'$ are determined by the levels before
$r+1$, so the induction hypothesis applies to each of them given these
levels. By \cref{lem:streaming-boundary-charge}, the expected number of
type~3 splits of the middle subpath, given the levels before $r$, is at
most $1/(4\gamma k\log n)$. Taking expectations given these levels, we
obtain
\[
 \E|\widehat Q|
 \le8+5\cdot14^{r_{\max}-r}+\frac{1+14^{r_{\max}-r}}{4\gamma k\log n}
 \le9+6\cdot14^{r_{\max}-r}\le14^{r_{\max}-r+1}.
\]

\underline{Combining the levels.} We replace the bridge subpaths by
star edges and the pieces of $P$ at the threshold level $r_0$ outside
them by the paths $\widehat Q$. The remaining boundary edges, which now
include those of type~3 splits, are kept. If the bound of
\cref{prop:cfr-pieces} holds with the larger constant, there are at
most $\sqrt{n/x}\,n^{O(1/\log\log n)}$ such star and boundary edges,
where the factor~$4$ in the definition of $r_0$ changes only constants.
The paths $\widehat Q$ for the at most $Z'_{r_0}$ remaining pieces have
$\E Z'_{r_0}\cdot14^{r_{\max}-r_0+1}=\sqrt{n/x}\,n^{O(1/\log\log n)}$
edges in expectation, and by Markov's inequality, at most $\log n$
times as many with probability $1-1/\log n$. If the constant
$C_h$ in the hopbound $h_x$ of~\eqref{eq:streaming-hopbound} is large
enough, our shortcut path $\widehat P$ thus has at most $h$ edges.
Moreover, $w_K(e)\le w_{K^{(\Delta)}}(e)$ for all boundary edges $e$,
and each path $\widehat Q$ has length at most
$w_K(Q)\le w_{K^{(\Delta)}}(Q)$, so \cref{prop:cfr-detours} and
\eqref{eq:stage-path-rounding} give
\[
 w_{K\cup H_\Delta}(\widehat P)\le(1+\eps)\,w_{K^{(\Delta)}}(P)
 \le(1+\eps)(1+\eps/4)\,w_K(P)\le(1+2\eps)\,w_K(P).
\]
Finally, the bound of \cref{prop:cfr-pieces} with the larger constant
fails with probability $o(1)$, the bounds of \cref{prop:cfr-balls} with
probability at most $n^{-10}$, and the Markov bound on the paths
$\widehat Q$ with probability $1/\log n$, while by
\cref{lem:streaming-searches}, the execution is aborted with
probability $O(1/\log n)$. For large $n$, with probability at least
$1/2$, none of these events happens. In this case,
$K\cup H^+\supseteq K\cup H_\Delta$ contains $\widehat P$.
\end{proof}

\Cref{lem:implicit-path-shortening} shows that the streaming trials
satisfy the condition~\eqref{eq:trial-event} of
\cref{lem:shared-hopset-iteration}, since $1+2\eps\le1+3\eps$. It
remains to bound the space and the passes of a streaming trial, which
we do next.

\subsection{Space and passes}
\label{sec:streaming-space}

In this subsection, we bound the space and the passes of our streaming
construction, and we prove \cref{lem:implicit-hopset-reduction}. Recall
from \cref{sec:cfr} that a vertex can lie in several subgraphs at the
same level of the recursion, a core subgraph and several fringe
subgraphs, and that we call each occurrence a copy of the vertex. A
streaming trial stores the vertex sets of the subgraphs at the current
level of the recursion, which take space proportional to the number of
copies at this level. It runs the searches of the level as one batch of
\cref{lem:core-batch}, whose space and passes grow with the number of
distance records, which a search creates when it settles a vertex. The
next lemma bounds the number of copies of each vertex and the number of
searches that settle it. It also bounds the probability that an execution is
aborted, which we used in the proof of
\cref{lem:implicit-path-shortening}.

\begin{lemma}
\label{lem:streaming-searches}
In the execution of \textnormal{\textsc{ImplicitTrial}} for a guess
$\Delta$, each vertex has $O(\log n)$ copies and is settled by
$\polylog n$ searches in expectation, and the execution is aborted with
probability $O(1/\log n)$.
\end{lemma}
\begin{proof}
Each copy of a vertex $v$ at a level $r<r_{\max}$ lies in at most one
core subgraph at level $r+1$ and, at an early level, by
\cref{prop:cfr-copies}, in an expected at most $2/(\gamma\log n)$
fringe subgraphs. Over the $O(\log n/\log\log n)$ levels, this gives
$O(1)$ copies of $v$ per level in expectation. Also by
\cref{prop:cfr-copies}, a copy of $v$ is settled by the searches of an
expected $O(\gamma k^{L+1}\log n)$ pivots and stars built in advance,
and each of the expected $O(\gamma k^L\log n)$ global stars settles $v$
at most once. With $k^L=\polylog n$, this proves the first claim.

When a search settles $v$, it creates one distance record of
\cref{lem:core-batch} plus one for each block in $\mathcal B$ that
contains $v$. Since these blocks do not depend on the random
choices of the trial, the expected number of distance records is at
most $\sum_v\polylog n\cdot(1+|\{B\in\mathcal B:v\in B\}|)
=\polylog n\cdot(n+\sum_{B\in\mathcal B}|B|)$, and
$\E\sum_p|B_p|=\widetilde O(n)$ by \cref{cor:small-balls}.
For $C$ large enough, Markov's inequality now shows that each of the
three counts in the abort rule exceeds the allowed value with
probability at most $1/\log n$. Lastly, the bounds of
\cref{prop:cfr-balls} fail with probability at most $n^{-10}$. If they
hold, then
$|B_p|\le2b_{r_0}\le x$ by our choice of $r_0$ and our assumption
$x\ge4\lceil\gamma k\log n\rceil$, as in the proof of
\cref{cor:small-balls}.
\end{proof}

We can now prove \cref{lem:implicit-hopset-reduction}. For
correctness, we combine \cref{lem:implicit-path-shortening} with
\cref{lem:shared-hopset-iteration} on building the hopset in stages.
For the space and the passes, we combine the abort rule, whose counts
\cref{lem:streaming-searches} bounds, with the cost of a batch of
searches in \cref{lem:core-batch}.

\begin{proof}[Proof of \cref{lem:implicit-hopset-reduction}]
Let $h=h_x$. After one pass to find $w_{\min}$ and $w_{\max}$, we run
\Cref{alg:stages} with \textnormal{\textsc{ImplicitTrial}} as the
trial and with $\eps/(10\log n)$ in place of $\eps$. The hopset $H_i$
of stage $i$ is stored as the pair $(H_0,\mathcal B)$ of star edges and
blocks output by the earlier stages, in the form of
\cref{def:implicit-hopset-intro}. The
edges of each trial are weighted by distances in graphs
$K_S^{(\Delta)}$ or $(G\cup H_0)[B_p]$, which are at least the
distances in $G\cup H_0$. By induction on $i$, it follows that all
edges of $H_i$ are noncontracting in $G$ and thus have positive
weights. Adding edges to $H_0$ can only decrease distances inside the
blocks, so $K_{i+1}$ contains all edges of $K_i$ and of the sets $H^+$
of the trials of stage $i$, none of them heavier than before. This and
\cref{lem:implicit-path-shortening} establish the hypotheses of
\cref{lem:shared-hopset-iteration}. With our choice of accuracy, that
lemma shows that the final pair represents an $(h,\eps)$-hopset of $G$
with high probability.

The abort rule caps the size of each block at $x$ and the number of
block vertices that an execution outputs at $n\polylog n$. Each star
edge comes from a vertex that a search settles, so an execution also
outputs at most $\polylog n\cdot(n+\sum_{B\in\mathcal B}|B|)$ star
edges. Over the $O(\log^3n)$ executions, this gives
$\sum_{B\in\mathcal B}|B|=\widetilde O(n)$ and then
$|H_0|=\widetilde O(n)$ at every stage. These polylogarithmic
factors do not compound over the stages, because the bound on the block
vertices does not depend on the stored pair.

We run the trials and their executions one after another, level by
level, and check the abort rule while they run. The vertex sets of the
searches of the batch of the current level are the vertex sets of the
subgraphs at this level, so the abort rule on the number of copies
bounds their total size by $\widetilde O(n)$. The stored pair, the
output found so far and these vertex sets thus take $\widetilde O(n)$
space. Before the
execution would be aborted, the batch creates $\widetilde O(n)$
distance records, so by \cref{lem:core-batch}, whose space bound holds
at any time, it takes $\widetilde O(n)$ space. Since we replaced $\eps$
by $\eps/(10\log n)$ in \eqref{eq:stage-radius}, the searches of the
batch have radius at most $R=O(h\log n/\eps)$, and they take
$\widetilde O(R+x)$ passes. Radii, labels, fringes and blocks
can be computed in memory. With $O(\log n)$ batches per execution and
$O(\log^3n)$ executions, we use $\widetilde O(n)$ space and
$\widetilde O(h_x/\eps+x)$ passes in total.
\end{proof}

\section{Conclusion and AI Disclosure}
\label{sec:conclusion}

We have shown that single-source reachability and shortest paths can be computed in the streaming model using $n^{1/3+o(1)}$ passes and $\widetilde O(n)$ space. 
To obtain this result, we introduced implicit shortcut sets and implicit hopsets, and we proved new bounds for these objects. There are several natural follow-up questions.

\begin{problem}
\label{prob:implicit-shortcut}
Does every $n$-vertex directed graph have a $(D,x)$-implicit shortcut set with diameter $D=(n/x)^{1/3}$?
\end{problem}

A positive answer would generalize the shortcut sets of Kogan and Parter~\cite{KP22}, which have $\widetilde O(n)$ edges and diameter $\widetilde O(n^{1/3})$. It would also be a natural first step towards improving the pass complexity of our algorithm.

Many problems admit simple yet optimal streaming algorithms. These algorithms are often much simpler than the best known algorithms for the same problem in parallel models. This motivates the following question.

\begin{problem}
\label{prob:simple}
Is there a simple semi-streaming algorithm for reachability with $n^{1/3}$ passes, or even with $n^{1/2}$ passes?
\end{problem}

All known semi-streaming algorithms for reachability with $n^{1/2+o(1)}$ or fewer passes, including ours, build on the recursive framework of~\cite{JLS19}. New algorithms for single-source reachability that depart from this approach would be exciting.

The open problems above are natural next steps, but the most important long-term goal is much more ambitious.

\begin{problem}
\label{prob:subpolynomial}
Is there a semi-streaming algorithm for single-source reachability with $n^{o(1)}$ passes, or does this problem require $n^{\Omega(1)}$ passes?
\end{problem}

We consider this to be a fundamental open problem in graph algorithms and streaming.

\paragraph{AI Disclosure.}

While developing the results in this manuscript, we used ChatGPT~5.5 Pro
and ChatGPT~5.6 Sol. At several points in our research, we gave these
tools the statement of a lemma or theorem together with our proof sketch,
and asked them to turn the sketch into a complete proof. Usually, the
resulting proof followed our sketch closely, as in the proofs of~
\cref{thm:abgs-implicit-shortcut,thm:implicit-hopset-intro,thm:core-distance,thm:hopset}. In two
cases, however, the AI-generated proof developed new ideas that we had not
anticipated: 
\begin{itemize}
  \item In \cref{sec:streaming-closures}, we build our implicit hopset in
  the streaming model in $O(\log n)$ stages. The hopset of the earlier
  stages is stored only implicitly, so a search inside a subgraph of the
  recursion sees the block closures of the earlier stages only within
  this subgraph. This can cause block-closure edges to be missing or heavier than in the
  original graph. A path that the later stages shortcut must then be
  split at such an edge. We call this a type~3 split, and the analysis of
  \cref{sec:sequential} does not account for the extra pieces it creates.
  The AI-generated proof handles this with a small change to the
  construction of \cref{sec:sequential}: each pivot shifts the radius of
  its ball by a small random amount, which makes type~3 splits rare. This
  change is what allows us to build implicit hopsets on top of the
  implicit hopsets of earlier stages. We describe it in
  \cref{sec:streaming-trial} and analyze it in
  \cref{lem:streaming-boundary-charge}.
  \item In \cref{sec:streaming-exact-reduction}, we prove a slightly
  stronger version of the reduction of~\cite{Rozhon} from exact to
  approximate SSSP (\cref{thm:streaming-exact-reduction}). The original
  reduction may need approximate distances in graphs with zero edge
  weights, but our approximate SSSP algorithms require positive weights.
\end{itemize}

We remark that all ideas and proofs in
\cref{sec:technical-overview,sec:streaming-reductions,sec:sequential} originated with us.

\vspace{3mm}
We also used Claude Opus~5.5 for writing and editing. Most of our 
technical material began as detailed outlines and proof sketches that we
gave to Claude, which expanded them into full text. We then improved this
text over many rounds, editing it ourselves and asking Claude for targeted rewrites. We take full responsibility for the manuscript's contents and the correctness of its results. 

\vspace{3mm}
Finally, we used GPT~6.1 Sol (in Github Copilot) to produce the Lean formalization of our main result, \cref{cor:streaming-exact-sssp}.

\bibliographystyle{alpha}
\bibliography{references}

@misc{ABGS,
  author        = {Ashvinkumar, Vikrant and Bernstein, Aaron and Probst Gutenberg, Maximilian and Saranurak, Thatchaphol},
  title         = {Parallel reachability and shortest paths on non-sparse digraphs: near-linear work and sub-square-root depth},
  year          = {2026},
  eprint        = {2605.03892},
  archivePrefix = {arXiv},
  primaryClass  = {cs.DS},
  note          = {arXiv preprint arXiv:2605.03892}
}

@misc{ABCGLR26,
  author        = {Azarmehr, Amir and Behnezhad, Soheil and Chung, Lily and Ghafari, Alma and Lange, Jane and Rubinfeld, Ronitt},
  title         = {Streaming Algorithms for Monotonicity Testing},
  year          = {2026},
  eprint        = {2608.07073},
  archivePrefix = {arXiv},
  primaryClass  = {cs.DS},
  note          = {arXiv preprint arXiv:2608.07073}
}

@inproceedings{AJJST22,
  author    = {Assadi, Sepehr and Jambulapati, Arun and Jin, Yujia and Sidford, Aaron and Tian, Kevin},
  title     = {Semi-Streaming Bipartite Matching in Fewer Passes and Optimal Space},
  booktitle = {Proceedings of the 2022 {ACM-SIAM} Symposium on Discrete Algorithms (SODA)},
  pages     = {627--669},
  year      = {2022}
}

@inproceedings{AR20,
  author    = {Assadi, Sepehr and Raz, Ran},
  title     = {Near-Quadratic Lower Bounds for Two-Pass Graph Streaming Algorithms},
  booktitle = {Proceedings of the 61st IEEE Annual Symposium on Foundations of Computer Science (FOCS)},
  pages     = {342--353},
  year      = {2020}
}

@inproceedings{CFR,
  author    = {Cao, Nairen and Fineman, Jeremy T. and Russell, Katina},
  title     = {Efficient construction of directed hopsets and parallel approximate shortest paths},
  booktitle = {Proceedings of the 52nd Annual {ACM} Symposium on Theory of Computing (STOC)},
  pages     = {336--349},
  year      = {2020}
}

@inproceedings{CKPSSY21,
  author    = {Chen, Lijie and Kol, Gillat and Paramonov, Dmitry and Saxena, Raghuvansh R. and Song, Zhao and Yu, Huacheng},
  title     = {Almost Optimal Super-Constant-Pass Streaming Lower Bounds for Reachability},
  booktitle = {Proceedings of the 53rd Annual {ACM} Symposium on Theory of Computing (STOC)},
  pages     = {570--583},
  year      = {2021}
}

@article{feigenbaum2005graph,
  title     = {On graph problems in a semi-streaming model},
  author    = {Feigenbaum, Joan and Kannan, Sampath and McGregor, Andrew and Suri, Siddharth and Zhang, Jian},
  journal   = {Theoretical Computer Science},
  volume    = {348},
  number    = {2-3},
  pages     = {207--216},
  year      = {2005},
  publisher = {Elsevier}
}

@inproceedings{Fineman18,
  author    = {Fineman, Jeremy T.},
  title     = {Nearly Work-Efficient Parallel Algorithm for Digraph Reachability},
  booktitle = {Proceedings of the 50th Annual {ACM} Symposium on Theory of Computing (STOC)},
  pages     = {457--470},
  year      = {2018}
}

@article{GO16,
  author  = {Guruswami, Venkatesan and Onak, Krzysztof},
  title   = {Superlinear Lower Bounds for Multipass Graph Processing},
  journal = {Algorithmica},
  volume  = {76},
  number  = {3},
  pages   = {654--683},
  year    = {2016}
}

@techreport{HRR98,
  author      = {Henzinger, Monika R. and Raghavan, Prabhakar and Rajagopalan, Sridhar},
  title       = {Computing on Data Streams},
  institution = {Digital Equipment Corporation, Systems Research Center},
  number      = {SRC-TN-1998-011},
  address     = {Palo Alto, CA},
  year        = {1998}
}

@book{JaJa92,
  author    = {J{\'a}J{\'a}, Joseph},
  title     = {An Introduction to Parallel Algorithms},
  publisher = {Addison-Wesley},
  year      = {1992}
}

@inproceedings{JLS19,
  author    = {Jambulapati, Arun and Liu, Yang P. and Sidford, Aaron},
  title     = {Parallel Reachability in Almost Linear Work and Square Root Depth},
  booktitle = {Proceedings of the 60th IEEE Annual Symposium on Foundations of Computer Science (FOCS)},
  pages     = {1664--1686},
  year      = {2019}
}

@article{KS,
  author  = {Klein, Philip N. and Subramanian, Sairam},
  title   = {A randomized parallel algorithm for single-source shortest paths},
  journal = {Journal of Algorithms},
  volume  = {25},
  number  = {2},
  pages   = {205--220},
  year    = {1997}
}

@inproceedings{Rozhon,
  author    = {Rozho{\v{n}}, V{\'a}clav and Haeupler, Bernhard and Martinsson, Anders and Grunau, Christoph and Žužić, Goran},
  title     = {Parallel breadth-first search and exact shortest paths and stronger notions for approximate distances},
  booktitle = {Proceedings of the 55th Annual {ACM} Symposium on Theory of Computing (STOC)},
  pages     = {321--334},
  year      = {2023}
}

@article{Zwick,
  author  = {Zwick, Uri},
  title   = {All pairs shortest paths using bridging sets and rectangular matrix multiplication},
  journal = {Journal of the {ACM}},
  volume  = {49},
  number  = {3},
  pages   = {289--317},
  year    = {2002}
}

@article{UY91,
  author  = {Ullman, Jeffrey D. and Yannakakis, Mihalis},
  title   = {High-Probability Parallel Transitive-Closure Algorithms},
  journal = {SIAM Journal on Computing},
  volume  = {20},
  number  = {1},
  pages   = {100--125},
  year    = {1991}
}

@inproceedings{Hesse03,
  author    = {Hesse, William},
  title     = {Directed graphs requiring large numbers of shortcuts},
  booktitle = {Proceedings of the 14th Annual {ACM-SIAM} Symposium on Discrete Algorithms (SODA)},
  pages     = {665--669},
  year      = {2003}
}

@inproceedings{HP18,
  author    = {Huang, Shang-En and Pettie, Seth},
  title     = {Lower Bounds on Sparse Spanners, Emulators, and Diameter-reducing shortcuts},
  booktitle = {16th Scandinavian Symposium and Workshops on Algorithm Theory (SWAT)},
  pages     = {26:1--26:12},
  year      = {2018}
}

@inproceedings{KP22,
  author    = {Kogan, Shimon and Parter, Merav},
  title     = {New Diameter-Reducing Shortcuts and Directed Hopsets: Breaking the $\sqrt{n}$ Barrier},
  booktitle = {Proceedings of the 2022 Annual {ACM-SIAM} Symposium on Discrete Algorithms (SODA)},
  pages     = {1326--1341},
  year      = {2022}
}

@inproceedings{BW23,
  author    = {Bernstein, Aaron and Wein, Nicole},
  title     = {Closing the Gap Between Directed Hopsets and Shortcut Sets},
  booktitle = {Proceedings of the 2023 Annual {ACM-SIAM} Symposium on Discrete Algorithms (SODA)},
  pages     = {163--182},
  year      = {2023}
}

@inproceedings{BH23,
  author    = {Bodwin, Greg and Hoppenworth, Gary},
  title     = {Folklore Sampling is Optimal for Exact Hopsets: Confirming the $\sqrt{n}$ Barrier},
  booktitle = {Proceedings of the 64th IEEE Annual Symposium on Foundations of Computer Science (FOCS)},
  pages     = {701--720},
  year      = {2023}
}

@inproceedings{HXX25,
  author    = {Hoppenworth, Gary and Xu, Yinzhan and Xu, Zixuan},
  title     = {New Separations and Reductions for Directed Preservers and Hopsets},
  booktitle = {Proceedings of the 2025 Annual {ACM-SIAM} Symposium on Discrete Algorithms (SODA)},
  year      = {2025}
}

@article{Cohen00,
  author  = {Cohen, Edith},
  title   = {Polylog-time and near-linear work approximation scheme for undirected shortest paths},
  journal = {Journal of the {ACM}},
  volume  = {47},
  number  = {1},
  pages   = {132--166},
  year    = {2000}
}

@inproceedings{EN16,
  author    = {Elkin, Michael and Neiman, Ofer},
  title     = {Hopsets with Constant Hopbound, and Applications to Approximate Shortest Paths},
  booktitle = {Proceedings of the 57th IEEE Annual Symposium on Foundations of Computer Science (FOCS)},
  pages     = {128--137},
  year      = {2016}
}

@inproceedings{MPVX15,
  author    = {Miller, Gary L. and Peng, Richard and Vladu, Adrian and Xu, Shen Chen},
  title     = {Improved Parallel Algorithms for Spanners and Hopsets},
  booktitle = {Proceedings of the 27th {ACM} Symposium on Parallelism in Algorithms and Architectures (SPAA)},
  pages     = {192--201},
  year      = {2015}
}

@inproceedings{CF23,
  author    = {Cao, Nairen and Fineman, Jeremy T.},
  title     = {Parallel Exact Shortest Paths in Almost Linear Work and Square Root Depth},
  booktitle = {Proceedings of the 2023 Annual {ACM-SIAM} Symposium on Discrete Algorithms (SODA)},
  pages     = {4354--4372},
  year      = {2023}
}

@inproceedings{Tho92,
  author    = {Thorup, Mikkel},
  title     = {On shortcutting digraphs},
  booktitle = {Graph-Theoretic Concepts in Computer Science, 18th International Workshop (WG)},
  series    = {Lecture Notes in Computer Science},
  volume    = {657},
  pages     = {205--211},
  year      = {1992}
}

@inproceedings{LVWX22,
  author    = {Lu, Kevin and Vassilevska Williams, Virginia and Wein, Nicole and Xu, Zixuan},
  title     = {Better Lower Bounds for Shortcut Sets and Additive Spanners via an Improved Alternation Product},
  booktitle = {Proceedings of the 2022 Annual {ACM-SIAM} Symposium on Discrete Algorithms (SODA)},
  year      = {2022}
}

@inproceedings{VXX24,
  author    = {Vassilevska Williams, Virginia and Xu, Yinzhan and Xu, Zixuan},
  title     = {Simpler and Higher Lower Bounds for Shortcut Sets},
  booktitle = {Proceedings of the 2024 Annual {ACM-SIAM} Symposium on Discrete Algorithms (SODA)},
  year      = {2024}
}

@inproceedings{KP22a,
  author    = {Kogan, Shimon and Parter, Merav},
  title     = {Beating Matrix Multiplication for $n^{1/3}$-Directed Shortcuts},
  booktitle = {49th International Colloquium on Automata, Languages, and Programming (ICALP)},
  year      = {2022}
}

@inproceedings{KP23a,
  author    = {Kogan, Shimon and Parter, Merav},
  title     = {Faster and Unified Algorithms for Diameter Reducing Shortcuts and Minimum Chain Covers},
  booktitle = {Proceedings of the 2023 Annual {ACM-SIAM} Symposium on Discrete Algorithms (SODA)},
  pages     = {212--239},
  year      = {2023}
}

@article{McG14,
  author  = {McGregor, Andrew},
  title   = {Graph Stream Algorithms: A Survey},
  journal = {{SIGMOD} Record},
  volume  = {43},
  number  = {1},
  pages   = {9--20},
  year    = {2014}
}

@inproceedings{HKN16,
  author    = {Henzinger, Monika and Krinninger, Sebastian and Nanongkai, Danupon},
  title     = {A Deterministic Almost-Tight Distributed Algorithm for Approximating Single-Source Shortest Paths},
  booktitle = {Proceedings of the 48th Annual {ACM} Symposium on Theory of Computing (STOC)},
  pages     = {489--498},
  year      = {2016}
}

@inproceedings{Sch08,
  author    = {Schudy, Warren},
  title     = {Finding Strongly Connected Components in Parallel Using {$O(\log^2 n)$} Reachability Queries},
  booktitle = {Proceedings of the 20th Annual ACM Symposium on Parallelism in Algorithms and Architectures (SPAA)},
  pages     = {146--151},
  year      = {2008}
}

@misc{BDHNR26,
  author        = {Bals, Ben and Dadush, Daniel and Hoppenworth, Gary and Nazari, Yasamin and {Rao K.N.}, Rajath},
  title         = {Faster Directed Hopsets, Distance Preservers, and Deterministic Shortcut Sets},
  year          = {2026},
  eprint        = {2610.12057},
  archivePrefix = {arXiv},
  note          = {arXiv preprint arXiv:2610.12057}
}

@inproceedings{EM21,
  author    = {Elkin, Michael and Matar, Shaked},
  title     = {Deterministic {PRAM} Approximate Shortest Paths in Polylogarithmic Time and Slightly Super-Linear Work},
  booktitle = {Proceedings of the 33rd {ACM} Symposium on Parallelism in Algorithms and Architectures (SPAA)},
  pages     = {198--207},
  year      = {2021}
}

@inproceedings{HKN14,
  author    = {Henzinger, Monika and Krinninger, Sebastian and Nanongkai, Danupon},
  title     = {Decremental Single-Source Shortest Paths on Undirected Graphs in Near-Linear Total Update Time},
  booktitle = {Proceedings of the 55th IEEE Annual Symposium on Foundations of Computer Science (FOCS)},
  pages     = {146--155},
  year      = {2014}
}

@inproceedings{BGWN20,
  author    = {Bernstein, Aaron and Probst Gutenberg, Maximilian and Wulff-Nilsen, Christian},
  title     = {Near-Optimal Decremental {SSSP} in Dense Weighted Digraphs},
  booktitle = {Proceedings of the 61st IEEE Annual Symposium on Foundations of Computer Science (FOCS)},
  pages     = {1112--1122},
  year      = {2020}
}

@inproceedings{LN22,
  author    = {{\L}{\k{a}}cki, Jakub and Nazari, Yasamin},
  title     = {Near-Optimal Decremental Hopsets with Applications},
  booktitle = {49th International Colloquium on Automata, Languages, and Programming (ICALP)},
  pages     = {86:1--86:20},
  year      = {2022}
}

@inproceedings{Elk17,
  author    = {Elkin, Michael},
  title     = {Distributed Exact Shortest Paths in Sublinear Time},
  booktitle = {Proceedings of the 49th Annual {ACM} Symposium on Theory of Computing (STOC)},
  pages     = {757--770},
  year      = {2017}
}

@inproceedings{ET22,
  author    = {Elkin, Michael and Trehan, Chhaya},
  title     = {$(1+\varepsilon)$-Approximate Shortest Paths in Dynamic Streams},
  booktitle = {Approximation, Randomization, and Combinatorial Optimization. Algorithms and Techniques (APPROX/RANDOM)},
  pages     = {51:1--51:23},
  year      = {2022}
}

@inproceedings{Nan14,
  author    = {Nanongkai, Danupon},
  title     = {Distributed Approximation Algorithms for Weighted Shortest Paths},
  booktitle = {Proceedings of the 46th Annual {ACM} Symposium on Theory of Computing (STOC)},
  pages     = {565--573},
  year      = {2014}
}

@inproceedings{CDKL19,
  author    = {Censor-Hillel, Keren and Dory, Michal and Korhonen, Janne H. and Leitersdorf, Dean},
  title     = {Fast Approximate Shortest Paths in the Congested Clique},
  booktitle = {Proceedings of the 38th {ACM} Symposium on Principles of Distributed Computing (PODC)},
  pages     = {74--83},
  year      = {2019}
}

@inproceedings{Naz19,
  author    = {Nazari, Yasamin},
  title     = {Sparse Hopsets in Congested Clique},
  booktitle = {23rd International Conference on Principles of Distributed Systems (OPODIS)},
  pages     = {34:1--34:16},
  year      = {2019}
}

@inproceedings{DN19,
  author    = {Dinitz, Michael and Nazari, Yasamin},
  title     = {Massively Parallel Approximate Distance Sketches},
  booktitle = {23rd International Conference on Principles of Distributed Systems (OPODIS)},
  pages     = {35:1--35:17},
  year      = {2019}
}

@misc{OpenAI26,
  author       = {{OpenAI}},
  title        = {An Upper Bound of 9/4 for the Matrix Multiplication Exponent},
  year         = {2026},
  howpublished = {Preprint, October 2, 2026. \url{https://github.com/openai/math/blob/adc7f1241b42e322a6451854ab7e4b4c146bf78a/preprints/Matrix-Multiplication-Nine-Fourths-October-2-2026/paper.pdf}}
}

@misc{Shp26,
  author = {Shpilka, Amir},
  title  = {Exposition of {OpenAI}'s Matrix Multiplication in Exponent 9/4 (and an Explicit Algorithm)},
  year   = {2026},
  note   = {Electronic Colloquium on Computational Complexity (ECCC), Report TR26-281}
}

\appendix
\crefalias{section}{appendix}
\crefalias{subsection}{appendix}
\section{Exact SSSP from Approximate SSSP}
\label{sec:streaming-exact-reduction}

In this appendix, we prove \cref{thm:streaming-exact-reduction} and
\cref{lem:parallel-exact-reduction}, which reduce exact SSSP to
approximate SSSP in the streaming and parallel models. Both lemmas
adapt the reduction of~\cite{Rozhon}, which reduces exact SSSP to
$O(\log^2n)$ instances of $(1+\eps)$-approximate SSSP, where
$\eps=\Theta(1/\log n)$.

We cannot use the reduction of~\cite{Rozhon} as a black box, for two
reasons. First, the authors of~\cite{Rozhon} implement their reduction
only in the parallel and distributed
models~\cite[Corollaries~1.8 and~1.9]{Rozhon}, and we need it in the
semi-streaming model. Implementing it in the semi-streaming model
requires a look at the graphs in which it needs approximate distances,
but it is otherwise fairly standard. Second, and more importantly, the
reduction of~\cite{Rozhon} reduces exact SSSP to $(1+\eps)$-approximate
SSSP in graphs with zero and fractional edge weights. Our approximate SSSP algorithms
(\cref{thm:streaming-sssp} and \cref{prop:parallel-oracle}), on the
other hand, require positive edge weights. Accordingly,
\cref{thm:streaming-exact-reduction} and
\cref{lem:parallel-exact-reduction} only assume an approximate SSSP
algorithm for graphs with positive, polynomially bounded integer
weights.

Appendix~C.1 of~\cite{Rozhon} sketches how to modify their reduction
so that it only needs approximate distances in graphs with integer
weights, and we assume this modified version. However, the modified
reduction may still need approximate distances in graphs with zero
weights. (This was not a problem for the authors of~\cite{Rozhon},
since the parallel algorithm of~\cite{CFR} that they use allows zero
weights.) We make a few fairly minor
changes to the reduction of~\cite{Rozhon} that remove the zero
weights.

We first summarize the reduction of~\cite{Rozhon}, and then we
describe our changes to it. In the last two subsections, we implement
the resulting algorithm in the streaming and parallel models.

\subsection*{The reduction of \texorpdfstring{\cite{Rozhon}}{[RHM+23]}}

In this subsection, we summarize the reduction of~\cite{Rozhon} and
describe the graphs in which it needs approximate distances. The
summary is condensed, and we refer to~\cite[Sections~3.1
and~5.2]{Rozhon} for the details. Let $G=(V,E,w)$ be a directed graph
with nonnegative, polynomially bounded integer weights and a source
$s$, and assume that every vertex is reachable from $s$. As usual, we
describe the reduction in terms of an \emph{oracle} that returns
$(1+\eps)$-approximate distances from a given source in a given
directed graph. We call the graphs in which the reduction queries the
oracle the \emph{oracle graphs}. The reduction has two levels. An outer loop
computes exact distances from smoothly approximate distances, which we
define below, and the procedure \textsc{Smoothing}
of~\cite[Algorithm~3]{Rozhon} computes smoothly approximate distances
with the oracle.

\paragraph{Smoothly approximate distances and potentials.}
A distance estimate $y\colon V\to\mathbb R_{\ge0}$ is \emph{smoothly
$2$-approximate} from $s$~\cite[Definition~1.6]{Rozhon} if $y(s)=0$ and
\[
 \dist_G(s,v)\le y(v)
 \quad\text{and}\quad
 y(v)-y(u)\le2\dist_G(u,v)
 \qquad\text{for all }u,v\in V.
\]
For $u=s$, the second condition says that $y(v)\le2\dist_G(s,v)$, so
$y$ is a $2$-approximation of the distances from $s$. The second
condition, called \emph{smoothness}, also bounds the difference
between the estimates of any two vertices by twice the distance
between them.

A \emph{potential} is a function $\pi\colon V\to\mathbb R$ with
$\pi(s)=0$ such that the \emph{reweighted} edge weights
\[
 w_\pi(u,v)=w(u,v)+\pi(u)-\pi(v)
\]
are nonnegative for all edges $(u,v)\in E$~\cite[Section~4.1]{Rozhon}.
Let $G^\pi=(V,E,w_\pi)$ denote the reweighted graph. Along a path from
$s$ to $v$, the terms $\pi(u)-\pi(v)$ telescope, so the length of the
path in $G^\pi$ is its length in $G$ minus $\pi(v)$. Hence $G^\pi$ has
the same shortest paths as $G$, and
$\dist_{G^\pi}(s,v)=\dist_G(s,v)-\pi(v)$.

\paragraph{The outer loop.}
The outer loop of~\cite[Algorithm~1]{Rozhon} is due to Klein and
Subramanian~\cite{KS}. It maintains a potential $\pi$, starting with
$\pi=0$. In each of $I=O(\log n)$ iterations, it computes a smoothly
$2$-approximate distance estimate $y$ from $s$ in $G^\pi$ with
\textsc{Smoothing}, and then it adds $y/2$ to $\pi$. By smoothness, we
have $y(v)-y(u)\le2w_\pi(u,v)$ for each edge $(u,v)$, so the new
reweighted weights $w_\pi(u,v)+y(u)/2-y(v)/2$ are nonnegative and
$\pi$ remains a potential. As $y(v)\ge\dist_{G^\pi}(s,v)$, one
iteration at least halves the distance from $s$ to $v$ in $G^\pi$.
The initial distances are polynomially bounded, so after $I=O(\log n)$
iterations, all of them are less than $1/3$. Then
$\dist_G(s,v)=\pi(v)+\dist_{G^\pi}(s,v)$ is the integer nearest to
$\pi(v)$~\cite[Lemma~3.1]{Rozhon}.

\paragraph{Smoothing.}
By~\cite[Theorem~5.1]{Rozhon}, the procedure \textsc{Smoothing}
computes the estimate $y$ with $O(\log n)$ oracle calls. Its first call
asks for $2$-approximate distances from $s$ in $G^\pi$. The resulting
estimate is a $2$-approximation, but it need not be smooth, since the
oracle can overestimate the distances to two nearby vertices by very
different amounts. \textsc{Smoothing} improves this estimate in
$t=O(\log n)$ rounds of the procedure \textsc{PartialSmoothing}
of~\cite[Algorithm~4]{Rozhon}. At the start of each round, the current
estimate $d$ is smooth up to a multiplicative factor and an additive
error~\cite[Definition~5.2]{Rozhon}. That is, for all $u,v\in V$, the difference $d(v)-d(u)$ is at
most this factor times $\dist_{G^\pi}(u,v)$ plus the additive error.
For the first estimate, the factor is $1$ and the additive error is
polynomially bounded. Each round halves the additive error and multiplies the factor by
$1+O(1/\log n)$~\cite[Proposition~5.12]{Rozhon}. After $t$ rounds, the
estimate is smoothly $2$-approximate, and \textsc{Smoothing} returns
it as $y$.

Each round of \textsc{PartialSmoothing} makes two oracle calls, on the
\emph{level graphs} $H_1$ and $H_2$ of~\cite[Definition~5.7]{Rozhon}.
To define them, we partition the real line into intervals whose length
is $\Theta(\log n)$ times the current additive error. For $H_2$, these
intervals are shifted by half their length. The level graph $H_i$ has
the vertex set $V\cup\{\sigma\}$, where $\sigma$ is a new vertex that
serves as the source of the oracle call. It contains each edge
$(u,v)\in E$ for which $d(u)$ and $d(v)$ lie in the same interval,
with its weight $w_\pi(u,v)$ multiplied by a factor that depends only
on the round. For each $u\in V$, it also contains an edge
$(\sigma,u)$, whose weight is the distance from the left endpoint of
the interval containing $d(u)$ to $d(u)$. Adding the left endpoint of
the interval containing $d(v)$ to the distance from $\sigma$ to $v$ in
$H_i$ thus gives the minimum, over all vertices $u\in V$, of $d(u)$
plus the distance from $u$ to $v$ in $H_i$ without $\sigma$.
\textsc{PartialSmoothing} computes this value approximately with the
oracle for $i=1,2$, and the new estimate of $v$ is the minimum of the
two results and $d(v)$. See~\cite[Section~5.1]{Rozhon} for why this
improves the smoothness of $d$.

\paragraph{Integer weights.}
The oracle graphs above can have fractional weights, since the
potential $\pi$ and the estimates $d$ can be fractional.
Appendix~C.1 of~\cite{Rozhon} sketches how to modify the reduction so
that every oracle graph has nonnegative, polynomially bounded integer
weights. The modification adds $1/\operatorname{poly}(n)$ to the
weight of each edge of an oracle graph, and then it scales and
discretizes these weights. The sketch only guarantees nonnegative
weights, and it does not specify the discretization, which may round
small weights down to zero. We therefore do not assume that the oracle
graphs have positive weights. We assume this version of the reduction.
With $I=O(\log n)$ iterations of the outer loop and $O(\log n)$ oracle
calls in each of them, it gives the following proposition.

\begin{proposition}[{\cite[Theorem~5.1, Lemma~3.1, and
Appendix~C.1]{Rozhon}}]
\label{prop:rhm-reduction}
Let $a\ge1$ be a constant. There is an algorithm that, given an
$n$-vertex directed graph $G$ with integer weights in
$\{0,1,\dots,n^a\}$ and a source $s$ from which every vertex is
reachable, computes the distances from $s$ in $G$ using $O(\log^2n)$
oracle calls. Each call asks for $(1+\eps)$-approximate distances from
a source in a directed graph with at most $n+1$ vertices and
nonnegative integer weights bounded by $n^{O(a)}$, for some
$\eps=\Theta(1/\log n)$ whose constant depends only on $a$.
\end{proposition}

\paragraph{Oracle graphs.}
Each oracle graph is a reweighted graph $G^\pi$ or a level graph, with
its weights modified edge by edge as in~\cite[Appendix~C.1]{Rozhon}.
Our implementations in the streaming and parallel models use two
features of the algorithm of \cref{prop:rhm-reduction}, which follow
from the summary above.
\begin{itemize}
 \item[(F1)] The vertex set of an oracle graph is $V$ or
 $V\cup\{\sigma\}$, and the graph contains at most one edge for each
 edge $(u,v)\in E$. Whether it contains this edge, and with which
 weight, depends only on $w(u,v)$, on the values $\pi(u)$, $\pi(v)$,
 $d(u)$ and $d(v)$ stored at $u$ and $v$, and on $O(1)$ global
 parameters, such as the interval length and the factor of the
 current round. All other edges of the oracle graph leave $\sigma$.
 \item[(F2)] Between oracle calls, the algorithm only updates $O(1)$
 vectors indexed by $V$, such as $\pi$ and $d$, and the $O(1)$ global
 parameters. It computes them from $n$, $a$ and the answers of the
 oracle, without reading $G$. The stored values are polynomially
 bounded multiples of $1/\operatorname{poly}(n)$, as in the
 discretization of~\cite[Appendix~C.1]{Rozhon}, so each of them takes
 $O(\log n)$ bits.
\end{itemize}
Before the modification of~\cite[Appendix~C.1]{Rozhon}, the oracle
graphs can have edges of weight zero. For example, an edge
$(\sigma,u)$ of a level graph has weight zero if $d(u)$ is the left
endpoint of its interval.

\subsection*{Our variant of the reduction}

Our approximate algorithms require positive weights, so they cannot
answer the oracle calls of \cref{prop:rhm-reduction} on graphs with
zero weights. \Cref{alg:exact-sssp} shows how we modify the reduction
to avoid such calls. It makes two changes to the algorithm of
\cref{prop:rhm-reduction}. First, it answers each oracle call with the
procedure \textsc{PositiveOracle}, which calls our approximate
algorithm on a graph with positive integer weights. Second, it runs
the reduction not on $G$ but on a graph $G^+$ with positive weights in
which every vertex is reachable from $s$. The positive weights of $G^+$
let us recover a shortest-path tree from the distances. We explain the
two changes in turn.

\begin{algorithm}[htb]
\caption{Our variant of the reduction of~\cite{Rozhon}. The input
graph $G$ has integer weights in $\{0,1,\dots,n^a\}$ and largest weight
$w_{\max}$, and $s$ is the source. \textsc{PositiveOracle} answers an
oracle call that asks for $(1+\eps)$-approximate distances from a
source $s_A$ in a graph $A$ with $n_A$ vertices and nonnegative
integer weights $w_A$.}
\label{alg:exact-sssp}
\begin{algorithmic}[1]
\Procedure{ExactSSSP}{$G,s$}
 \State $w_\infty\gets n(nw_{\max}+1)$
 \State $G^+\gets G$ with weights $w^+(u,v)=n\,w(u,v)+1$, plus an edge
 $(s,v)$ of weight $w_\infty$ for each $v\ne s$
 \State Compute $\dist_{G^+}(s,v)$ for all $v$ with the algorithm of
 \cref{prop:rhm-reduction}, answering each oracle call $(A,s_A,\eps)$
 by \Call{PositiveOracle}{$A,s_A,\eps$}
 \State $V_s\gets\{v\in V:\dist_{G^+}(s,v)<w_\infty\}$
 \Comment{the vertices reachable from $s$ in $G$}
 \State For each $v\in V_s\setminus\{s\}$, choose a parent edge
 $(u,v)\in E$ with $\dist_{G^+}(s,u)+w^+(u,v)=\dist_{G^+}(s,v)$
 \State \Return $\dist_G(s,v)=\lfloor\dist_{G^+}(s,v)/n\rfloor$ for
 $v\in V_s$, $\dist_G(s,v)=\infty$ for $v\notin V_s$, and the parent
 edges
\EndProcedure
\Statex
\Procedure{PositiveOracle}{$A,s_A,\eps$}
 \State $\theta\gets\lceil3n_A/\eps\rceil$, and $A^+\gets A$ with
 weights $\theta w_A+1$
 \State $\widehat d^+\gets(1+\eps/3)$-approximate distances from
 $s_A$ in $A^+$ \Comment{our approximate algorithm}
 \State \Return $\widehat d(v)=0$ if $\widehat d^+(v)<\theta$, and
 $\widehat d(v)=\widehat d^+(v)/\theta$ otherwise
\EndProcedure
\end{algorithmic}
\end{algorithm}

\paragraph{Answering oracle calls with positive weights.}
Consider an oracle call that asks for $(1+\eps)$-approximate distances
from a source $s_A$ in a graph $A$ with $n_A$ vertices and
nonnegative integer weights $w_A$, where $\eps\le1$. The obvious way to
remove zero weights is to add a small amount $\eta>0$ to each edge
weight. This increases the distance to each vertex by less than
$n_A\eta$. However, a vertex $v$ with $\dist_A(s_A,v)=0$ may then
receive a positive estimate, which is not a $(1+\eps)$-approximation
of zero. We avoid this problem by using that the distances in $A$ are
integers. \textsc{PositiveOracle} first multiplies all weights by the
large integer $\theta=\lceil3n_A/\eps\rceil$, and then it adds one to
them, which gives the graph $A^+$ with weights $\theta w_A+1$. On a
simple path, the added ones contribute less than $n_A\le\eps\theta/3$
to the length in $A^+$. A vertex at distance $\ell$ from $s_A$ in $A$
is therefore at distance between $\theta\ell$ and $\theta\ell+n_A$ in
$A^+$. The estimates of the vertices at distance zero are thus less
than $\theta$, while all other estimates are at least $\theta$, so we
can recognize the vertices at distance zero.

\begin{lemma}
\label{lem:exact-positive-oracle}
Let $A$ be a directed graph with $n_A$ vertices and nonnegative
integer weights, let $s_A$ be a vertex of $A$, and let
$0<\eps\le1$. If the approximate algorithm in
$\textnormal{\textsc{PositiveOracle}}(A,s_A,\eps)$ returns
$(1+\eps/3)$-approximate distances from $s_A$ in $A^+$, then
\textnormal{\textsc{PositiveOracle}} returns $(1+\eps)$-approximate distances
from $s_A$ in $A$.
\end{lemma}

\begin{proof}
Let $v$ be a vertex of $A$. If $s_A$ does not reach $v$, then both
distances and both estimates of $v$ are infinite. Otherwise, let
$\ell=\dist_A(s_A,v)$. Each path $P$ from $s_A$ to $v$ has length
$\theta w_A(P)+|P|\ge\theta\ell$ in $A^+$, where $|P|$ is the hop length
of $P$. A shortest simple path from $s_A$ to $v$ in $A$ has hop
length less than $n_A$. Hence
\[
 \theta\ell\le\dist_{A^+}(s_A,v)<\theta\ell+n_A,
\]
and the estimate of the approximate algorithm satisfies
$\theta\ell\le\widehat d^+(v)<(1+\eps/3)(\theta\ell+n_A)$. If $\ell=0$, then
$\widehat d^+(v)<(4/3)n_A<\theta$, as $\eps\le1$, so we return $\widehat d(v)=0$. If
$\ell\ge1$, then $\widehat d^+(v)\ge\theta$, and we return
$\widehat d(v)=\widehat d^+(v)/\theta$. Using $n_A/\theta\le\eps/3\le
\eps\ell/3$, we get
\[
 \ell\le\widehat d(v)<(1+\eps/3)(\ell+\eps\ell/3)=(1+\eps/3)^2\ell\le(1+\eps)\ell,
\]
where the last step again uses $\eps\le1$.
\end{proof}

\paragraph{Preprocessing the input graph.}
\Cref{prop:rhm-reduction} assumes that every vertex is reachable from
$s$, and it computes only distances. If all weights are positive, then
the distances determine a shortest-path tree. For a vertex $v\ne s$
that is reachable from $s$, we can then choose any edge $(u,v)$ with
$\dist_G(s,u)+w(u,v)=\dist_G(s,v)$ as the parent edge of $v$, since the distance
from $s$ strictly decreases as we follow parent edges backwards from
$v$. With zero weights, such edges can form a cycle.
\Cref{alg:exact-sssp} therefore first replaces each weight $w(u,v)$ by
the positive weight $w^+(u,v)=n\,w(u,v)+1$. A simple path $P$ has hop
length $|P|<n$. Its new length $w^+(P)=n\,w(P)+|P|$ thus determines its
old length as $w(P)=\lfloor w^+(P)/n\rfloor$, and a shortest path from
$s$ under $w^+$ is also a shortest path under $w$. The algorithm also
adds an edge $(s,v)$ of weight $w_\infty=n(nw_{\max}+1)$ for each
vertex $v\ne s$, where $w_{\max}$ is the largest weight of $G$. Every
simple path of $G$ has length less than $w_\infty$ under $w^+$, so the
weight $w_\infty$ acts as an infinite weight. The new edges make every
vertex reachable from $s$ without changing the distances to the
vertices that are reachable from $s$ in $G$. Let $G^+$ denote the
resulting graph.

The following lemma shows that \cref{alg:exact-sssp} is correct if
all calls of our approximate algorithm succeed. It also records the
graphs on which these calls are made.

\begin{lemma}
\label{lem:exact-positive-weights}
Let $a\ge1$ be a constant, and let $\eps=\Theta(1/\log n)$ be the
accuracy of \cref{prop:rhm-reduction} for the constant $a+3$. Suppose
that $G$ has integer weights in $\{0,1,\dots,n^a\}$, and that every
call of the approximate algorithm in \textnormal{\textsc{PositiveOracle}} returns
$(1+\eps/3)$-approximate distances. Then \cref{alg:exact-sssp} outputs
$\dist_G(s,v)$ for all vertices $v$ and a shortest-path tree rooted at
$s$. It makes $O(\log^2n)$ calls of the approximate algorithm, each on
a graph with at most $n+1$ vertices, at most $m+2n$ edges, and positive
integer weights bounded by $n^{O(a)}$.
\end{lemma}

\begin{proof}
The graph $G^+$ has integer weights between $1$ and
$w_\infty\le n^{a+3}$, and every vertex is reachable from $s$ in
$G^+$. By \cref{lem:exact-positive-oracle}, \textsc{PositiveOracle}
answers each oracle call with the required accuracy. Thus the
algorithm of \cref{prop:rhm-reduction} for the constant $a+3$ computes
$\dist_{G^+}(s,v)$ for all $v$ with $O(\log^2n)$ oracle calls. Each of
them makes one call of the approximate algorithm on a graph $A^+$ with
at most $n+1$ vertices. By (F1), the graph $A^+$ has at most one edge
for each of the $m+n-1$ edges of $G^+$, plus at most $n$ edges out
of $\sigma$.
Its weights are positive integers bounded by
$\theta n^{O(a)}+1=n^{O(a)}$, as $\theta=O(n\log n)$.

It remains to show that the output is correct. Every simple path $P$
of $G$ has length $w^+(P)=n\,w(P)+|P|\le(n-1)(nw_{\max}+1)<w_\infty$,
while a path of $G^+$ that uses a new edge $(s,v)$ has length at
least $w_\infty$. The vertex $s$ has $\dist_{G^+}(s,s)=0$, so $s\in V_s$
and we output $\dist_G(s,s)=0$. Consider a vertex $v\ne s$. If $v$ is not reachable
from $s$ in $G$, then every path from $s$ to $v$ in $G^+$ uses a new
edge. Hence $\dist_{G^+}(s,v)\ge w_\infty$, and we output
$\dist_G(s,v)=\infty$. Otherwise, $\dist_{G^+}(s,v)$ is the minimum of
$n\,w(P)+|P|$ over all simple paths $P$ from $s$ to $v$ in $G$. As
$0\le|P|<n$, a path $P$ with $w(P)>\dist_G(s,v)$ has
$n\,w(P)+|P|\ge n\dist_G(s,v)+n$, which is larger than the value of a
shortest simple path. The minimum is therefore attained by a path with
$w(P)=\dist_G(s,v)$, and $\lfloor\dist_{G^+}(s,v)/n\rfloor=\dist_G(s,v)$.
In particular, $\dist_{G^+}(s,v)<w_\infty$, so $V_s$ is exactly the set
of vertices reachable from $s$ in $G$.

Finally, let $v\in V_s\setminus\{s\}$. The last edge $(u,v)$ of a
shortest path from $s$ to $v$ in $G^+$ is not a new edge, since this
path has length less than $w_\infty$. Hence $(u,v)\in E$ is a valid
parent edge, so the algorithm can choose a parent edge for $v$. For
the chosen parent edge $(u,v)$, we have
$\dist_{G^+}(s,u)<\dist_{G^+}(s,v)$, as $w^+(u,v)\ge1$, and thus $u\in
V_s$. Following parent edges backwards from $v$ therefore gives a
simple path $Q$ of $G$ from $s$ to $v$. Its length is
$w^+(Q)=\dist_{G^+}(s,v)$, as the equalities of the parent edges
telescope. Hence $w(Q)=\lfloor w^+(Q)/n\rfloor=\dist_G(s,v)$, and the
parent edges form a shortest-path tree rooted at $s$.
\end{proof}

In the next two subsections, we answer the calls of the approximate
algorithm with our streaming and parallel algorithms.

\subsection*{The streaming model}

We now implement \cref{alg:exact-sssp} in the streaming model. This
proves the following lemma from \cref{sec:streaming-reductions}.

\exactreduction*

\begin{proof}
Let $\eps=\Theta(1/\log n)$ be the accuracy of \cref{prop:rhm-reduction}
for the constant $a+3$, and choose the constant $c=c(a)>0$ such that
$c/\log n\le\eps/3$. We run \cref{alg:exact-sssp} and answer each call
of the approximate algorithm in \textsc{PositiveOracle} with the given
streaming algorithm, using fresh random bits. Its estimates are
$(1+c/\log n)$-approximate and hence $(1+\eps/3)$-approximate. By
\cref{lem:exact-positive-weights}, the calls are on graphs with $O(n)$
vertices and positive, polynomially bounded integer weights, so the
streaming algorithm can answer them. The same lemma shows that
\cref{alg:exact-sssp} computes the distances from $s$ and a
shortest-path tree if all calls succeed.

Next, we bound the space and the number of passes. We store the $O(1)$
vectors indexed by $V$ and the $O(1)$ global parameters of (F2), which
takes $\widetilde O(n)$ space and no passes. We
run the oracle calls one at a time. During a call, the streaming
algorithm reads the graph $A^+$ as a stream, and we simulate each pass
over $A^+$ by one pass over the input stream. By (F1), each edge
$(u,v)$ of $G$
gives at most one edge of $A$, and whether it does and with which
weight depends only on $w(u,v)$, the values stored at $u$ and $v$, and
$O(1)$ stored parameters. The weight of the corresponding edge of
$A^+$ follows from its weight in $A$. Thus when we read $(u,v)$ from
the stream, we compute the corresponding edge of $A^+$ and pass it to
the streaming algorithm. The at most $2n$ other edges of $A^+$ leave
$s$ or $\sigma$, and we keep them in memory. Hence each call uses
$\widetilde O(n)$ space and $T_{\rm apx}$ passes. With $O(\log^2n)$
calls, plus one pass to find $w_{\max}$ and one pass to choose the
parent edges, we use $\widetilde O(n)$ space and
$O(T_{\rm apx}\log^2n)$ passes.

Finally, suppose that the approximate algorithm succeeds with high
probability. Then each call succeeds with high probability, given the
outcomes of the earlier calls. By a union bound over the $O(\log^2n)$
calls, all of them succeed with high probability.
\end{proof}

\subsection*{The parallel model}

The authors of~\cite{Rozhon} already implement their reduction in the
parallel model, where~\cite[Corollary~1.8]{Rozhon} answers the oracle calls
with the algorithm of~\cite{CFR}. We nevertheless need
\cref{lem:parallel-exact-reduction}, since our parallel approximate
algorithm (\cref{prop:parallel-oracle}) also requires positive weights.
We therefore use \cref{alg:exact-sssp} in the parallel model as well,
which proves the following lemma from \cref{sec:parallel-exact}.

\parallelreduction*

\begin{proof}
We run \cref{alg:exact-sssp} with $c$ chosen as in the proof of
\cref{thm:streaming-exact-reduction}, and we answer each call of the
approximate algorithm with the given parallel algorithm. By
\cref{lem:exact-positive-weights}, the $O(\log^2n)$ calls are on
graphs with $O(n)$ vertices, $O(m)$ edges (as $m\ge n$) and positive, polynomially
bounded integer weights, so each of them takes $W_{\rm apx}$ work and
$D_{\rm apx}$ depth. By (F1), we can build each graph $A^+$ from $G$ and the
stored vectors with $O(m)$ work and $O(1)$ depth. Updating the stored
vectors between calls, building $G^+$, and choosing the parent edges
take $\widetilde O(m)$ work and $\polylog n$ depth. In total, we use
$\widetilde O(W_{\rm apx}+m)$ work and $\widetilde O(D_{\rm apx})$
depth. The failure probability is bounded as in the proof of
\cref{thm:streaming-exact-reduction}.
\end{proof}

\end{document}